\documentclass[12pt]{iopart}
\usepackage{iopams}  

\expandafter\let\csname equation*\endcsname\relax

\expandafter\let\csname endequation*\endcsname\relax
\usepackage{amsmath}
\usepackage{graphicx}
\usepackage{fancyhdr}
\usepackage{natbib}

\usepackage{footnote}
\usepackage{threeparttable}

\usepackage{graphicx}  
\usepackage{float}
\usepackage{url}

\usepackage{braket}
\newcommand{\updownarrows}{\mathbin\uparrow\hspace{-.3em}\downarrow}
\newcommand{\downuparrows}{\mathbin\downarrow\hspace{-.3em}\uparrow}
\renewcommand{\upuparrows}{\mathbin\uparrow\hspace{-.3em}\uparrow}
\renewcommand{\downdownarrows}{\mathbin\downarrow\hspace{-.3em}\downarrow}

\usepackage{wasysym} 

\begin{document}\sloppy

\title[Trapped Rydberg ions]{Trapped Rydberg ions: a new platform for quantum information processing}

\author{A. Mokhberi$^{1}$, M. Hennrich$^{2}$, F. Schmidt-Kaler$^{1, 3}$}
\address{$^{1}$QUANTUM, Institut f\"ur Physik, Johannes Gutenberg-Universit\"at Mainz, Staudingerweg 7, 55128 Mainz, Germany\\
$^{2}$Department of Physics, Stockholm University, SE-106 91 Stockholm, Sweden\\
$^{3}$Helmholtz-Institut Mainz, Staudingerweg 18, 55128 Mainz, Germany}

\date{\today}
\ead{\mailto {arezoo.mokhberi@uni-mainz.de}, \mailto {fsk@uni-mainz.de}}

\begin{abstract}
In this chapter, we present an overview of experiments with trapped Rydberg ions and outline the advantages and challenges of developing applications of this new platform for quantum computing, sensing and simulation. Trapped Rydberg ions feature several important properties, unique in their combination: they are tightly bound in a harmonic potential of a Paul trap, in which their internal and external degrees of freedom can be controlled in a precise fashion. High fidelity state preparation of both internal and motional states of the ions has been demonstrated, and the internal states have been employed to store and manipulate qubit information. Furthermore, strong dipolar interactions can be realised between ions in Rydberg states and be explored for investigating correlated many-body systems. 
By laser coupling to Rydberg states, the polarisability of the ions can be both enhanced and tuned. This can be used to control the interactions with the trapping fields in a Paul trap as well as dipolar interactions between the ions. Thus, trapped Rydberg ions present an attractive alternative for fast entangling operations as compared to those mediated by normal modes of trapped ions, which are advantageous for a future quantum computer or a quantum simulator.  
\end{abstract}

{\pagestyle{plain}
\tableofcontents
\cleardoublepage}

\maketitle

\section{Introduction}\label{sec:intro}
Cold atoms and ions are currently attracting great interest for {\bf applications in quantum information processing (QIP), quantum simulation and sensing}. Trapped Rydberg ions offer a unique opportunity for combining advantages of precisely controllable trapped-ion qubits~\citep{haeffner08a} with long-range and tunable Rydberg interactions~\citep{saffman10a}. 
One decade after the first theoretical proposal for exploring Rydberg ions as a new platform for such applications~\citep{mueller08a}, experimental work has succeeded in demonstrating major milestones. These include the first excitation of trapped Ca$^{+}$ ions to Rydberg states in a radiofrequency (RF) ion trap~\citep{feldker15a}, the observation and characterisation of the trapping fields' effects on Rydberg ions and spectroscopy of Rydberg transitions for Sr$^{+}$~\citep{higgins17a, higgins19a} and Ca$^{+}$~ions~\citep{bachor16a, mokhberi19a}. Further progress has been achieved by coherent manipulation of a Rydberg state of single trapped Sr$^{+}$ ion~\citep{higgins17b, higgins19a}. More recently, sub-microsecond entangling operations have been demonstrated for two Rydberg ions using strong dipolar interactions between them~\citep{zhang20a}. Correspondingly, several theory proposals have been published that take benefit from the spectacular features of trapped Rydberg ions.

\subsection{Trapped ions for quantum technology experiments and for fundamental studies}\label{sec:introI}
Today, trapped ions are one of the most promising physical systems for QIP~\citep{bermudez17a, blatt08a}. Following the pioneering ideas of Cirac and Zoller for {\bf quantum computing based on trapped-ion qubits}~\citep{cirac95a}, quantum gates and building blocks of a quantum processor based on trapped ions have been realised~\citep{bruzewicz19a, roos08a, schmidtkaler03a}.
The fidelity for two-ion logic gate operations at this time is better than few parts in thousand~\citep{ballance16a, gaebler16a}, which meets one of the requirements for a future quantum computer, according to the DiVincenzo criteria~\citep{divincenzo00a,knill05a}. Multi-ion entanglement~\citep{friis18a, kaufmann17a, monz11a}, the Deutsch Josza quantum algorithm~\citep{gulde03a, olmschenk09a}, teleportation~\citep{riebe04a}, free-programmable gate sequences~\citep{figgatt17a} and quantum error correction~\citep{schindler11a} have been experimentally demonstrated.
Quantum gate operations are typically implemented using spin-dependent light forces or magnetic gradients in combination with excitation of collective motional modes of the ion crystal. The challenge is to scale up this architecture; as the number of ions in the common confining potential well grows, it becomes increasingly difficult to address a single vibrational mode without parasitic coupling to other modes, and hence the fidelity of gates decreases. To address this issue, two approaches are currently pursued. Wineland proposed scalable QIP using a ``quantum-CCD''~\citep{kielpinski02a}. This is a trapping device in which ions are shuttled around in a large microfabricated array of traps~\citep{blakestad09a} such that quantum gate operations are excecuted sequentially, but only on a few ions in a common potential. This approach has potential for fault-tolerant QIP~\citep{bermudez17a}. An alternative is a modular architecture of small traps connected via photonic links for establishing the long-range entanglement~\citep{monroe14a}.

One of the key prerequisites for both approaches is to {\bf implement fast logic gates} such that many operations are possible within the memory coherence time of the qubit register without loss in gate fidelity. In the quantum-CCD approach, additionally the shuttling time needed for reconfiguration operations has to be in addition taken into account~\citep{kaufmann17b}. Gates need to be robust against fluctuations of experimental parameters including electric fields, magnetic fields, laser field intensities and laser field frequencies. This is a daunting, but achieveable task as demonstrated using ultra-fast laser pulses~\citep{wongcampos17a}, amplitude-modulated high-intensity laser beams~\citep{schaefer18a} and mixed-frequency laser pulses~\citep{shapira18a}. 

In this context, {\bf trapped  Rydberg ions offer an exciting alternative for fast gate operations} based on their strong and tunable dipolar interactions. Rydberg mediated gate operations and collective encoding of multi-qubits have been implemented in ultracold neutral gases~\citep{gross17a, hofstetter02a, labuhn16a, saffman10a}. Entangling operations in neutral atomic gases are preferably performed using the Rydberg blockade mechanism~\citep{jaksch00a, lukin01a, saffman05a}, as demonstrated in pioneering experiments for pairs of single atoms individually held in optical tweezers~\citep{isenhower10a, wilk10a}. For trapped ions, the Rydberg blockade mechanism has been recently demonstrated and employed for implementing a sub-microsecond entangling operation~\citep{zhang20a}. Theoretical studies propose using Rydberg ions for shaping the spectrum of motional modes in a linear crystal of trapped ions and for parallel execution of quantum gates~\citep{li13a, li14a}. For implementing quantum logic operations within a few nanoseconds, a protocol has been proposed that uses impulsive electric pulses to shuttle ions~\citep{walther12a} while Rydberg states undergo a state-dependent force. Because of the large polarisability of Rydberg states, these kicking forces gives rise to geometric phases that are used for a controlled phase gate between two ions~\citep{vogel19a}. 
 
In {\bf quantum simulation}, a well-controlled multi-particle quantum system mimics the behaviour of complex solid state, high-energy or biological system. Some of the most advanced experimental demonstrations for quantum simulations have been realised in trapped ion crystals~\citep{blatt12a} and in cold neutral gases~\citep{bloch12a}. The former system usually employs the Coulomb interation between ions~\citep{zhang17a} whereas the latter takes advantage of dipolar interactions between Rydberg atoms~\citep{weimer10a}. Rydberg atoms confined in optical lattices in linear and two-dimensional arrays~\citep{bernien17a, gross17a, labuhn16a} are known as flexible platform for quantum simulation. Experimental advances in neutral atomic gases include the application of defect-free two-dimensional arrays~\citep{ohl19a}, arbitrarily-shaped three-dimensional arrays~\citep{barredo18a} and the use of a quantum gas microscope that allows for the manipulation of a large number of spin systems with single-site resolution~\citep{schauss15a}. In a system of Rydberg ions, the advantages of both systems merge to offer possibilities for novel experiments~\citep{mueller08a, nath15a}. 
The coupling between electronic and vibrational degrees of freedom of atoms in such optical lattices can be engineered and used to tailor exotic multibody interactions~\citep{gambetta20a}.

Furthermore, Rydberg dipolar interactions between atoms and ions are powerful tools for {\bf exploring strongly correlated many-body systems}, novel quantum phases~\citep{gambetta19a, henkel10a, pupillo10a, weimer08a} and coherent dynamics in such systems~\citep{ lesanovsky10a, olmos10a, weimer08a}. Symmetry-breaking mechanisms in non-equilibrium systems transversing a phase transition have been theoretically and experimentally explored using trapped ions~\citep{delcampo10a, fishman08a, jurcevic17a, pyka13a, retzker08a, ulm13a} and cold atoms in optical lattices~\citep{bernien17a, keesling19a}.     
On the other hand, a symmetry-protected phase transition has recently been demonstrated using precisely engineered Rydberg interactions between ultracold neutral atoms in optical tweezers~\citep{leseleuc19a}. 
Rydberg excitation of a single ion in a linear ion array modifies the ion motional mode spectrum mainly because of the large polarisability of Rydberg states. This effect can be used to trigger structural phase transitions in a coherent fashion~\citep{bultrusch11a, li12a} and to observe quantum signatures like superposition in arrays of cold trapped ions~\citep{bultrusch12a, silvi16a}. 

{\bf Tuning the interactions between the qubits of trapped Rydberg ions} uses both the Coulomb and the Rydberg dipolar interactions and in addition the inter-particle distance may be set independently by adjusting the parameters of the Paul trap. Such a flexibility offers opportunities for simulating many-body systems with coupling parameters in different ranges. For instance, having the internal dynamics of Rydberg states of ions mapped to an effective spin model, one can simulate the dynamics of the spin excitation transfer along a chain of ions~\citep{mueller08a}. 

{\bf Quantum sensing and metrology} are applications which are close-to-market and among the first quantum technologies resulting in commercial prototypes~\citep{bao18a, hinton17a, maletinsky12a}. Cold atoms in circular Rydberg states~\citep{hulet83a} are highly susceptible to the electric and magnetic fields~\citep{gallagher05a}, and thus are well-suited for precise electromagnetic field sensing~\citep{brune94a, facon16a, penasa16a} and for sensing photonic fields inside a cavity quantum electrodynamics~(QED) setup~\citep{brune96a,raimond01a}. For magnetometry, high sensitivity and spatial resolution were achieved in atomic vapours~\citep{kominis03a, wasilewski10a}, ultracold atomic gases~\citep{koschorreck11a, vengalattore07a} and for single trapped ions ~\citep{baumgart16a, kotler11a}. Entangled states of two trapped ions have been used to precisely map an inhomogeneous magnetic field over a distance that is about~$10^5$~times larger than the size of the ions' motional wavepacket~\citep{ruster17a}. It would be of great technological impact to combine the sensor capabilities of Rydberg states with the controllability of trapped ions. 
 
\subsection{Properties of trapped Rydberg ions}\label{sec:introII} 
Rydberg states are highly-excited bound states of a Coulomb potential, whose properties scale with the principal quantum number~$n$. {\bf Significantly enhanced properties} of atoms and ions in Rydberg states, as compared to ground or near-ground states, are essential for the applications mentioned above. The scaling relations of some key properties for atomic Rydberg states are given in Table~\ref{table:properties}. The enhanced orbital size of Rydberg states implies a high electric polarisability. This boosts the interaction strength between ions, as the van der Waals and the dipole-dipole interaction coefficients are enhanced by several orders of magnitude for ions in Rydberg states as compared to near ground states. The lifetimes of Rydberg states depend on the orbital angular momentum~\citep{merkt94b}, and can reach several milli-seconds for circular states~\citep{cantat20a}. 
\begin{table}
\renewcommand{\arraystretch}{1.5}
\centering
\scalebox{0.75}{
\begin{threeparttable}
\begin{tabular}{@{}l l l l p{2cm} p{2cm} p{2cm} p{2cm}}
\hline
\mr
Property& \textit{n}-scaling & $\mathcal{Z}$-scaling & $50$S, $^{88}$Sr$^{+}$\\
\mr
Binding energy $E_{n}$&$n^{-2}$&$\mathcal{Z}^2$&$3.8 \times 10^{-21}$~J~$\sim 5.88$~THz\\
Energy separation $E_{n+1}-E_{n}$&$n^{-3}$&$\mathcal{Z}^2$&$1.6\times 10^{-22}$~J~$\sim 241$~GHz\\
Fine structure splitting &$n^{-3}$&$\mathcal{Z}^4$&$2.5\times 10^{-24}$~J~$\sim 3.8$~GHz~\tnote{(1)}\\
Orbital size $\langle \boldsymbol r \rangle$&$n^{2}$&$\mathcal{Z}^{-1}$&$89$~nm\\
Electric quadrupole moment $\langle e {\boldsymbol r}^{2} \rangle$&$n^{4}$&$\mathcal{Z}^{-2}$&$3.10 \times 10^{-34}$~C m$^{2}$~\tnote{(2)}\\
Natural lifetime $\tau_{nat}$&$n^{3}$&$\mathcal{Z}^{-4}$&$6.7$~$\mu$s~\tnote{(3)}\\
Blackbody radiation limited lifetime $\tau_{BBR}$&$n^{2}$&$\mathcal{Z}^{-4}$&$5.6$~$\mu$s\\
Transition dipole moment $\langle g \vert e{\boldsymbol r} \vert nLJ  \rangle$&$n^{-3/2}$&$\mathcal{Z}^{-1}$&$9.33 \times 10^{-32}$~C m~$\sim 0.028$~D~\tnote{(4)}\\
Transition dipole moment $\langle n L^{\prime} J^{\prime} \vert e{\boldsymbol r} \vert nLJ  \rangle$&$n^{2}$&$\mathcal{Z}^{-1}$&$4.07 \times 10^{-27}$~C m~$\sim 1220.04$~D~\tnote{(5)}\\
Electric polarisability $\alpha$ &$n^{7}$&$\mathcal{Z}^{-4}$&$1.02 \times 10^{-30}$~C$^{2}$ m$^{2}$ J$^{-1}$~$\sim 96.93$~MHz/(V/cm)$^2$\\
Dipole-dipole interaction strength &$n^{4}$&$\mathcal{Z}^{-2}$&$2.05 \times 10^{-27}$~J~$\sim 3.1$~MHz~\tnote{(6)}\\
Van der Waals interaction strength &$n^{11}$&$\mathcal{Z}^{-6}$&$5.63 \times 10^{-30}$~J~$\sim 8.5$~kHz~\tnote{(7)}\\
\mr
\hline
\end{tabular}
 \begin{tablenotes}
     \item[(1)] For splitting between 50P$_{3/2}$ and 50P$_{1/2}$.
     \item[(2)] For 50D$_{3/2}$.
     \item[(3)] For 50P$_{1/2}$ state of Sr$^{+}$ $\tau_{nat}\approx 88$~$\mu$s, for 50P$_{1/2}$ state of Ca$^{+}$ $\tau_{nat} \approx 192$~$\mu$s~\citep{glukhov13a}. Longer lifetimes have been observed for circular states of neutral atoms, e.g., a circular state of Rb with $n=52$ in a 4-Kelvin cryostat setup exhibits $\tau_{nat}=3.7$~ms~\citep{cantat20a}.  
     \item[(4)] For $6$P$_{1/2}\rightarrow50$S$_{1/2}$.
     \item[(5)] For $50$S$_{1/2}\rightarrow50$P$_{1/2}$.
     \item[(6)] For $50$S-$50$P Microwave-dressed states and 4~$\mu$m inter-ion distance.
     \item[(7)] For $50$S and 4~$\mu$m inter-ion distance.
   \end{tablenotes}
    \end{threeparttable}
    }
    \centering
    \caption{\label{table:properties}Properties of the Rydberg atoms and ions scaling with the principal quantum number~$n$ and the core charge~$\mathcal{Z}$~\protect\citep{higgins18a}. $\mathcal{Z}$ is equal to $+1$ for neutral Rydberg atoms and to $+2$ for singly-charged Rydberg ions, which have overall charge of one unit of elementary positive charge $+e$. The last column shows values calculated for the $50$S state of the $^{88}$Sr$^{+}$ ion unless specified in the footnotes~\protect\citep{higginsprivate20a}.}
\end{table}

\begin{figure}
\centering
\includegraphics[width=0.7\textwidth]{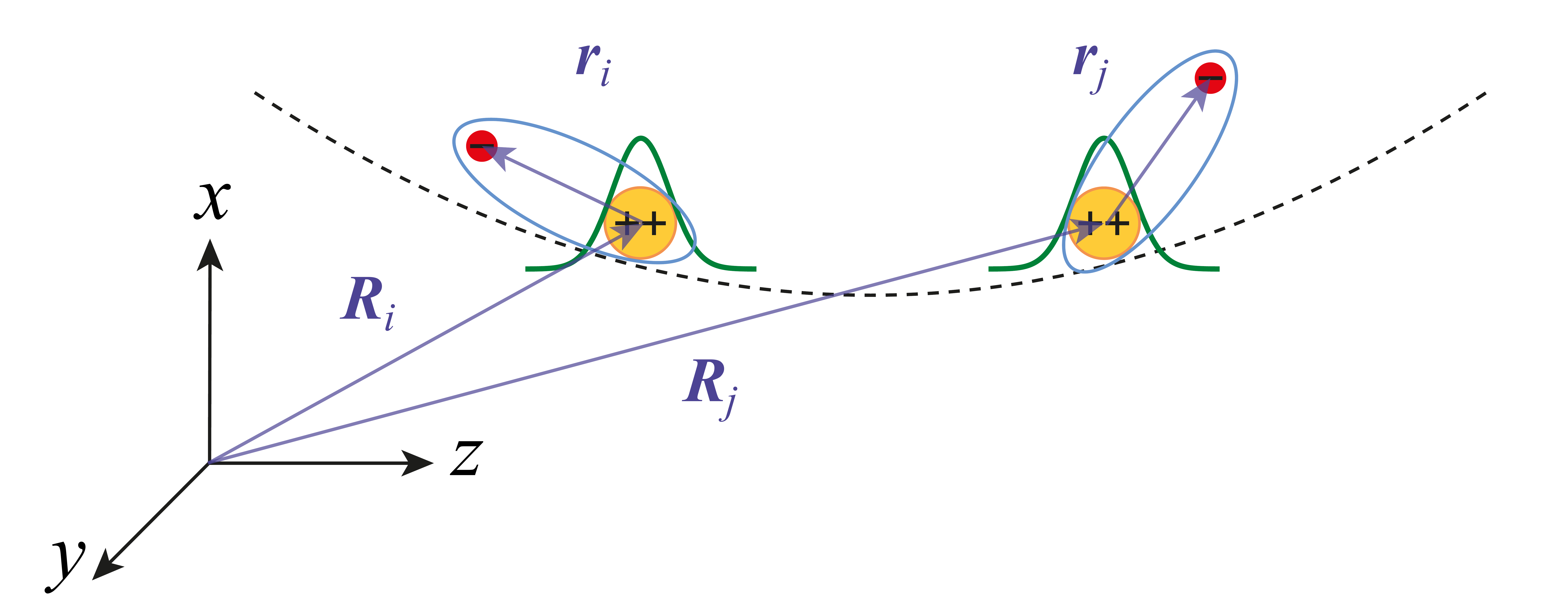}
\caption{Schematic of interacting singly-charged trapped Rydberg ions in a single harmonic potential. ${\boldsymbol R}_{i}$~(${\boldsymbol R}_{j}$) and ${\boldsymbol r}_{i}$~(${\boldsymbol r}_{j}$) are the positions of the $i$~($j$)-th ionic cores (yellow circle) and those of the Rydberg electrons (red circle) respectively. Strong Coulomb repulsion between the cores is balanced by the trapping force. The dashed line depicts the harmonic trapping potential that confines the charges. }
\label{fig:interacting2ions}
\end{figure}

The most prominent aspect for a system of trapped ions in Rydberg states is their mutual interactions. Here, we consider {\bf the interaction potential for two singly-charged ions}, each with a single valence electron in a Rydberg orbital (Fig.~\ref{fig:interacting2ions}) 
\begin{equation}
\begin{aligned}
V &= \frac{e^2}{4 \pi \epsilon_{0}} \Big( \frac{4}{\vert {\boldsymbol R}_{i}-{\boldsymbol R}_{j} \vert}-\frac{2}{\vert {\boldsymbol R}_{i}-({\boldsymbol R}_{j}+{\boldsymbol r}_{j}) \vert}-\frac{2}{\vert ({\boldsymbol R}_{i}+{\boldsymbol r}_{i})-{\boldsymbol R}_{j} \vert}\\ &+\frac{1}{\vert ({\boldsymbol R}_{i}+{\boldsymbol r}_{i})-({\boldsymbol R}_{j}+{\boldsymbol r}_{j}) \vert} \Big),
\label{eq:V1interacting2ions}
\end{aligned}
\end{equation}
where $e$ is one unit of elementary positive charge, ${\boldsymbol R}_{i}$~(${\boldsymbol R}_{j}$) and ${\boldsymbol r}_{i}$~(${\boldsymbol r}_{j}$) are the positions of the $i$~($j$)-th ionic cores and of the Rydberg electron bonded to it respectively. 
In a typical experiment, the inter-ion distance is about a few micrometres, which is in most cases much larger than the Rydberg electron orbital size, e.g., about 100~nm for the 53S$_{1/2}$ state of Sr$^{+}$ ions~\citep{higgins18a}, see Table~\ref{table:properties} for scaling with the principal quantum number~$n$. From the multipole expansion of this potential, assuming that the Rydberg orbital radius is much smaller than the inter-ion distance, i.e., $\vert {\boldsymbol r}_{i} \vert, \vert {\boldsymbol r}_{j} \vert \ll \vert {\boldsymbol R}_{i}-{\boldsymbol R}_{j} \vert$, one obtains 
\begin{equation}
\begin{aligned}
V &= \frac{e^2}{4 \pi \epsilon_{0}} \Big( \frac{1}{R_{ij}}+\frac{({\boldsymbol R}_{i}-{\boldsymbol R}_{j})\cdot ({\boldsymbol r}_{i}-{\boldsymbol r}_{j})}{R^{3}_{ij}}+\frac{r^2_{i}-3({\bf n}_{ij} \cdot {\boldsymbol r}_{i})^2+r^2_{j}-3({\bf n}_{ij} \cdot {\boldsymbol r}_{j})^2}{2R^{3}_{ij}}\\ &+\frac{{\boldsymbol r}_{i} \cdot {\boldsymbol r}_{j}-3({\bf n}_{ij} \cdot {\boldsymbol r}_{i})({\bf n}_{ij} \cdot {\boldsymbol r}_{j})}{R^{3}_{ij}} \Big),
\label{eq:V2interacting2ions}
\end{aligned}
\end{equation}
where $\vert {\boldsymbol R}_{i}-{\boldsymbol R}_{j} \vert= R_{ij}$ and ${\bf n}_{ij}=({\boldsymbol R}_{i}-{\boldsymbol R}_{j})/R_{ij}$.
The first three terms describe the Coulomb interaction, dipole-charge and quadrupole-charge interactions respectively, and are absent between interacting Rydberg neutral atoms. The fourth term describes the dipole-dipole interaction and is common to both neutral and charged atoms in Rydberg states.
Each of these interaction terms as well as inter-ion distances can be engineered to tune multi-particle interactions.
As an interesting example, the charge-dipole term plays a significant role for designing a fast two-ion gate operation which is performed by impulsive electric pulses~(Sec.~\ref{sec:future_prospects}).  

This review is structured as follows. First, we consider the challenges and approaches for experiments with Rydberg states of trapped ions in Sec.~\ref{sec:exp_approaches}. Ion trapping techniques relevant for control and manipulation of trapped Rydberg ions are discussed in Sec.~\ref{sec:initialization}. Experimental results for spectroscopy of trapped Rydberg ions as well as effects due to the trapping electric fields are presented in Sec.~\ref{sec:spectro}. Coherent manipulation of Rydberg state of a trapped ion and experimental work for generating entanglement using Rydberg dipolar interactions between trapped ions are discussed in Sec.~\ref{sec:coh_spectroscopy}. We conclude with prospects and the outlook for this research field in Sec.~\ref{sec:future_prospects} and Sec.~\ref{sec:outlook}. 

\section{Experimental approaches}\label{sec:exp_approaches}
Experiments with Rydberg states of trapped ions involve specific challenges in addition to working with trapped ions in a Paul trap~(Sec.~\ref{subsec:iontrapping}). Currently, two groups pursue experimental work on trapped Rydberg ions using $^{40}$Ca$^{+}$ ions and $^{88}$Sr$^{+}$ ions. Both species feature a single valance electron and the isotope with nuclear spin $F$=0, thus no Hyperfine structure. Three main requirements are usually considered: 

\begin{itemize}
\item Rydberg excitation of hydrogen-like alkaline earth ions requires high-energy photons due to the large binding energy of the electron to the doubly-charged core. Ionization energies of Be$^{+}$, Mg$^{+}$, Ca$^{+}$, Sr$^{+}$ and Ba$^{+}$ ions correspond to energies of photons at about 68.1, 82.5, 104.4, 112.4 and 123.9~nm respectively~\citep{nistasd18}. The generation of vacuum ultra-violet (VUV) radiation, beam steering and its integration into an experimental setup is challenging. 
For the cases of Ca$^{+}$, Sr$^{+}$ and Ba$^{+}$ ions, a long-lived metastable D state can be excited before Rydberg excitation, reducing the remaining excitation energies to wavelengths of $121.9$, $134.9$ and $133.3$~nm respectively~\citep{nistasd18}. Two schemes have been thus far applied for bridging this energy gap to drive Rydberg transitions of trapped ions: {\bf a single-step approach} using VUV laser radiation for the first observation of Rydberg resonances in trapped Ca$^{+}$ ions~\citep{feldker15a} and {\bf a two-step approach} using UV lasers for the coherent manipulation of Rydberg levels in trapped Sr$^{+}$ ions~\citep{higgins17b}. Excitation schemes using three or more steps (e.g.~\citep{lange91a} for Rydberg ions in free space) involve complications for coherent manipulation of quantum states, and thus are not of interest for the applications discussed in this review. In the single-step laser excitation from metastable D states of alkaline-earth ions, electric dipole transition selection rules allow transitions to P and F levels. In the two-step approach, the excitation from metastable D states is carried out via an intermediate P state, and thus S and D states can be directly driven~(Sec.~\ref{subsec:exp_approaches_ca40} and~Sec.~\ref{subsec:exp_approaches_sr88ca40}).

\begin{figure}
\centering
\includegraphics[width=1.0\textwidth]{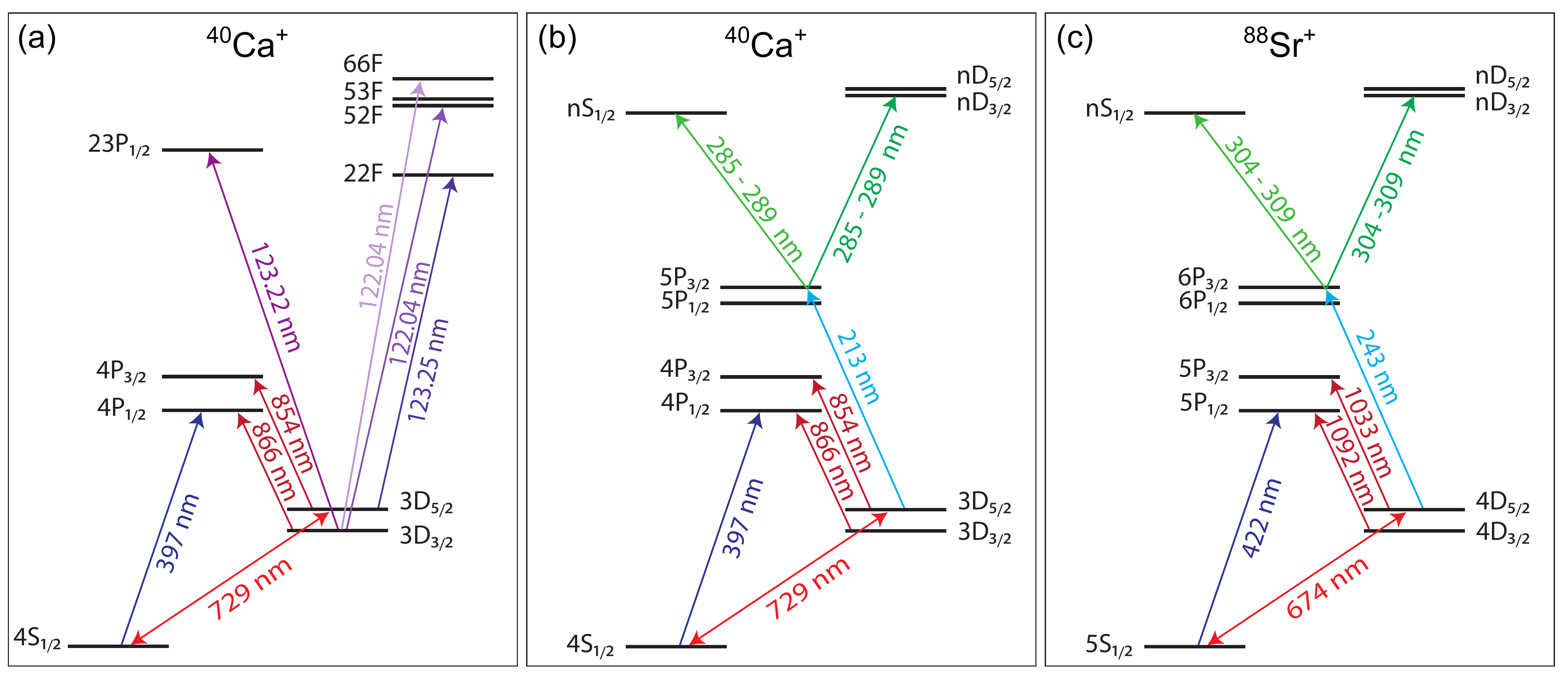}
\caption{Energy level diagrams for $^{40}$Ca$^{+}$ (a, b) and for $^{88}$Sr$^{+}$ ions (c) and laser wavelengths for Rydberg excitation, laser cooling and detection. Rydberg states are excited from the long-lived metastable D states. (a) Single-step excitation scheme for P or F levels of Ca$^{+}$~(purple arrows). (b), (c) Two-step excitation scheme via an intermediate P state to S or D states of Ca$^{+}$ and Sr$^{+}$ (light blue and green arrows). In (b) and (c), the lasers are detuned from the intermediate, short-lived P$_{3/2}$ states, see a detailed example in Fig.~\ref{fig:coh_level_scheme}. Laser beams for a standard trapped ion experiment in $^{40}$Ca$^{+}$ and $^{88}$Sr$^{+}$ ions (blue and red arrows).}
\label{fig:leveldiagrams}
\end{figure}  

\item {\bf The detection of the population in Rydberg states} of single trapped ions requires the implementation and adaptation of electron shelving techniques. Observing the state-dependent fluorescence~(Sec.~\ref{subsec:detection}) is a non-destructive efficient detection method, in contrast to photoionization or electric field ionisation methods that have been widely used in large samples of cold neutral atomic gases~\citep{gallagher05a}.

\item A key requirement for experiments with Rydberg ions is the ability to precisely {\bf minimise parasitic stray electric fields} at the position of ions. Such electric fields can displace ions from the RF-field nodal line~(Sec.~\ref{subsec:iontrapping}) to new equilibrium positions at which the amplitude of the oscillating electric field is non-zero (Sec.~\ref{subsec:micromotion}). Various methods have been developed for determining the voltages that are to be applied on trap electrodes~\citep{berkeland98a}. Minimisation of such stray fields is essential to avoid energy shifts of Rydberg resonances and coupling effects due to the trapping electric fields~(Sec.~\ref{subsec:micromotion}).  Coherent excitation of Rydberg levels requires such control over electric fields at the ion position.

\end{itemize}  

\subsection{Single-step Rydberg excitation of $^{40}${\rm Ca}$^{+}$ ions}\label{subsec:exp_approaches_ca40}
\begin{figure}
\centering
\includegraphics[width=1.0\textwidth]{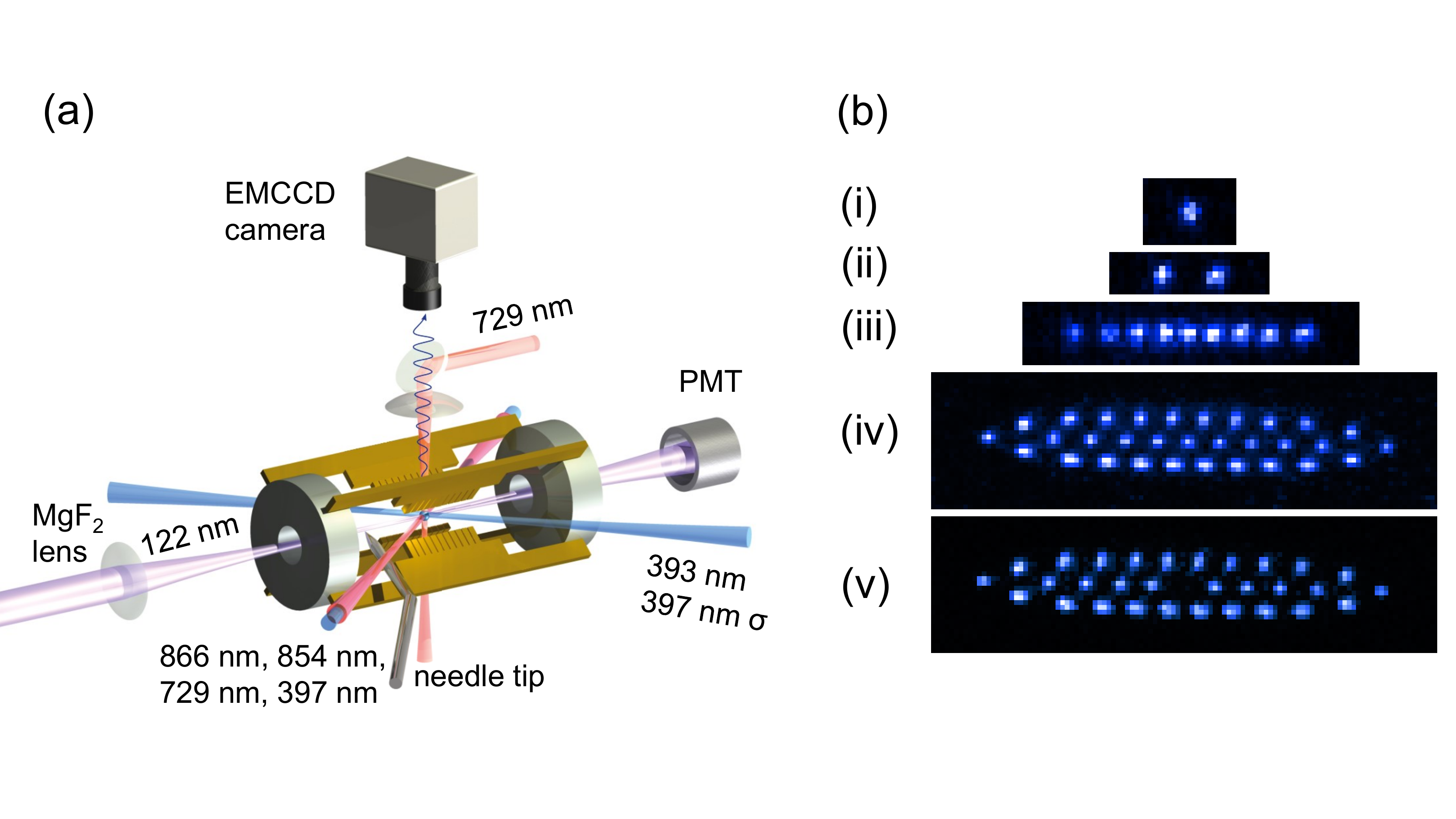}
\caption{(a) Schematic of the experimental setup for Rydberg excitation of trapped $^{40}$Ca$^{+}$ using a 122~nm laser beam. The ion trap features four gold-coated blade electrodes and two endcaps with through holes that allow for the 122~nm beam optical access. The ion fluorescence is imaged on an EMCCD camera, each bright spot corresponds to the fluorescence of one single ion as the dipole allowed transitions near 397~nm and 866~nm are driven, see the energy level diagram for $^{40}$Ca$^{+}$ ions in Fig.~\ref{fig:leveldiagrams}(a). Adapted from~\protect\citep{feldker15a}. (b)~(i-v)~False colour fluorescence images of linear and two-dimensional arrays of cold trapped Ca$^{+}$ ions in a linear Paul trap, ion distances are about few~micrometres. The central ion in (iv), which is in the 4S$_{1/2}$ state (bright state), is optically pumped into the 3D$_{5/2}$ state (dark state) in (v) using a tightly focused 729~nm beam. Note that here an anisotropy of the radial confinement allows for trapping two-dimensional crystals such that single ion addressability is assured.}
\label{fig:setup_Ca1ph}
\end{figure}

The first Rydberg excitation of trapped ions was demonstrated for $^{40}$Ca$^{+}$ ions in a linear Paul trap~\citep{feldker15a}~(Fig.~\ref{fig:setup_Ca1ph}(a)).  
A continuous-wave, vacuum ultra-violet (VUV) laser source at 122~nm was used to excite Rydberg F states from metastable 3D$_{3/2}$ or 3D$_{5/2}$ states with about one second lifetimes~(Fig.~\ref{fig:leveldiagrams}(a)).
This setup was used to observe $3$D$_{3/2}\rightarrow 52$F, $3$D$_{3/2}\rightarrow 53$F and $3$D$_{5/2}\rightarrow 66$F~\citep{feldker15a}, $3$D$_{5/2}\rightarrow 22$F~\citep{bachor16a} and $3$D$_{3/2}\rightarrow 23$P$_{1/2}$~\citep{mokhberi19a} transitions (Fig.~\ref{fig:leveldiagrams}(a)). 
The VUV beam was generated using four-wave frequency mixing in mercury vapour~\citep{bachor16a, eikema99a, kolbe12a, schmidtkaler11a}.
The efficiency of this mixing process is a sensitive function of the wavelength generated, and thus the output power varies between 0.2 to 10 $\mu$W~\citep{schmidtkaler11a}.
The Rabi frequency estimated for the $3$D$_{3/2}\rightarrow 53$P$_{1/2}$ transition using the beam parameters in that setup is about $20$~kHz~\citep{feldker16a}.   

\subsection{Two-step Rydberg excitation of $^{88}${\rm Sr}$^{+}$ and $^{40}${\rm Ca}$^{+}$ ions}\label{subsec:exp_approaches_sr88ca40}
The two-photon Rydberg excitation of trapped ions has been first demonstrated for $^{88}$Sr$^{+}$ ions~\citep{higgins17a}. Rydberg S and D transitions were driven from the 4D$_{5/2}$ state (or optionally from the 4D$_{3/2}$ state) via the intermediate 6P$_{3/2}$ state using laser light at 243~nm and 305~nm~(Fig.~\ref{fig:leveldiagrams}(c)). A similar approach has been used for Rydberg excitation of $^{40}$Ca$^{+}$ ions using two UV lasers at 213~nm and 286~nm~(Fig.~\ref{fig:leveldiagrams}(b)). Laser sources at these wavelengths are commercially available with output powers of tens to hundreds of milliwatts, which allow for fast coherent manipulation of ions in Ryberg states (Sec.~\ref{sec:coh_spectroscopy}). 
Another major advantage of this approach is that the Doppler broadening of resonances can be mitigated using a counter-propagating setup for the two UV laser beams (Sec.~\ref{subsec:lamb_dicke}). Moreover, control of the polarisation of such UV beams allows for addressing Zeeman sublevels of Rydberg states~(Sec.~\ref{sec:coh_spectroscopy}).

\section{Initalization of trapped ion crystals}\label{sec:initialization}
\subsection{Trapping ions}\label{subsec:iontrapping}
In this section, the theory for trapping ions in a Paul trap and methods for manipulating their electronic and motional degrees of freedom are discussed. 
Paul traps have been used for a wide-range of applications from quantum optics, quantum simulation~\citep{blatt12a} and quantum computing~\citep{haeffner08a} to precision measurements~\citep{asvany15a}, metrology~\citep{huntemann16a} and controlled chemistry~\citep{willitsch17a}. 
In a Paul trap, {\bf three-dimensional confinement of charged particles is achieved by applying a static electric field in combination with a time-varying electric field} oscillating at radio frequency (RF)~\citep{major05a}. 

In this review, we consider RF traps that generate a nearly quadrupolar electric potential close to the trap centre, which can be written as:
\begin{equation}
\Phi({\boldsymbol R},t)=(\gamma^{\prime}_{x} X^2+\gamma^{\prime}_{y} Y^2+\gamma^{\prime}_{z} Z^2)\cos (\Omega_{\rm RF}t)+(\gamma_{x} X^2+\gamma_{y} Y^2+\gamma_{z} Z^2),
\label{eq:paultrapgeneral}
\end{equation}
where ${\boldsymbol R}\equiv X \hat{x}+Y \hat{y}+Z \hat{z}$ is the ion position vector  decomposed in Cartesian coordinates and $\Omega_{\rm RF}$ is the RF drive frequency. $\gamma_{i}$ and $\gamma^{\prime}_{i}$, with $i\in\lbrace x,y,z \rbrace$, are geometric factors which depend on the geometry of trap electrodes and voltages applied to them.  
The potential given in Eqn.~\ref{eq:paultrapgeneral} has to fulfil the Laplace equation~$\Delta \Phi=0$ at every instant in time, and thus the geometric factors are constrained by $\gamma^{\prime}_{x}+\gamma^{\prime}_{y}+\gamma^{\prime}_{z}=0$ and $\gamma_{x}+\gamma_{y}+\gamma_{z}=0$. For this reason, there is no three-dimensional local minimum in free space for a charge acted on by only static electric fields~(Earnshow's theorem)~\citep{foot05a}. Charged particles can be trapped by combination of static electric fields with either dynamical electric fields (Paul traps) or magnetic fields (Penning traps)~\citep{dehmelt69a, paul90a}. For positive charges, the common choice for the geometric factors is $\gamma^{\prime}_{x}=-\gamma^{\prime}_{y}=\gamma^{\prime}$, $\gamma^{\prime}_{z}=0$ and $-(\gamma_{x}+\gamma_{y})=\gamma_{z}=2 \gamma >0$.
This configuration is referred to as {\bf the linear Paul trap} with translational symmetry along the $z$ axis, also called the trap axis. The confining force along this axis provided by the static electric field is typically weaker as compared to the dynamical trapping force in any direction in the $x$-$y$ plane. In a linear Paul trap, Eqn.~\ref{eq:paultrapgeneral} is written as 
\begin{equation}
\Phi({\boldsymbol R},t)=\gamma^{\prime} (X^2-Y^2) \cos (\Omega_{\rm RF}t)-\gamma( (1+\epsilon) X^2+ (1-\epsilon)Y^2-2Z^2).
\label{eq:paultrap}
\end{equation}
Here, $\epsilon$ is a dimensionless parameter that breaks the axial symmetry and removes the degeneracy of the two radial modes. In most experiments, once the trap geometry is fixed, $\gamma^{\prime}$ can be varied using a single RF voltage source, whereas $\gamma$ can be controlled via different voltages applied to segmented trap electrodes~\citep{hucul08a}.

Stable solutions of the equations of motion can be found from solutions to the Matthieu equations~\citep{major05a}, for an in-depth discussion see~\citep{leibfried03a, major05a, wineland98a}. In most RF traps, the motion of the ion can be described by two components ${\boldsymbol R}={\boldsymbol R_{\rm sec}}+{\boldsymbol R}_{\rm mm}$, where ${\boldsymbol R_{\rm sec}}$ and ${\boldsymbol R_{\rm mm}}$ denote the secular slow motion of the ion and the fast driven motion called ``micromotion'' respectively. 
However, one should distinguish unavoidable micromotion due to secular oscillation of the ion around the RF node from ``excess micromotion'' caused by stray static electric fields or by a phase difference between RF voltages applied to trap electrodes~\citep{berkeland98a}~(Sec.~\ref{subsec:micromotion}). 
The time-dependent potential in Eqn.~\ref{eq:paultrap} is treated within an adiabatic approximation and is time averaged over one RF period~\citep{major05a}. This approximation corresponds to an effective and mass-dependent pseudopotential~\citep{dehmelt90a}, which is given by $\Phi_{\rm pseudo}({\boldsymbol R},M)=\frac{Z e {\gamma^{\prime}}^2(X^2+Y^2)}{M \Omega_{\rm RF}^2}$.
The frequencies of the ion oscillation along the three axes~(Fig.~\ref{fig:setup_Ca1ph}) are thus given by
\begin{eqnarray}
\label{eq:secularfreqX}
\omega_{X}=\sqrt{\frac{2e^2 \mathcal{Z}^2 {\gamma^{\prime}}^2 }{M^2 \Omega_{\rm RF}^2}-\frac{2\mathcal{Z} e \gamma (1+\epsilon)}{M}}, \\
\label{eq:secularfreqY}
\omega_{Y}=\sqrt{\frac{2e^2 \mathcal{Z}^2 {\gamma^{\prime}}^2 }{M^2 \Omega_{\rm RF}^2}-\frac{2\mathcal{Z} e \gamma (1-\epsilon)}{M}},\\
\label{eq:secularfreqZ}
\omega_{Z}=2\sqrt{\frac{\mathcal{Z} e\gamma}{M}}.
\end{eqnarray}
Here, $M$ and $\mathcal{Z}$ are the mass and the charge of the ion, and $e$ is one elementary positive charge. Under typical operation conditions, the secular frequencies range between a few 100~kHz and 10~MHz.

\subsection{Normal mode analysis}\label{subsec:normalmodes}
Now we consider $N$ ions in a harmonic potential given in Eqn.~\ref{eq:paultrap} and describe a general approach to {\bf calculate their equilibrium positions and vibrational modes}. Here, we assume that the kinetic energy of ions $E_{\rm kin}$($\sim k_{\rm B}T$) is reduced by laser cooling (Sec.~\ref{subsec:lasercooling}) such that $\frac{E_{\rm pot}}{E_{\rm kin}}=\frac{\mathcal{Z}^2 e^2}{4 \pi \epsilon_{0} a_{\rm WS} k_{\rm B}T} \gtrsim 170 $, where $E_{\rm pot}$ is the potential energy of the system and $a_{\rm WS}$ is the Wigner-Seitz radius~\citep{pollock73a, slattery80a}. Under this condition, trapped ions undergo a phase transition into translationaly cold and spatially organised structures called ``Coulomb crystals''~\citep{bollinger94a}~(Fig.~\ref{fig:setup_Ca1ph}~(b)).  In such ion crystals, the amplitude of the ions' oscillation around their equilibrium positions are much smaller than the inter-ion separation, which ranges between 2~$\mu$m and 20~$\mu$m under typical trapping conditions.

The total energy $E$ of the system is given by~\citep{james98a}
\begin{equation}
E=\sum_{i=1}^{N} \mathcal{Z} e \Phi({\boldsymbol R}_{i}, M_{i})+\frac{1}{2}\sum\limits_{\substack{i,j \\ i \neq j}}^N \frac{(\mathcal{Z} e)^2}{4 \pi \epsilon_{0} \vert {\boldsymbol R}_{i}-{\boldsymbol R}_{j} \vert}+\sum_{i=1}^N \frac{M_{i}}{2}\dot{{\boldsymbol R}^2_{i}}.
\label{eq:Nionenergy}
\end{equation}
Here, $M_{i}$ and $R_{i}$ ($R_{j}$) are the mass and coordinate of the $i$- ($j$)-th ion and $\Phi(R, M)$ is the trapping potential given in~Eqn.~\ref{eq:paultrap}, where the time-dependent term is approximated by the mass-dependent pseudopotential. The equilibrium positions of ions $\boldsymbol{R}^0_{i}$ are calculated by solving a set of linear equations $\frac{\partial E_{\rm pot}}{\partial R_{i}}=0$, where $E_{\rm pot}$ is given by the first two terms in Eqn.~\ref{eq:Nionenergy}.
In a typical linear Paul trap, $\gamma^{\prime}$ is about two orders of magnitude larger than $\gamma$ (Eqn.~\ref{eq:paultrap}), and hence $\omega_{X,Y}\gg\omega_{Z}$ (Eqns.~\ref{eq:secularfreqX}-\ref{eq:secularfreqZ}). Thus ions form a chain along the trap axis $z$~(Fig.~\ref{fig:setup_Ca1ph}(b-iii)).  
For a chain of $N$ ions with equal masses $M$, the minimum inter-ion distance is given by $d_{\rm min} \approx (\frac{\mathcal{Z}^2 e^2}{4 \pi \epsilon_{0} M \omega^2_{Z}})^{1/3} \frac{2.018}{N^{0.559}}$~\citep{james98a}.
By increasing the number of the ions in the trap, a structural phase transition from a linear to a two-dimensional structure might occur depending on the anisotropy of the trapping potential which is given by~$\mathcal{A}_{X,Y}=(\frac{\omega_{Z}}{\omega_{X,Y}})^2$~\citep{dubin93a,kaufmann12a}, where $\omega_{X}$, $\omega_{Y}$ and $\omega_{Z}$ are the secular trapping frequencies in Eqns.~\ref{eq:secularfreqX}-\ref{eq:secularfreqZ}. Images of such a two-dimensional Coulomb crystal is shown in Figs.~\ref{fig:setup_Ca1ph}(b-iv). 

By solving the Lagrangian equations of the motion and computing the Hessian, one can obtain the normal modes eigenvectors $e^{i}_{\alpha}$ and their corresponding frequencies $\omega_{\alpha}$~\citep{james98a}. 
Each normal mode accounts for an independent oscillation, which can be quantised and given in the form of the momentum and position operators
\begin{equation}
\hat{P}_{\alpha}=i \sqrt{\frac{\hslash M \omega_{\alpha}}{2}} (\hat{a}_{\alpha}-\hat{a}^{\dagger}_{\alpha} ),\hspace{3mm}
\hat{X}_{\alpha}=\sqrt{\frac{\hslash}{2 M \omega_{\alpha}}} (\hat{a}_{\alpha}+\hat{a}^{\dagger}_{\alpha} ),
\label{eq:Xoperator}
\end{equation} 
where $\hat{a}_{\alpha}$ and $\hat{a}^{\dagger}_{\alpha}$ are the lowering and raising ladder operators of the normal mode~$\alpha$ at the motional frequency~$\omega_{\alpha}$. Using this notation, the quantised form of the $i$-th ion's excursion around its equilibrium position as a function of ion coordinates is given by
\begin{equation}
\hat{X}_{i}=\sum_{\alpha=1}^{3N} ({\boldsymbol e}^{i}_{\alpha})^{-1} \sqrt{\frac{\hslash}{2 M \omega_{\alpha}}} (\hat{a}_{\alpha}+\hat{a}^{\dagger}_{\alpha} ).
\label{eq:quantisedX}
\end{equation} 
The size of the ground state motional wavepacket of the ion in a typical setup is about 5--20~nm, much smaller than its orbit size of about 100~nm in a given Rydberg state. 

\subsection{Light-ion interaction in the Lamb-Dicke regime}\label{subsec:lamb_dicke}
\begin{figure}
	\centering
	\includegraphics[trim={3cm 11cm 2.5cm 0},clip,width=1.0\textwidth]{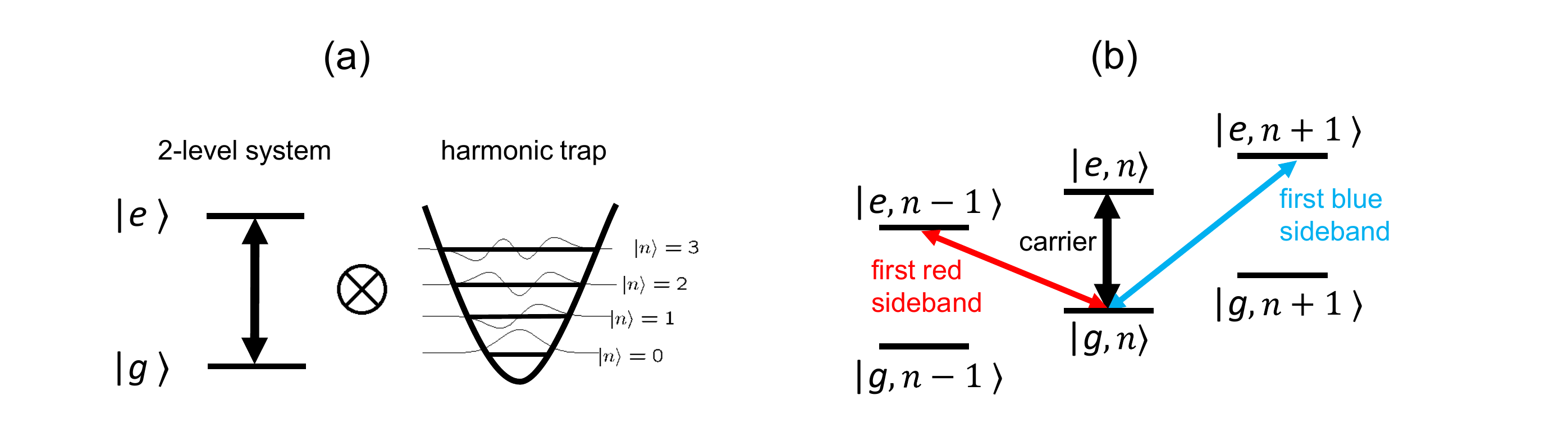}
	\caption{(a) Schematic of the electronic  two-level system with $\ket{g}$ and $\ket{e}$ states coupled by laser light (black arrow). The confinement of the ion in a harmonic well leads to quantized motional degrees of freedom, also shown is an ion wavepacket in position space for each phonon state from $n=0$~to~$3$ (grey). (b) Coupling between internal (electronic) and external (vibrational) degrees of freedom via carrier, first red and first blue sideband transitions.}
	\label{fig:qubit}
\end{figure}

In a trapped-ion qubit, {\bf electronic levels can be coherently coupled to motional degrees of freedom} by applying suitable laser pulses~(Fig.~\ref{fig:qubit}). 
The transition that couples the $\ket{g}$ and $\ket{e}$ electronic states with no change in the number of motional quanta known as phonons, is termed the ``carrier transition''. 
For a tightly confined ion wavepacket and for suitable tuning of the laser frequency on a narrow optical transition, the transition from $|g,n\rangle$ to $|e,n-1\rangle$ is described by the Jaynes-Cummings Hamiltonian~\citep{leibfried03a} referred to as ``red sideband''~(Fig.~\ref{fig:qubit}(b)). In addition, since the laser field acts as a drain or source of energy, the atom-photon coupling does not satisfy energy conservation. The $|g,n\rangle$ to $|e,n+1\rangle$ transition is described by the anti-Jaynes-Cummings Hamiltonian and is coined as ``blue sideband''. 
For red sidebands and blue sidebands the number of phonons decreases (or increases) due to the photon recoil momentum kick~(Fig.~\ref{fig:qubit}(b)). 
Beyond these cases, there are many other possibilities for interchanging energy of multiples of the motional quantum. Including also micromotion effects, a set of transitions can be driven at frequencies of integer multiples of the RF frequency, the so-called ``micromotion sidebands'', and at a sum over integer multiples of the RF frequency and secular frequencies.

The coupling strength for a given sideband transition depends on the ratio of the photon recoil energy to one quanta of the ion oscillation, and is given by {\bf the Lamb-Dicke parameter}~\citep{haroche06a}
\begin{equation}
\eta=\sqrt{\frac{\hslash}{2 M \omega_{\alpha}}} \boldsymbol{k}\cdot \hat{{\boldsymbol e}}_{\alpha}.
\label{eq:Lamb-Dicke effective}
\end{equation}
Here, $\boldsymbol{k}$ is the wavevector of the laser light that drives the transition and $\hat{{\boldsymbol e}}_{\alpha}$ is the unit vector of the normal mode eigenvector at $\omega_{\alpha}$. 
For strongly confined ions, the extension of the atomic wavepacket is much smaller than the transition wavelength. This interaction regime is referred to as {\bf the Lamb-Dicke regime} that is defined by $\eta \sqrt{\langle (\hat{a}+\hat{a}^\dagger)^2 \rangle} \sim \eta \sqrt{n} \ll  1$, where $n$ denotes the average phonon number of the corresponding mode. The excitation strength for the carrier and first red and blue sidebands are respectively given by
\begin{equation}
\Omega_{n \rightarrow n}=(1-\eta^2 n) \Omega_{0},
\label{eq:Lamb-Dicke Rabi}
\end{equation}
\begin{equation}
\Omega_{n \rightarrow n-1}=\eta \sqrt{n} \Omega_{0},
\label{eq:Lamb-Dicke Rabi}
\end{equation}
\begin{equation}
\Omega_{n \rightarrow n+1}=\eta \sqrt{n+1} \Omega_{0},
\label{eq:Lamb-Dicke Rabi}
\end{equation}
where $\Omega_{0}$ is the Rabi frequency of the carrier transition. Within the Lamb-Dicke regime, the transitions that change the motional quantum number by more than one are strongly suppressed.

In a two-photon process, the effective Lamb-Dicke parameter $\eta_{\rm eff}$ is thus identified by replacing $\boldsymbol{k}$ with $\boldsymbol{\Delta k}$, where the difference wavevector is given by ${\boldsymbol \Delta k}={\boldsymbol k}_{1}-{\boldsymbol k}_{2}$. Note that $\eta_{\rm eff}$ depends on the alignment of ${\boldsymbol k}_{1}$ and ${\boldsymbol k}_{2}$ with respect to the direction of a given normal mode.
The use of short wavelength laser light for driving Rydberg transitions of trapped ions may potentially give rise to a large Lamb-Dicke parameter. 
Lowering the effective Lamb-Dicke parameter by using counter-propagating beams is an advantage of the two-step excitation setup (Sec.~\ref{subsec:exp_approaches_sr88ca40}).  In contrast, Rydberg excitation using a single VUV beam or two co-propagating UV beams lies outside of the Lamb-Dicke regime. For instance, for $^{40}$Ca$^{+}$ ions confined at an axial trapping frequency of $1$~MHz, for the $3$D$_{5/2}\rightarrow 66$F transition driven by single photon at $122$~nm, one obtains $\eta \approx 0.58$ for an ion in the motional ground state $|n=0\rangle$, while for the $3$D$_{5/2}\rightarrow 60$S transition driven via the $5$P$_{3/2}$ state using two counter-propagating UV beams at $213$~and~$285$~nm, $\eta_{\rm eff} \approx 0.08$ can be achieved. In the latter case, even for thermal average quantum numbers up to $n_{\rm mean} \approx 20$, the ion excitation remains within the Lamb-Dicke regime.  

\subsection{Rydberg ions in the dynamical trapping field of a linear Paul trap}\label{subsec:rydbergionintrap}
In this section, a detailed account of {\bf the dynamics of a single Rydberg ion} in a linear Paul trap is presented. 
The Hamiltonian of this system is given by 
\begin{equation}
\mathcal{H}=\frac{\boldsymbol{P}^2}{2M}+\frac{\boldsymbol{p}^2}{2m}+V(|{\boldsymbol r}-{\boldsymbol R}|)+V_{ls}({\boldsymbol r}-{\boldsymbol R})+2e\Phi({\boldsymbol R},t)-e\Phi({\boldsymbol r},t),
\label{eq:Hlab}
\end{equation}
where $\boldsymbol{P}$ ($\boldsymbol{p}$) and $M$ ($m$) are the momentum and mass of the ion core (of the electron), and $\boldsymbol{R}$ ($\boldsymbol{r}$) is the position vector of the ion core (of the electron). $V(|{\boldsymbol r}-{\boldsymbol R}|)$ is an angular-momentum-dependent model potential for the Coulomb interaction between the ionic core and the Rydberg electron, and $V_{ls}({\boldsymbol r}-{\boldsymbol R})$ stands for the spin-orbit coupling. The last two terms in Eqn.~\ref{eq:Hlab} account for the coupling of the ionic and the electronic charge to the trap electric potential given in Eqn.~\ref{eq:paultrap}. 

In the centre-of-mass frame, the Hamiltonian of the system can be written as
\begin{equation}
H= H_{\rm{I}}+H_{\rm{e}}+H_{\rm{Ie}},
\label{eq:Hcom}
\end{equation}
where 
\begin{eqnarray}
\label{eq:HCM}
H_{\rm{I}}&=& \frac{\boldsymbol{P}^2}{2M}+e\Phi({\boldsymbol R},t), \\
\label{eq:Helectron}
H_{\rm{e}} &=& \frac{\boldsymbol{p}^2}{2m_{\rm{e}}}+V(|{\boldsymbol r}|)+V_{ls}({\boldsymbol r})-e\Phi({\boldsymbol r},t), \\
\label{eq:Hion-electron}
H_{\rm{Ie}}&=& -2e\gamma^{\prime} (Xx-Yy) \cos (\Omega_{\rm RF}t)+2e\gamma(Xx+Yy-2Zz).
\end{eqnarray}
with ${\boldsymbol R}\equiv X \hat{x}+Y \hat{y}+Z \hat{z}$ and ${\boldsymbol r}\equiv x \hat{x}+y \hat{y}+z \hat{z}$. 
Here, corrections due to the finite nuclear mass are neglected.
$H_{\rm{I}}$ is the free Hamiltonian of the external motion of the ionic core, which can be approximated by
\begin{equation}
\label{eq:HIeffective}
H_{\rm{I}}^{\rm{eff}} = \frac{\boldsymbol{P}^2}{2M} +\frac{M}{2}\sum_{\rho=X,Y,Z}\omega_{\rho}^2\rho^2,
\end{equation}
where $\omega_\rho$ is the secular trapping frequency~(Eqns.~\ref{eq:secularfreqX}-\ref{eq:secularfreqZ}).

The Hamiltonian of the Rydberg electron, denoted with $H_{\rm{e}}$ in Eqn.~\ref{eq:Helectron}, includes the Coulomb and spin-orbit coupling terms $V(|{\boldsymbol r}|)$ and $V_{ls}({\boldsymbol r})$, which can be calculated from multi-channel quantum defect theory~\citep{aymar96a}.
The last term in Eqn.~\ref{eq:Helectron} accounts for the interaction between the Rydberg electron and trapping electric fields, which can be written as
\begin{equation}
\label{eq:eTrapspherical}
H_{\rm{e-Trap}}=-2\sqrt{\frac{2\pi}{15}}e\gamma^{\prime} r^2 (Y_2^{2}(\theta,\phi)+Y_2^{-2}(\theta,\phi))\cos(\Omega_{\rm RF} t)-4\sqrt{\frac{\pi}{5}}e\gamma r^2Y_2^0(\theta,\phi),
\end{equation}
where $\theta$ and $\phi$ are the polar and azimuthal angles with respect to the trap axis~$z$ and $Y_l^{m}(\theta,\phi)$ are Laplace spherical harmonics.
The energy shifts due to quadrupole static and RF electric fields are calculated for a Rydberg state $\vert n, L, J, m_{J} \rangle$ with $n$, $L$ and $J$ the principal, angular and total angular quantum numbers and $m_{J}$ the projection of $J$ on the quantization axis. Here, the time $t$ is treated merely as a parameter, since the dynamics of the Rydberg electron occurs in a much faster time scale as compared with the period of the RF potential. 
Note that in this review we discuss only the cases in which the magnetic field direction, the ``quantization'' axis, coincides with the trap axis $z$. Such a configuration allows for reducing the number of coupling terms arising from the electron-trap effect.

The first-order energy shifts due to the static trapping potential can be calculated assuming that the ion is located perfectly at the trap centre, at which the minimum of the static and the RF quadrupole electric fields overlap. These energy shifts are given by
\begin{equation}
\Delta E_{{\rm st},L,J, m_{J}}= \gamma Q_{L,J} \frac{J(J+1)-3m^{2}_{J}}{J(2J-1)}
\label{eq:generalSTFirstshifts}
\end{equation}
where the corresponding quadrupole moment is given by
\begin{equation}
Q_{L,J}= -e \frac{2J-1}{2J+2} \langle n, L, J \vert r^2 \vert n, L, J \rangle.
\label{eq:generalQLJ}
\end{equation}
As an example, for the D$_{3/2}$ state and principal quantum number $n\gg 1$, it is approximated by $Q_{D, 3/2}\approx 2 e a^2_{0} n^2/ 5 (4 \mathcal{Z}+2)(5n^2+1-3L(L+1))$, with ionic core charge $\mathcal{Z}=+2$ and Bohr radius $a_{0}$~\citep{higgins17a}. States with $J=1/2$ do not have a permanent quadrupole moment, and therefore do not experience a first order energy shift, i.e., $\Delta E_{{\rm st, S}_{1/2}}= 0$ and $\Delta E_{{\rm st, P}_{1/2}}= 0$.  

\begin{figure}
\centering
\includegraphics[width=0.65\textwidth]{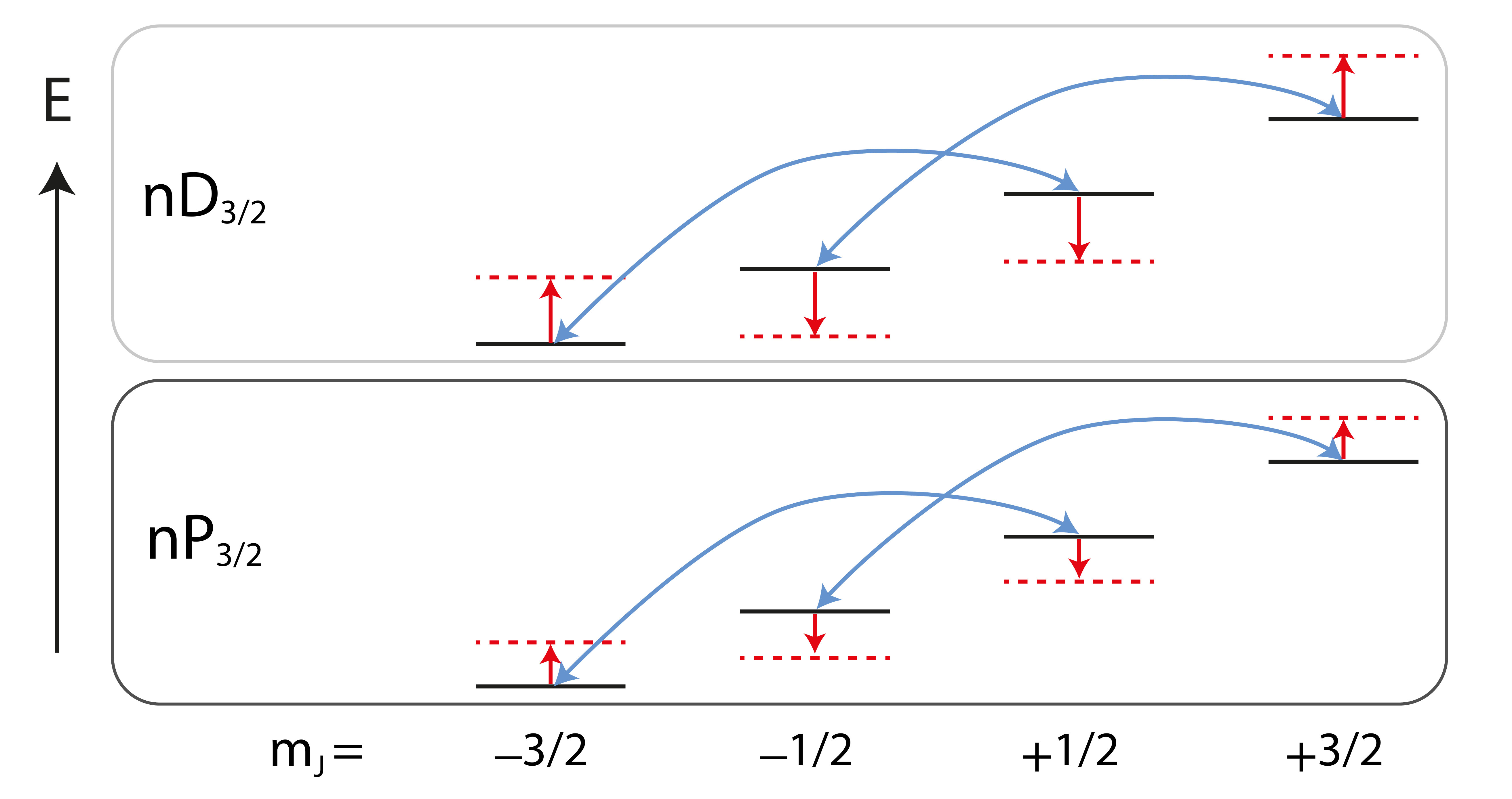}
\caption{First-order energy shifts due to the static quadrupole trapping field~(red arrows) and level couplings due to the RF quadrupole trapping field~(blue arrows) for $n$D$_{3/2}$ and $n$P$_{3/2}$ states of an alkaline-earth ion. Note that in these cases the quantization axis, given by the magnetic field direction, is along the trap axis.}
	\label{fig:electron_Etrap_coupling}
\end{figure}

The time-dependent terms of the Hamiltonian given in Eqn.~\ref{eq:eTrapspherical} lead to the coupling between Zeeman states with $\Delta m_{J}=\pm 2$. For most experiments, it is reasonable to assume that the fine structure splitting is much larger than the RF drive frequency, and thus only the coupling to the states of the same Zeeman manifold should be taken into account. The coupling due to the RF field is characterised by an effective Rabi frequency defined by~$C=-2Q_{L, J}\gamma^{\prime}/(5 \sqrt{3}\hslash) $~\citep{higgins17a}. The above effects are schematically depicted in Fig.~\ref{fig:electron_Etrap_coupling}.
It is important to notice that higher-order terms due to the electron-trap coupling are not negligible for $n>10$. 
Moreover, these terms become noticeable when the minima of the static and the RF quadrupolar fields do not overlap~(Sec.~\ref{subsec:micromotion} and Sec.~\ref{subsec:trapeffects}). Note that only higher-order terms are relevant for $n$S$_{1/2}$ and $n$P$_{1/2}$ states.
  
The electron-ion Hamiltonian, $H_{\rm{Ie}}$ in Eqn.~\ref{eq:Hion-electron}, is a function of the size of the electron orbit, which is significantly large for a highly excited state in comparison to a low-lying state.
This gives rise to additional energy shifts proportional to the polarisability of the Rydberg state excited, which modify the ion oscillation frequencies by~\citep{schmidtkaler11a}
\begin{eqnarray}
\label{eq:omegaXryd}
\Delta \omega_{X}&=& \sqrt{\frac{(4e^2 {\gamma^\prime}^2+8e^2\gamma^2(1+\epsilon)^2)\mathtt{\nu}^{(2)}}{M}}, \\
\label{eq:omegaYryd}
\Delta \omega_{Y}&=& \sqrt{\frac{(4e^2 {\gamma^\prime}^2+8e^2\gamma^2(1-\epsilon)^2)\mathtt{\nu}^{(2)}}{M}}, \\
\label{eq:omegaZryd}
\Delta \omega_{Z}&=& \sqrt{\frac{8e^2\gamma^2\mathtt{\nu}^{(2)}}{M}}.
\end{eqnarray}
Here, $\mathtt{\nu}^{(2)}=\sum_{m \neq n} \vert \langle \Psi_{m}\vert {\boldsymbol r} \vert \Psi_{n} \rangle \vert^2/(E_{n}-E_{m})$, which is proportional to the polarisability of the excited state and scales with $n^7$. 
Note that this frequency shift is about two orders of magnitude larger for both radial modes as compared to the axial one because of the larger field gradients of the trapping field in the $x$-$y$ plane, under typical operation parameters. Further analysis of these effects is considered in relation with the treatment of parasitic stray electric fields in Sec.~\ref{subsec:micromotion}.

\subsection{Loading and laser cooling of ions}\label{subsec:lasercooling}
In experiments with Rydberg ions, the use of {\bf a fast and reliable loading method} is necessary, since the ion loss probability increases by Rydberg excitations~(Sec.~\ref{subsec:doubleioni}). 
An atomic beam is produced either by using a resistively heated oven or by laser ablation from a solid state target~\citep{leibrandt07a}. Ions are generated inside the trap by photoionization which is effective and can be used for isotope selective loading~\citep{hendricks07a, lucas04a, wolf18a}.
One approach for effective loading is to photoionise laser-cooled neutral atoms inside the trap volume~\citep{bruzewicz16a, cetina07a}. Alternatively, in ion traps with segmented electrodes, a reservoir of ions stored in a separate trapping zone can be used~\citep{blakestad09a}. In this method, ions are transported from such a loading zone into the experimental zone at which the laser pulses for qubit control are applied. Such a ``remote loading'' has been applied in experiments with Rydberg Ca$^{+}$ ions in the Mainz blade trap with segmented electrodes~\citep{mokhberi19a}, see Fig.~\ref{fig:setup_Ca1ph}. An additional advantage of remote loading is that slight contamination of trap electrodes due to deposition of the atomic beam or the build-up of charges from photoionization laser beams occur only near the loading zone, and the electrodes near the experimental zone are not affected. To improve the loading efficiency and to reduce ion heating rates, appropriate design of the trap electrode geometry and optimisation of applied static voltages and their time sequences are important~\citep{clercq16a, home06a}.   

The likelihood of double ionisation events might increase when using a thermal oven, which can be inferred to the black-body radiation effect~\citep{higgins18a}. A good low-temperature alternative is cryogenic Paul traps~\citep{brandl16a}, in which the atomic flux is generated by laser ablation~\citep{leibrandt07a}. Ion loading by laser ablation was reported to improve experiments with Rydberg Sr$^{+}$ ions~\citep{higgins18a}~(Sec.~\ref{subsec:doubleioni}). 

After loading, ions are {\bf Doppler-cooled on a dipole-allowed transition} to an equivalent temperature of about a few millikelvin~\citep{eschner03a, leibfried03a, metcalf99a}. 
In this temperature regime, the mean phonon number in motional eigenstates~$n_{\rm mean} \leq 1$, and sub-Doppler cooling techniques are applied for preparing the ground state as a starting point for coherent manipulation of electronic and motional quantum states~\citep{wineland98a}. Different techniques have been employed for {\bf sub-Doppler cooling}, namely, cooling on a dipole-forbidden quadrupole-allowed transition, cooling on stimulated Raman transitions~\citep{leibfried03a}, cooling using electromagnetically induced transparency~\citep{morigi00a} and polarization gradient cooling~\citep{dalibard89a}.
 
In this review, we focus on the first cooling method, which is typically implemented by using {\bf an optical transition whose radiative lifetime is narrow} as compared to the period of the ion oscillations in the trap. This method is efficient in the Lamb-Dicke regime~(Sec.~\ref{subsec:lamb_dicke}), in which spectrally resolved sidebands of the ion motion occur (so called strong coupling limit). This technique has enabled nearly perfect motional ground state cooling ($>99$\%)~\citep{eschner03a, leibfried03a, wineland98a}. As an example, $^{40}$Ca$^{+}$ ions are Doppler cooled on the $(4s)^{2}$S$_{1/2}\rightarrow(4p)^{2}$P$_{1/2}$ transition while pumping on the $(3d)^{2}$D$_{3/2}\rightarrow(4p)^{2}$P$_{1/2}$ and $(3d)^{2}$D$_{5/2}\rightarrow(4p)^{2}$P$_{3/2}$ transitions using laser beams at 397, 866 and 854~nm, respectively~(Fig.~\ref{fig:detection}(a)). 
The laser frequency is ``red-detuned'' from the cooling transition by approximately $\Gamma_{\rm c}/2$, where $\Gamma_{\rm c}$ is the natural linewidth of the transition. 
Sideband spectroscopy of motional modes is performed using ultra-stable laser light at 729~nm that drives all different transitions between sublevels on the $(4s)^{2}$S$_{1/2} \rightarrow(3d)^{2}$D$_{5/2}$ quadrupole transition ($\Gamma_{\rm q}=0.14$~Hz~\citep{chwalla09b}). The 4S$_{1/2}$, $m_{J}=-1/2 \rightarrow$~3D$_{5/2}$, $m_{J}=-5/2$ transition is a good choice for sideband cooling~\citep{roos00a}, and serves for ion initialisation and state-dependent fluorescence detection (Sec.~\ref{subsec:initial}~and~Sec.~\ref{subsec:detection}). Similar schemes and techniques are applicable to Sr$^{+}$ ions~(Figs.~\ref{fig:detection}(b,c)).

Ground state cooling of ions may be required for coherent manipulation of Rydberg states as the thermal distribution of motional quanta in normal modes can cause frequency shifts and asymmetrical broadening of Rydberg lines~\citep{higgins17a}. In addition, phonon number dependent transitions might be strongly driven and hinder coherent control of Rydberg states~\citep{higgins19a}. These effects are described in terms of significant Stark shifts that lead to the modification of trapping frequencies as given in Eqns.~\ref{eq:omegaXryd}-\ref{eq:omegaZryd} in Sec.~\ref{subsec:micromotion}.

\subsection{Initialization of electronic states}\label{subsec:initial}
As starting point for driving a Rydberg transition, it is essential to initialise the ion in a pure state by {\bf optical pumping}. Here, we describe initialisation techniques for four cases that are relevant for experiments with Rydberg $^{40}$Ca$^{+}$ ions, noting that there is a one-to-one similarity to those applied to $^{88}$Sr$^{+}$ ions~(Fig.~\ref{fig:leveldiagrams}). Spectroscopic investigations rely upon a cycle of consecutive steps of cooling, initialisation, optical pumping, Rydberg excitation and detection. Such a sequence is performed typically in 5--20~ms on a single ion or ions in a crystal and is repeated 100 times. 

\begin{itemize}
\item Initialisation in 4S$_{1/2}$, $m_{J}=-1/2$ state:
Any population in metastable 3D$_{3/2}$ and 3D$_{5/2}$ states is transferred to the 4S$_{1/2}$ state using laser light at 866~nm and 854~nm. Both ground state Zeeman sublevels are populated, but the population in 4S$_{1/2}$, $m_{J}=+1/2$ state is frequency-selectively excited to $3$D$_{5/2}$, $m_{J}=-3/2$, from where it is quenched again by radiation near 854~nm. After repeating the cycle ten times, the 4S$_{1/2}$, $m_{J}=-1/2$ state is prepared with better than 97$\%$ probability.
\item Initialisation in metastable 3D$_{5/2}$, $m_{J}=-5/2$ state:
Starting with 4S$_{1/2}$, $m_{J}=-1/2$,  the ion is coherently transferred into the 3D$_{5/2}$, $m_{J}=-5/2$ state using a $\pi$-pulse of laser light at 729~nm. The efficiency of this ``electron shelving'' typically exceeds 99$\%$, verified by state-dependent fluorescence detection~(Sec.~\ref{subsec:detection}).
\item Initialisation in a superposition state of 4S$_{1/2}$, $m_{J}=-1/2$ and 3D$_{5/2}$, $m_{J}=-5/2$:
Applying a $\pi/2$-pulse of laser light at 729~nm initialises the ion in an equal superposition. Controlling the laser phase, frequency and the laser pulse duration allows for generating arbitrary superposition states of Zeeman sublevels~\citep{ruster16a}. 
\item Initialisation in 3D$_{3/2}$ states:
The population is optically pumped to 4S$_{1/2}\rightarrow$~4P$_{1/2}$ transition followed by a decay with a probability of 6\% into the 3D$_{3/2}$ state~\citep{roos00a}. Note that in this way an incoherent mixture of Zeeman sublevels is populated.
\end{itemize}

To enable population transfer with {\bf improved robustness}, square laser pulses at constant frequency and intensity are replaced by optimised laser pulses such as rapid adiabatic passage~(RAP)~\citep{noel12a, poschinger09a, poschinger12a, wunderlich07a, yamazaki08a} and stimulated Raman adiabatic passage~(STIRAP)~\citep{gebert16a, sorensen06a}. In RAP, the adiabatic transfer of a population between two atomic states is implemented by a frequency and amplitude modulated laser pulse as the dynamics of the atomic state dressed by this light field evolves. In a typical experiment, the laser intensity is shaped with Gaussian time dependence, whereas the frequency is varied linearly in time across an atomic resonance.
A sequence of initialisation in 4S$_{1/2}$, $m_{J}=-1/2$ state (the first item above) combined with a RAP pulse allows for initialisation with 99.8\% probability, e.g, used in four-ion entanglement generation demonstarated in~\citep{kaufmann17a}.   
In STIRAP, usually a three-level $\Lambda$-system is used such that the adiabatic transfer of the population from the initial state to the final state occurs with neglible population of a short-lived intermediate state. The application of STIRAP for manipulation of Rydberg states driven in a two-step excitation process is discussed in Sec.~\ref{sec:coh_spectroscopy:STIRAP}. 

\subsection{Detection and readout schemes}\label{subsec:detection}
\begin{figure}
\centering
\includegraphics[width=1.0\textwidth]{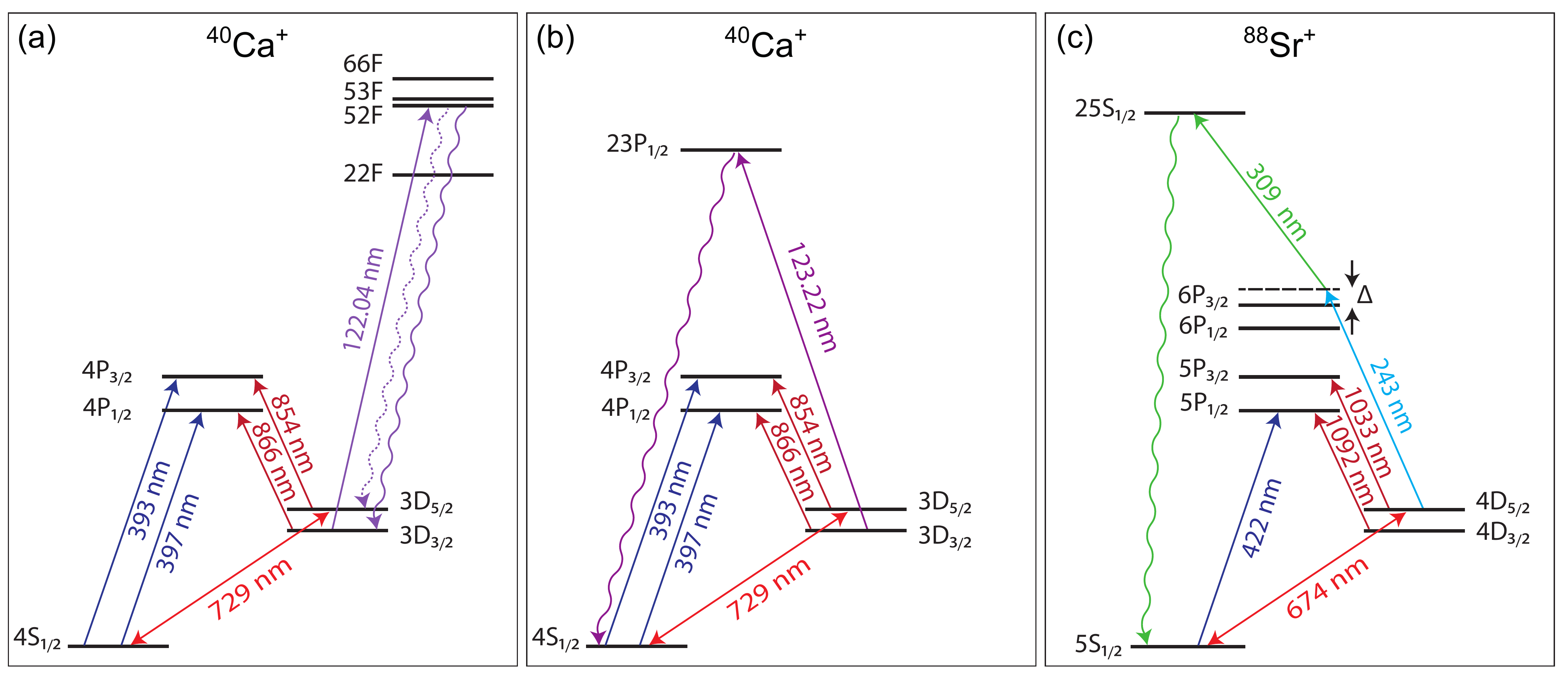}
\caption{Successful Rydberg excitations are measured based on electron shelving and state-dependent fluorescence detection using the 4S$_{1/2}$, 4P$_{1/2}$ and 3D$_{5/2}$ states in Ca$^{+}$ ions, and the 5S$_{1/2}$, 5P$_{1/2}$ and 4D$_{5/2}$ states in Sr$^{+}$ ions. Energy level diagrams and laser wavelengths relevant for detecting Rydberg states are shown for (a) 52F$_{7/2}$ and (b) 23P$_{1/2}$ states of $^{40}$Ca$^{+}$ and (c) for 25S$_{1/2}$ state of $^{88}$Sr$^{+}$ ions. In the two-step excitation in (c), the laser at 243~nm: (i) is detuned by $\Delta$ from the intermediate, short-lived 5P$_{3/2}$ state for the direct detection scheme and (ii) is on resonance with the 5P$_{3/2}$ state for the EIT detection scheme, see text for details.}
\label{fig:detection}
\end{figure}

For {\bf non-destructive and highly efficient detection} of Rydberg excitations of trapped ions, electron shelving techniques are used. The ion fluorescence from the electric dipole-allowed transition used for Doppler cooling is detected by an electron-multiplying charge-coupled device (EMCCD) camera (see, e.g., the setup in Fig.~\ref{fig:setup_Ca1ph}(a)) or a photomultiplier detector (PMT). Relevant states and transitions for such detection schemes are illustrated for Ca$^{+}$ Fig.~\ref{fig:detection}(a, b) and Sr$^{+}$ ions in Fig.~\ref{fig:detection}(c). For Ca$^{+}$ ions, a typical choice is the 4S$_{1/2}\rightarrow$~4P$_{1/2}$ transition with 1~ns lifetime of the upper level and transition wavelength near $397$~nm (Fig.~\ref{fig:detection}(a)). 
The ion scatters millions of photons per second when this transition is driven, i.e., 4S$_{1/2}$ state is populated, and thus this state is counted as a ``bright'' state. If the ion is initialised in 3D$_{5/2}$ state, which does not scatter photons at that frequency, it is counted as a ``dark'' state. This state is independently coupled to the S ground state by an electric quadrupole allowed transition. Such a V-type three-level system allows for electron shelving and state-dependent fluorescence detection (Fig.~\ref{fig:detection}). Measuring collected photons for few hundred~microseconds allows to distinguish between the bright and dark states with better than 99.99\% efficiency~\citep{myerson08a}.
 
A successful excitation from a Zeeman sublevel of the 3D$_{5/2}$ state to a Rydberg state is revealed by discriminating between the remaining population in the initial state(s), which needs to be corrected only by the small probability that the Rydberg state decays back into the initial level~\citep{schmidtkaler11a}. Successful events are thus detected using subsequent state-dependent fluorescence detection in dark and bright states. 
For a given Rydberg transition to be investigated spectroscopically, the detection scheme has to be adapted according to rates for possible decay channels. Excitation to states with high principal quantum numbers are preferred for taking advantage of enhanced Rydberg properties~(Table~\ref{table:properties}). To avoid coupling between Rydberg levels with $j\geq 1/2$ due to trapping electric fields~(Sec.~\ref{subsec:rydbergionintrap}), we employ quantum states with total angular momentum~$j=1/2$. For coherent manipulation, S transitions are well suited~\citep{higgins17b}, whereas for quantum gate operations, P states are preferred because of longer lifetimes~\citep{vogel19a, zhang20a}. If the decay rate of a Rydberg level is much faster as compared to the excitation rate, the process acts similar to optical pumping.

Here, we give {\bf three examples for detecting Rydberg} F and P states in Ca$^{+}$ ions and S states in Sr$^{+}$ ions (Fig.~\ref{fig:detection}). (i) For the 3D$_{3/2}\rightarrow$~52F$_{1/2}$ transition, the excited state population decays into $3$D$_{5/2}$ and $3$D$_{3/2}$ states with about 93\% and ~7\% probabilities respectively~\citep{feldker16a}, where the latter causes a reduction on the detection efficiency. The successful Rydberg excitation is proven from the ion fluorescence on the S$_{1/2}\rightarrow$~P$_{1/2}$ transition (bright state). The efficiency of this scheme can be increased by using Zeeman sublevels of the D$_{5/2}$ state~\citep{bachor16a}. (ii) For the 3D$_{3/2}\rightarrow$~23P$_{1/2}$ transition, one should take into account that P and S states of the Ca$^{+}$ and Sr$^{+}$ ions decay in multiple steps into the S$_{1/2}$ ground state of the ion with high probabilities~\citep{glukhov13a}. Thus, shortly after a successful Rydberg excitation, the ion appears in the bright state, see Fig.~\ref{fig:detection}(b). (iii) For the 4D$_{5/2}\rightarrow$~25S$_{1/2}$ transition, the Rydberg state decays into the ground state~(Fig.~\ref{fig:detection}(c)), and the excitation probability is deduced from the population of these initial~(dark) and final~(bright) states. 

In Rydberg excitation using the two-step scheme~(Sec.~\ref{sec:exp_approaches}), an alternative detection technique based on the {\bf electromagnetically-induced-transparency}~(EIT) effect~\citep{harris97a} has been developed in experiments with Sr$^{+}$ ions~\citep{higginszhang20a}. EIT is a quantum interference effect that is observed as the absorption reduction of a weak light field, which is on resonance with an atomic transition, in the presence of a second light field that near-resonantly couples the upper level to a third level~\citep{boller91a}.
Further examples of such coherent phenomena in a three-level $\Lambda$-system used for two-step Rydberg excitation of ions are discussed in Sec.~\ref{sec:coh_spectroscopy}.  
The three-level $\Lambda$-system for a $^{40}$Ca$^{+}$ ion is represented by the initial 3D$_{5/2}$ state, the intermediate 5P$_{3/2}$ state and a certain Rydberg state, which can be any state in the $n$S$_{1/2}$, $n$D$_{3/2}$ or $n$D$_{5/2}$ series. The probe laser at 213~nm is set at the 3D$_{5/2} \rightarrow 5$P$_{3/2}$ resonance, while the pulse laser at 287~nm is scanned near the 5P$_{3/2} \rightarrow n$S$_{1/2}$ transition (Fig.~\ref{fig:detection}). 
Ions are initialised in one of the Zeeman levels of the 3D$_{5/2}$ (dark) state. Since the pump laser is in resonance with the 5P$_{3/2}$ state with a lifetime of about 10~ns~\citep{safronova11a}, a bright signal is counted. Once the probe laser is on resonance with the Rydberg state, the 5P$_{3/2}$ state energy level is shifted due to the coupling between this intermediate state and the Rydberg state. Consequently, the P state becomes transparent to the probe laser at 213~nm and the population of the 3D$_{5/2}$ state increases. Such peaks in the dark state population are used to determine the transition frequencies with the advantage of avoiding losses or effects from the trapping field. This technique was applied to measure S-and D-series resonances in Ca$^{+}$ ions (Fig.~\ref{fig:qdefect}).

\subsection{Controlling electric fields and minimising stray fields}\label{subsec:micromotion}
Because of the large polarisability of Rydberg ions, minimising stray electric fields at the ion position is crucial to avoid significant Stark shifts.
For trapped Rydberg ions, the Stark effect can be treated as a perturbation to the trap potential since only time scales exceeding the RF period $\Omega_{\rm RF}$ and energy shifts much smaller than~$\hslash \Omega_{\rm RF}$ are relevant. For a state with polarisability $\alpha$, the Stark shift is given by $\Delta E_{\rm Stark}=-\frac{1}{2} \alpha \langle {\boldsymbol E({\boldsymbol R},t)}^2 \rangle_{t}$, where the electric field near the trap centre is given by  
\begin{equation}
\begin{aligned}
{\boldsymbol E}({\boldsymbol R},t)= & -2\gamma^{\prime} (X \hat{x}-Y \hat{y}) \cos (\Omega_{\rm RF}t)+2\gamma( (1+\epsilon) X \hat{x}+ (1-\epsilon)Y \hat{y}-2Z \hat{z}) \\
& +{\boldsymbol E}_{\rm 0} \varphi_{\rm RF} \sin (\Omega_{\rm RF}t)+ {\boldsymbol E}_{\rm st}.
\end{aligned}
\label{eq:micromotionE}
\end{equation}
Here, ${\boldsymbol R}\equiv X \hat{x}+Y \hat{y}+Z \hat{z}$ is the ion position vector, $\varphi_{\rm RF}$ is the phase difference between the RF potentials applied to trap electrodes with amplitude of ${\boldsymbol E}_{\rm 0}$ near the trap centre, and $\gamma^{\prime}$, $\gamma$ and $\epsilon$ identify the trapping potential as given in Eqn.~\ref{eq:paultrap}.
${\boldsymbol E}_{\rm st}$ accounts for an stray electric field that shifts the ion off the RF centre and causes extra oscillation, ``excess micromotion''. In most experiments, $\varphi_{\rm RF}$ is nulled using an appropriate trap design, and thus the main contribution to excess micromotion is due to stray electric fields. An additional source of fluctuating electric fields is photo electrons from stray light of the UV or VUV beams~(Sec.~\ref{sec:exp_approaches}). The off-centre shift of a singly-charged ion in a linear RF trap due to ${\boldsymbol E}_{\rm st}$ is given by~\citep{berkeland98a}
\begin{equation}
{r}_{\rm d}= \frac{e {\boldsymbol E}_{\rm st} \cdot \hat{{\boldsymbol e}}_{\alpha}}{M \omega^{2}_{\alpha}}, 
\label{eq:micromotionX}
\end{equation}
where $\hat{{\boldsymbol e}}_{\alpha}$ is the unit vector of the mode eigenvector at $\omega_{\alpha}$. 

{\bf Depending on the spatial distance between the minima of the static and the RF trapping electric fields}, two different cases are considered. {\bf In case both minima overlap} within the extension of the vibrational ground state wavefunction, which is typically about 10-50~nm, the stray electric field and the ion displacement are negligible, i.e., ${\boldsymbol E}_{\rm st}=0$ and ${\boldsymbol r}_{\rm d}=0$. Assuming the vibrational components of the wavefunction remain approximately unchanged during the Rydberg transition, the energy shift can be written as~\citep{higgins19a}
\begin{equation}
\Delta E_{\rm Stark}= (n_{X}+\frac{1}{2}) \hslash (\omega^{\prime}_{X}(\alpha_{r})-\omega^{\prime}_{X}(\alpha_{g}))+ (n_{Y}+\frac{1}{2}) \hslash (\omega^{\prime}_{Y}(\alpha_{r})-\omega^{\prime}_{Y}(\alpha_{g})),
\label{eq:EphononchangingWEAK}
\end{equation}
where $n_{X(Y)}$ is the phonon number in the radial mode of oscillation along the $X(Y)$ axis and $\alpha_{r}$ ($\alpha_{g}$) is the polarisability of the $\ket{r}$ (or $\ket{g}$) state, respectively. The trapping frequency $\omega^{\prime}_{X(Y)}=\omega_{X(Y)}+\Delta \omega_{X(Y)}$ is given in Eqns.~\ref{eq:secularfreqX}-\ref{eq:secularfreqZ} and Eqns.~\ref{eq:omegaXryd}-\ref{eq:omegaZryd}. Note that the effect arises from a slightly stiffer or shallower harmonic trapping potential as seen by the ionic core depending on the polarisability of the Rydberg state~$\alpha_{r}$. This effect may be enhanced by 7 to 9 orders of magnitude as compared to the near-ground state ion~$\alpha_{g}$~(Eqns.~\ref{eq:omegaXryd}-\ref{eq:omegaZryd}). 
Electronic wavefunctions of initial and highly-excited states are not necessarily orthogonal, and thus the phonon conservation is not strictly applied. Such phonon-number-changing transitions are most likely driven {\bf in the case of imperfect micromotion compensation}, in which the minima of the static and RF trapping electric fields do not overlap, i.e., ${\boldsymbol E}_{\rm st}\neq0$ and ${\boldsymbol r}_{\rm d}\neq0$. 
The equilibrium position of the ion is consequently shifted by
\begin{equation}
X^{\prime}_{\rm d}= X_{\rm d} (1-\frac{2 \alpha {\gamma^{\prime}}^2}{M \omega^{2}_{X}})^{-1}.
\label{eq:Xdphononchanging}
\end{equation}
In this case, the Stark shift results in~\citep{higgins19a}  
\begin{equation}
\Delta E_{\rm Stark}= -\frac{1}{2} \alpha \langle {\boldsymbol E({\boldsymbol r}_{\rm d},t)}^2 \rangle_{t}= -\alpha {\gamma^{\prime}}^2 (X^{2}_{\rm d}+Y^{2}_{\rm d}).
\label{eq:EphononchangingSTRONG}
\end{equation}

\begin{figure}
\centering
\includegraphics[trim={0 3cm 4cm 0},clip,width=1.0\textwidth]{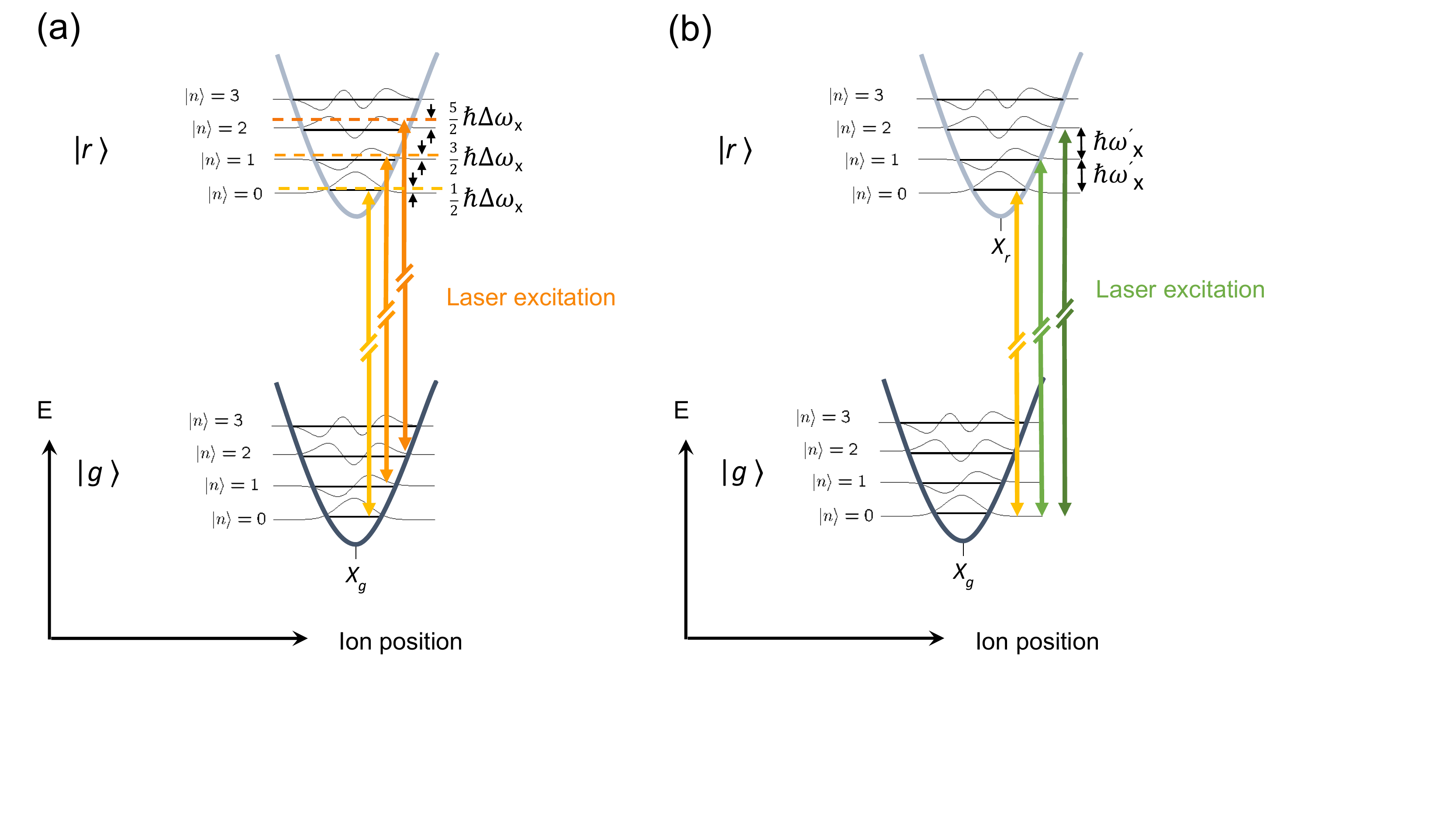}
\caption{Illustration of transition frequency shifts in Rydberg excitation of trapped ions owing to the Stark effect. (a) Phonon-number-preserving transitions occur in case of non-zero motional quanta in a given mode. (b) Phonon-number-changing transitions are driven strongly in the case of ions subjected to stray electric fields. Note that in this case the equilibrium position of the Rydberg ion $X_{r}$ is different from that of the ground-state $X_{g}$. For the phonon states shown in (a) and (b), wavepackets of the ion in a harmonic potential well in the $x$ direction are depicted in grey.}
\label{fig:phononchanging}
\end{figure}

In both cases of phonon-number-dependent energy shifts (Fig.~\ref{fig:phononchanging}), the shifts can be seen as higher-order terms contributing to the ion-trapping field coupling given in Eqn.~\ref{eq:eTrapspherical}. This effect is more pronounced when stray electric fields cause excess micromotion. Micromotion effects alter the shape of the atomic transition resonances, due to Stark shifts induced by the RF electric field and from second-order Doppler shifts. Even with nearly perfect minimisation of excess micromotion, asymmetric broadening of Rydberg resonances may occur if the ion(s) are only Doppler-cooled. This effect is understood in terms of the broadening of the thermal wavepacket, with phonon distributions in radial and axial modes, when the Stark shift is taken into account~(Sec.~\ref{subsubsec:trapeffect-Polar}). 
  
On the positive side, such state-dependent modification of trapping frequencies plays an important role. The effect can be understood as a difference in the effective mass for the ion in the Rydberg state as compared to an ion in the low-lying or ground state, which modifies the frequencies of the common modes of vibration as well as the ion equilibrium positions in a state-depending fashion. This unique property in ion crystals consisting of Rydberg ions has rich applications in quantum simulation~(Sec.~\ref{subsec:excitationtransfer}), quantum computation~(Sec.~\ref{subsec:fastgatewithEfield}) and structural phase transitions~(Sec.~\ref{subsec:phasetransition}).

\section{Spectroscopy of Rydberg transitions}\label{sec:spectro}
In this section, we discuss experimental results for {\bf incoherent Rydberg spectroscopy} of single trapped ions, in which the laser-interaction time exceeds the lifetime of excited states. These spectroscopy results are important for understanding Rydberg state properties~(Sec.~\ref{subsec:polarisability} and Sec.~\ref{subsec:Qdefect}), for revealing the effect of trapping fields on Rydberg spectral line properties~(Sec.~\ref{subsec:trapeffects}) and for applications in sensing and metrology~(Sec.~\ref{subsec:polarisability}).
Note that in the relatively new experiments with Rydberg ions only few atomic properties have been thus far measured with required precisions, and therefore new spectroscopy data are essential. 


\subsection{Determination of the electric polarisability}\label{subsec:polarisability}
\begin{figure}
\centering
\includegraphics[width=0.75\textwidth]{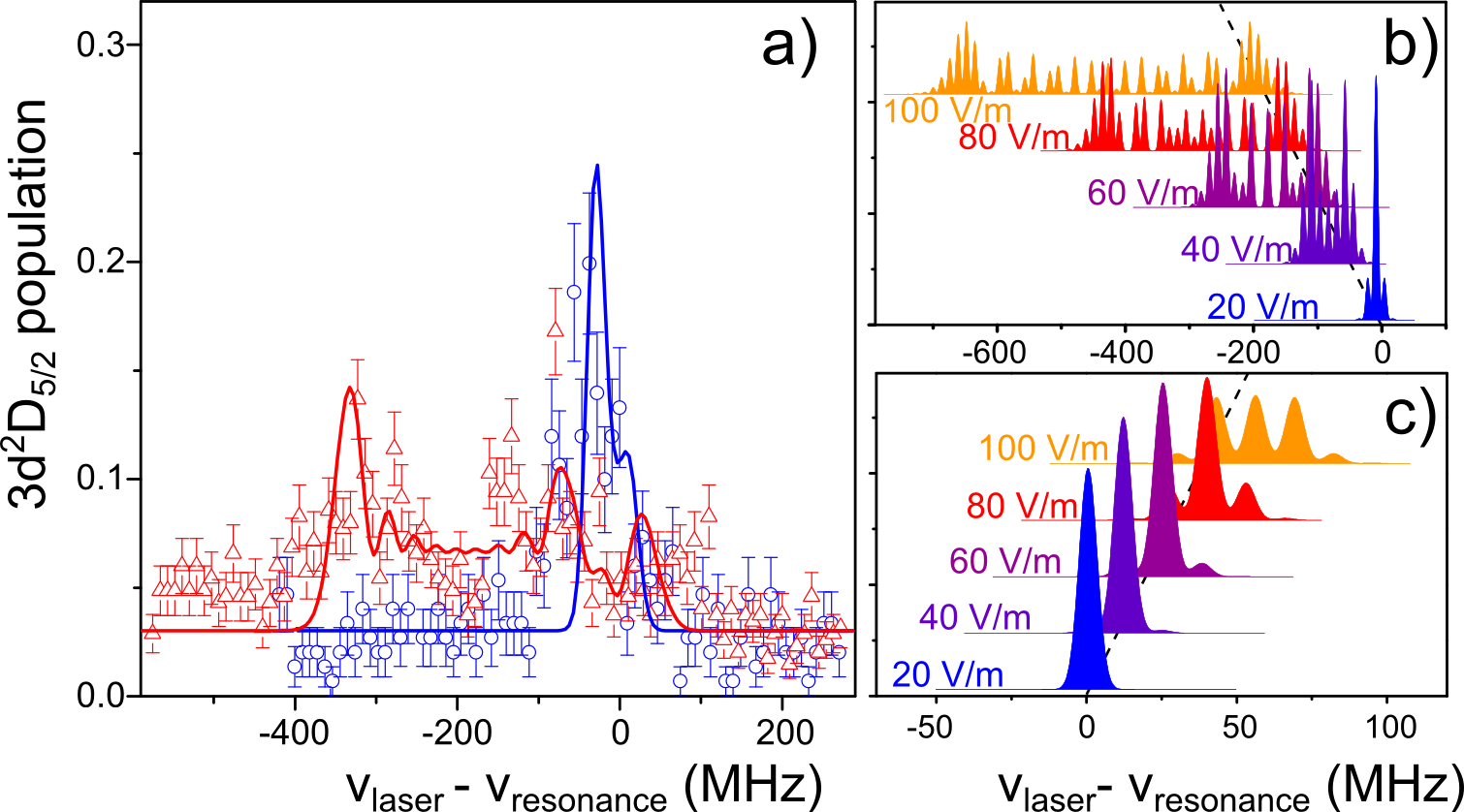}
\caption{Rydberg lines for trapped $^{40}$Ca$^{+}$ ions. (a) The $3$D$_{3/2}\rightarrow 52$F transition measured at $\vert {\boldsymbol E}_{res} \vert=24$~V/m~(in blue) and $84$~V/m~(in red) with a laser pulse of 30~ms duration and power of about 0.5~$\mu$W with a beam waist of about 15~$\mu$m. (b) and (c) Calculations for 51F and 51P resonances excited form $3$D$_{3/2}$ state at different residual electric fields respectively, for which cooling to the Doppler limit and the trap drive frequency~$\Omega_{\rm RF}=2\pi \times 6.5$~MHz is assumed. Adapted from~\protect\citep{feldker15a}.}
\label{fig:Fline}
\end{figure}

Rydberg atoms and ions exhibit exaggerated electric polarisability~(Table~\ref{table:properties}), which can be deduced from the modification of their line shapes in the presence of an external electric field. F states of hydrogen-like ions feature large polarisability, see for instance calculated values in~\citep{kamenski14a}, and are ideal for such studies. The trap was operated such that a single Ca$^{+}$ ion undergoes micromotion along the trap axis. Ions were initialised in 3D$_{3/2}$ states and were excited to the $52$F state using laser light at $122$~nm. The spectrum is shown at two different oscillating electric field strengths at the ion position~\citep{feldker15a}, Fig.~\ref{fig:Fline}(a). The linewidth of the 52F transition varies between 60 and 400 MHz full width at half maximum (FWHM), depending on the trap control parameters, and widens out at higher electric field strengths.

To describe the line shapes quantitatively, the Stark shift due to the oscillating electric field of the trap and the Doppler effect resulting from the driven motion of the ion were modelled. The time-dependent resonance frequency of the transition can be written as~\citep{feldker15a}
\begin{equation}
\omega(t)=\omega_{0}+\boldsymbol{k}\cdot\boldsymbol{R}_{\rm mm}{\rm \Omega_{\rm RF}} {\rm sin}({\rm \Omega_{\rm RF}}t)-\frac{\alpha {\boldsymbol{E}_{\rm res} ^{2}}}{2 \hslash} {\rm cos}^{2}({\rm \Omega_{\rm RF}}t).
\label{eq:freq}
\end{equation} 
Here, $\omega_{0}$ is the unaffected resonance frequency, and $\boldsymbol k$ and ${\boldsymbol R}_{\rm mm}$ are the wavevector of the VUV laser and the ion micromotion amplitude. $\alpha$ is the polarisability of the Rydberg state, and $\boldsymbol{E}_{\rm res}$ is a residual RF electric field at the ion position. Thus, the modulated laser field seen by the ion is~\citep{feldker15a}
\begin{equation}
E_{\rm laser}(t)\propto e^{-i \omega_{0} t} e^{i 2 \beta_{\alpha} \rm \Omega_{\rm RF} t}  \sum_{m^{\prime}}{J_{m^{\prime}}(\beta_{\rm mm})} e^{i m^{\prime} (\rm \Omega_{\rm RF} t+\frac{\pi}{2})} \sum_{m}{ (-1)^m J_{m} (\beta_{\alpha}) e^{2i m (\rm \Omega_{\rm RF} t)}}.
\label{eq:lineshape}
\end{equation} 
Here, $\beta$ denotes the modulation index of the Bessel function $J_{n}(\beta)$ with $\beta_{\rm mm}=\boldsymbol{k}\cdot\boldsymbol{R}_{\rm mm}$ and $\beta_{ \alpha}=\alpha {\boldsymbol{E}_{\rm res} ^{2}}/8 \rm \Omega_{\rm RF}$, respectively. 
Micromotion sidebands caused by the Doppler effect appear at $m \times \rm \Omega_{\rm RF}$, whereas micromotion sidebands due to the quadratic Stark effect occur at $2 m \times \rm \Omega_{\rm RF}$, where $m$ is an integer number~(see Figs.~\ref{fig:Fline}(b,c) and Fig.~\ref{fig:23pline}(b)). A Floquet analysis of such spectra has been recently computed~\citep{pawlak20a}. 

By measuring $\vert \boldsymbol{E}_{\rm res} \vert$ independently, see Sec.~\ref{subsubsec:trapeffect-doppler} and Fig.~\ref{fig:23pline}(c), one can estimate the polarisability of an excited state from fitting the observed resonances to the model given in Eqn.~\ref{eq:lineshape}. Note that such a simple measurement allows for a quick identification of the angular momentum of an unknown Rydberg state from the sign and magnitude of  $\alpha$. The result is~$\alpha_{52 \rm F}=10^{+7}_{-3} \times 10^2$~MHz/ (V/cm)$^{2}$~\citep{feldker15a} for the 52F state, a value consistent with the predicted value of $\alpha_{52 \rm F}=8 \times 10^2$~MHz/ (V/cm)$^{2}$~\citep{kamenski14a} obtained in second-order perturbation theory when spin-orbit coupling is neglected.

\subsection{Resolved Zeeman sublevels of Rydberg states}\label{subsec:zeeman}
The coupling terms due to the trapping electric fields (Eqn.~\ref{eq:eTrapspherical}) are negligible for $n$S states up to $n<50$~\citep{mueller08a}, and therefore $m_{J}$ remains a ``good'' quantum number for low-lying states as well as for Rydberg S-states. In a two-step laser excitation of ions, single Zeeman sublevels are driven using circularly polarised UV beams that drive a given Rydberg transition (Sec.~\ref{sec:exp_approaches}). Depending on the polarisation of the light, each laser beam may drive $\sigma^{+}$ transitions, $\sigma^{-}$ transitions or both transitions. For instance, see resolved resonances for Zeeman sublevels of Rydberg S levels of Sr$^{+}$ ions on the 4D$_{3/2} \rightarrow $~25S$_{1/2}$ transition in~\citep{higgins17a}.

\subsection{Rydberg series and determination of the quantum defect}\label{subsec:Qdefect}
To assign observed Rydberg resonances, measured lines are fitted to a line model, e.g., the one described in Sec.~\ref{subsec:polarisability}, and resulting level energies are fitted to the Rydberg-Ritz formula~\citep{ritz1908a, rydberg1890a}
\begin{equation}
E_{n,l,j}=I^{++}-\frac{R^{*} \mathcal{Z}^{2}}{(n-\mu(E))^{2}}+\frac{R^{*} \mathcal{Z}^{4} \alpha_{ls} ^{2}}{(n-\mu(E))^{3}}{\Big [}\frac{3}{4(n-\mu(E))}-\frac{1}{(j+1/2)}{\Big ]}
\label{eq:Rydberformula}
\end{equation}
Here, $\alpha_{ls}$ and $R^{*}$ are the fine structure constant and the reduced Rydberg constant and $\mathcal{Z}=+2$ is the charge of the ionic core. $I^{++}$, $\mu(E)$ and $n$ denote the double ionization limit, the quantum defect and the principal quantum number respectively.
The third term in Eqn.~(\ref{eq:Rydberformula}) accounts for the fine-structure splitting. 
The energy dependence of the quantum defect $\mu(E)$ is approximated up to second order by a Taylor expansion
\begin{equation}
\mu(E_{n,l,j})= \mu ^{0}_{l,j}(I^{++})-\frac {\partial \mu}{\partial E_{n,l,j}} \frac{R^{*}}{(n-\mu^{1}_{l,j})^{2}}+O [(E_{n,l,j}-I^{++})^{2}],
\label{eq:Qdefect}
\end{equation}
where $\mu ^{0}_{l,j}$, $\mu ^{1}_{l,j}$, $I^{++}$ and ${\partial \mu}/{\partial E_{n,l,j}}$ are treated as fit parameters. 
The model compiled from Eqns.~(\ref{eq:Rydberformula}-\ref{eq:Qdefect}) is usually simplified by replacing $\mu^{1}$ by $\mu^{0}$ which can provide sufficiently good results, as reported for $ns,np,nd$ series of Rb atoms~\citep{li03a} and for $ns,nd,nf,ng$ series of $^{88}$Sr$^{+}$ ions in free space~\citep{lange91a}. However, this simplification spoils the meaning of the quantum defect as discussed by Drake and Swainson \citep{drake91a} and can ultimately limit the accuracy of the quantum defect method for precision measurements.
The truncation of higher-order terms in Eqn.~\ref{eq:Qdefect} must be verified based on the fit results, e.g., see~\citep{deiglmayr16a}.

\begin{figure}
\centering
\includegraphics[width=0.6\textwidth]{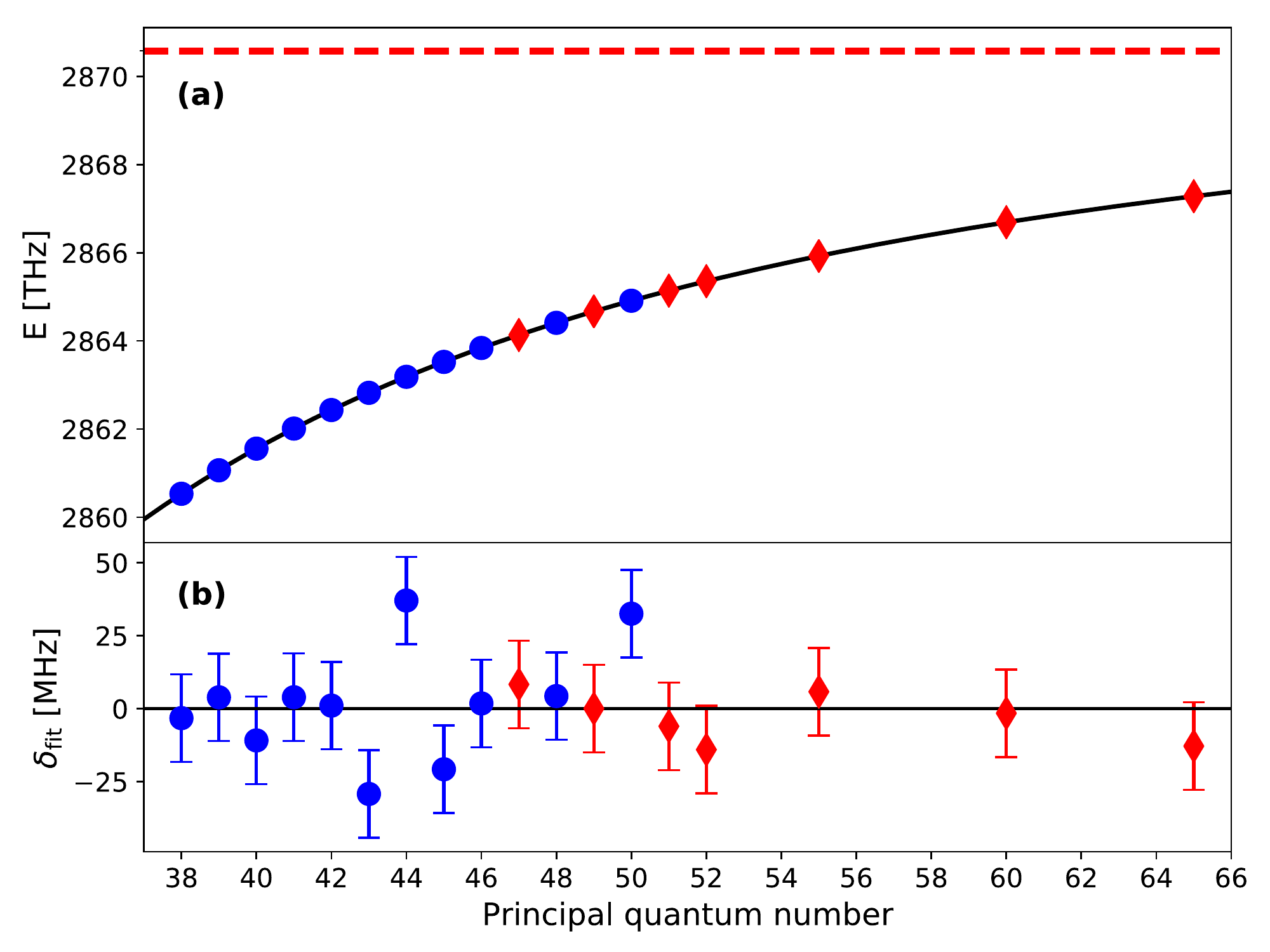}
\caption{(a)~and~(b)~S-state level energies of $^{40}$Ca$^{+}$ ions fitted to the Rydberg-Ritz formula and the corresponding fit residuals. 
Measurements using a two-step excitation scheme~(blue dots) and using EIT spectroscopy~(red diamonds) are plotted versus the principal quantum number.
In both cases, a single ion was initialised from the 3D$_{5/2}$, $m_{J}=-1/2$ state~\citep{chwalla09a}. The red dashed line shows the double ionization limit~$I^{++}$ deduced from the fit of measured level energies to Eqn.~\ref{eq:Qdefect}~(dashed red line). }
\label{fig:qdefect}
\end{figure}

Rydberg state spectroscopy in combination with Rydberg-series extrapolation provides the most precise method for determining the ionization threshold~$I^{++}$~\citep{deiglmayr16a, peper19a}. 
For trapped Rydberg ions, S and D Rydberg series of $^{88}$Sr$^{+}$ ions~\citep{higgins18a} and S, P, D and F series of $^{40}$Ca$^{+}$ ions~\citep{andrijauskas20a, bachor16a, feldker15a, mokhberi19a} were used to determine the quantum defects and their ionization energies.
The most precise measurement has been done for the S states of a single Ca$^+$ ion using the two-step Rydberg excitation scheme~(Sec.~\ref{sec:exp_approaches}), in which states with principal quantum number $n$ ranging between 38 and 65 were excited~\citep{andrijauskas20a}, see Fig.~\ref{fig:qdefect}. Two methods were applied to detect Rydberg resonances; the direct excitation and the EIT techniques described in Sec.~\ref{subsec:detection}. The EIT technique does not yield any population transfer to the Rydberg state, and therefore has enabled the observation of highly excited states of trapped ions with $n>50$, beyond the theoretical limit for instability of Rydberg ions inside a Paul trap~\citep{mueller08a}, see also Sec.~\ref{subsec:doubleioni}. 

\subsection{Spectral line effects due to trapping electric fields}\label{subsec:trapeffects}
For Rydberg states of ions in the dynamic potential of a linear Paul trap, two classes of effects due to static and RF quadrupolar electric fields have been observed, arising either {\bf from high polarisability or from large quadrupole moments} of ions in Rydberg states. The former emerge as a result of the Rydberg electron interaction with the ionic core, described with $H_{\rm{e-Trap}}$ in Eqn.~\ref{eq:eTrapspherical}~(Sec.~\ref{subsec:rydbergionintrap}). The effect is noticeable for highly-excited states as the polarisability scales with~$n^7$~(Sec.~\ref{subsubsec:trapeffect-Polar}). The second effect is the coupling between states with large electric quadrupole moments in Rydberg states, which scales with $n^4$.
The coupling in this case is driven by the time-dependent electric quadrupole trapping field, and occurs between Rydberg states with $J>\frac{1}{2}$~(Sec.~\ref{subsubsec:trapeffect-Quadru}). $n$S$_{1/2}$ and $n$P$_{1/2}$ states (with $J=\frac{1}{2}$) have negligible quadrupole moments, and therefore are not affected, and are employed for coherent manipulation and for entangling operations~~(Sec.~\ref{sec:coh_spectroscopy}). Note that also the Zeeman sublevels of the 3D$_{5/2}$ state of Ca$^{+}$ ions exhibit a small, quadrupolar differential shift of about 30~Hz, which was observed using entangled states of a two-ion crystal~\citep{roos06a}. However, for Rydberg states this effect is significant. For instance, for the 27D$_{3/2}$ state of Sr$^{+}$ ion, the coupling is larger by a factor of about 10$^6$, corresponding to a calculated shift of about $30$~MHz~\citep{higgins18a}. 

\subsubsection{Micromotion sidebands due to the Doppler effect}\label{subsubsec:trapeffect-doppler}
\begin{figure}[h!]
\centering
\includegraphics[width=1\textwidth]{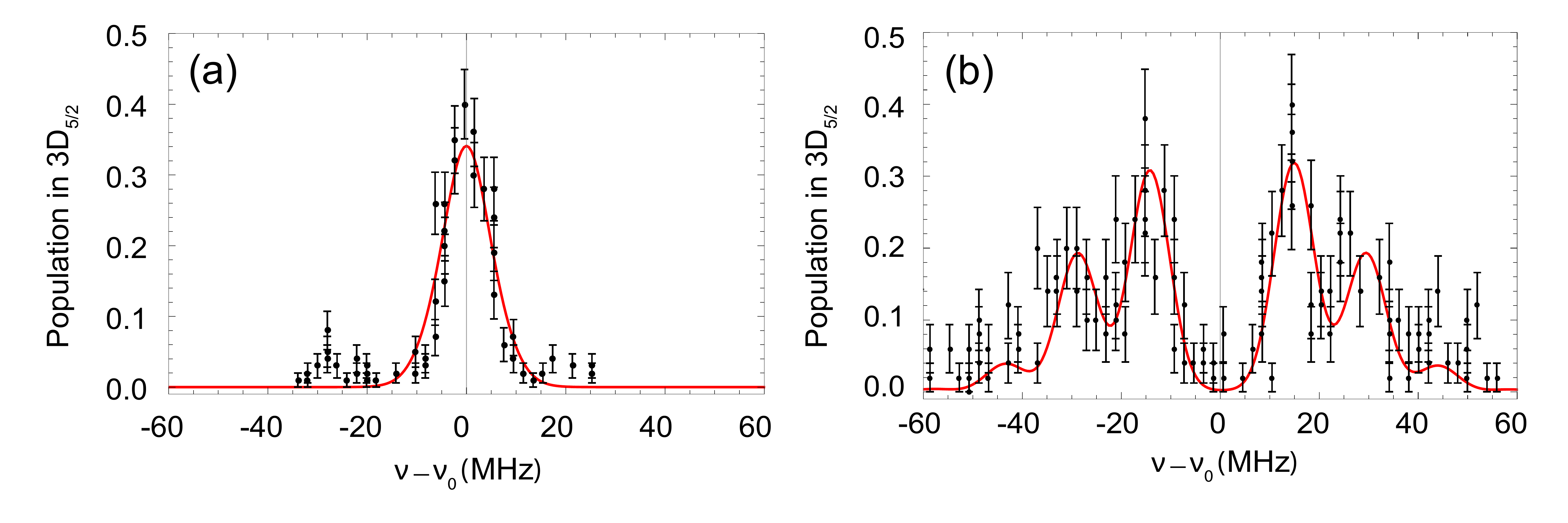}
\caption{Rydberg resonances for the 3D$_{3/2} \rightarrow $~23P$_{1/2}$ transition in a Ca$^{+} $ ion measured at the residual oscillating electric field $\vert \boldsymbol{E}_{\rm res} \vert$: (a) $\vert \boldsymbol{E}_{\rm res} \vert <10$~V/m with the RF voltage $V_{\rm RF}=120$ and ${\rm \Omega}_{\rm RF}=2\pi\times 5.98$~MHz, and (b) $\vert \boldsymbol{E}_{\rm res} \vert=160$~V/m with $V_{\rm RF}=280$~V and ${\rm \Omega}_{\rm RF}=2\pi\times 14.56$~MHz. Error bars stem from quantum projection noise for 100 measurement repetition for each data point. Calculated line shapes (in red) are obtained from Eqn.~\ref{eq:lineshape}. Adapted from~\protect\citep{mokhberi19a}.}
\label{fig:23pline}
\end{figure}
--~As discussed in Sec.~\ref{subsec:micromotion}, excess micromotion causes line broadening due to the Doppler effect. This effect is pronounced when a given transition is driven by a short wavelength as observed for the 3D$_{3/2} \rightarrow$~23P$_{1/2}$ transition of trapped Ca$^{+}$ ions, which is driven by laser light near 123.17~nm~\citep{mokhberi19a}. For this transition and typical laser parameters, Doppler broadening of the resonance shape dominates largely over polarisability effects. 

The excitation probabilities for this transition are detected by electron shelving in the 3D$_{5/2}$ state~(Fig.~\ref{fig:23pline}). 
If excess micromotion is minimised such that the residual RF electric field at the ion position $\vert \boldsymbol{E}_{\rm res} \vert < 10$~V/m, the spectrum do not exhibit additional sidebands.
By moving the ion along the trap axis~$z$, ions are subjected to $\vert \boldsymbol{E}_{\rm res} \vert=160$~V/m, and excess micromotion becomes noticeable. 
Note that for these sidebands the modulation depth $\beta_{\rm mm}$ depends on the angle between $\boldsymbol k$ and ${\boldsymbol R}_{\rm mm}$ as given in Eqn.~\ref{eq:lineshape}. 
The amplitude of the oscillating electric field, $\vert \boldsymbol{E}_{\rm res}\vert$, were precisely mapped out along the trap axis using resolved sideband spectroscopy on the 4S$_{1/2}, m_{J}=-1/2 \rightarrow$~3D$_{5/2}, m_{J}=-5/2$ transition at 729~nm~\citep{roos00a}. 
In this method, the Rabi frequency on the carrier transition and the first micromotion sideband are measured, and are used to determine $\vert \boldsymbol{E}_{\rm res}\vert$ and correspondingly $\beta_{\rm mm}$ (Eqn.~\ref{eq:lineshape})~\citep{roos00a}.


\subsubsection{Stark effect on highly polarisable Rydberg ions}\label{subsubsec:trapeffect-Polar}
\begin{figure}[h!]
\centering
\includegraphics[width=0.9 \textwidth]{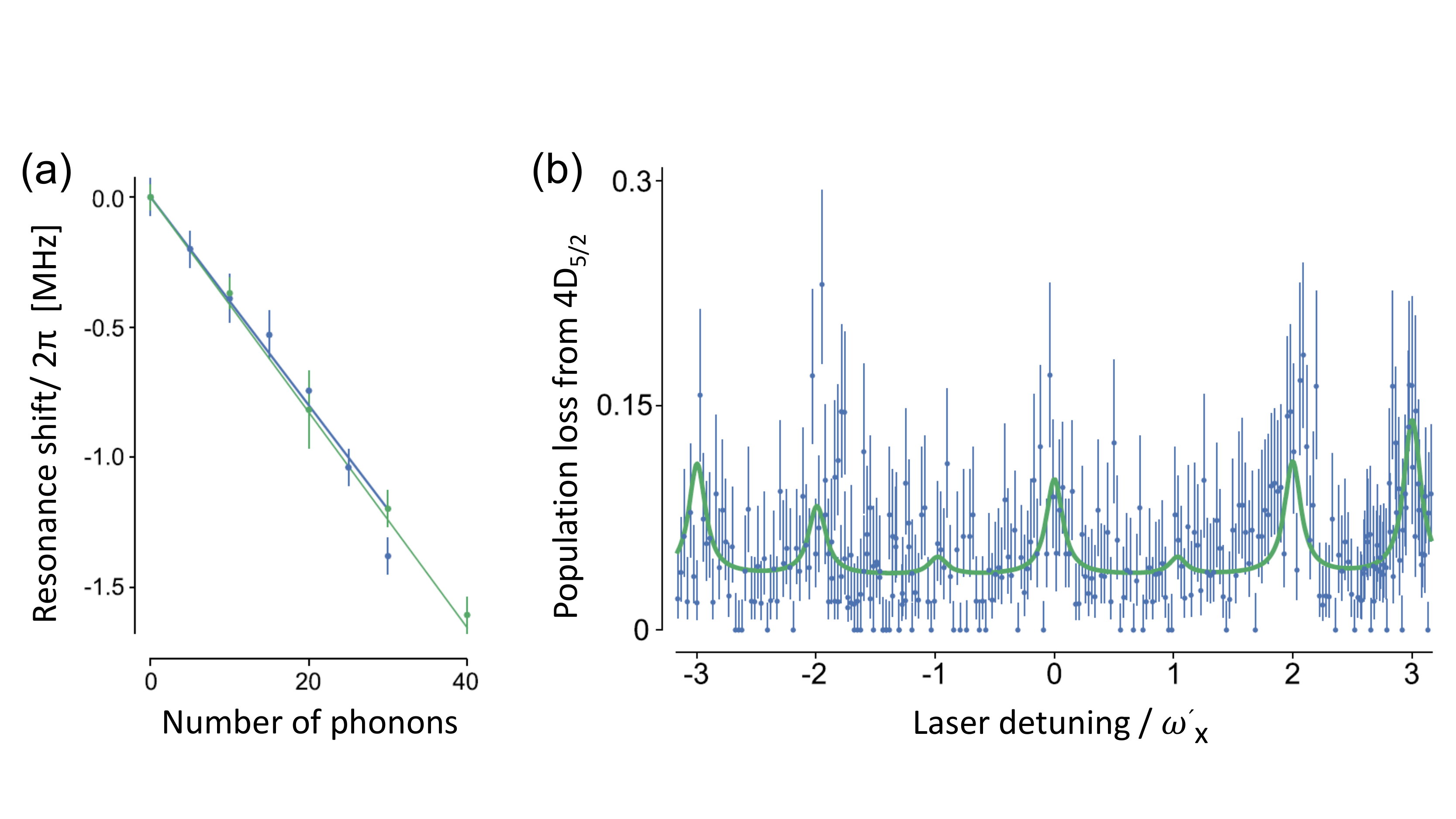}
\caption{Observations of the trap effects on highly polarisable Rydberg ions. (a)~The 4D$_{5/2} \rightarrow$~46S$_{1/2}$ resonance frequency is measured for a single Sr$^{+}$ ion as the phonon number in the $x$-~(blue) and $y$-modes~(green) increases. Note that only the phonon number of one mode is varied for each data set, while the other radial mode is cooled to near zero phonon number. The lines are obtained from Eqn.~\ref{eq:EphononchangingWEAK} with~$\Delta \omega_{x}=\omega^{\prime}_{x}-\omega_{x}=-2 \pi \times 40.1$~kHz and ~$\Delta \omega_{y}=\omega^{\prime}_{y}-\omega_{y}=-2 \pi \times 41.4$~kHz, see the illustriation in Fig.~\ref{fig:phononchanging}(a). (b)~Rydberg excitation spectra for an ion prepared with the phonon number $n_{x}=20$ and $n_{y} \simeq 0$. Resonances at integer multiples of $\omega^{\prime}_{X}$ correspond to phonon-number-changing transitions, illustrated in Fig.~\ref{fig:phononchanging}(b). Measured data~(blue dots) was modelled by Lorentzian absorption lines~(green). Adapted from~\protect\citep{higgins19a}.
\label{fig:SHIFTSPolarisability}}
\end{figure}

--~Significant energy shifts in Rydberg ions owing to the quadratic Stark effect arising from the trapping electric fields~(Sec.~\ref{subsec:micromotion}) were observed in Sr$^{+}$ ions.
As the phonon number of motional states increases, the size of the ion wavepacket becomes larger, and  thus the ion experiences larger interaction with the dynamic electric trapping field.
In the experiment, the 4D$_{5/2} \rightarrow$~46S$_{1/2}$ transition frequency was measured as a function of the phonon number for a single ion prepared in the number states of the radial modes~($x$ and $y$ modes)~(Fig.~\ref{fig:SHIFTSPolarisability}(a)).
These observed Stark shifts are modelled using the mean-squared RF electric field~(Sec.~\ref{subsec:micromotion}).
We note that calculations in which intrinsic micromotion has been taken into account showed only 2\% deviations from the models given by Eqn.~\ref{eq:EphononchangingWEAK}~and~\ref{eq:EphononchangingSTRONG}~\citep{higgins19a}.
Calculations in Fig.~\ref{fig:SHIFTSPolarisability}(a,b) use polarisability of the Rydberg state $\alpha_{r}\equiv\alpha_{46{\rm S}_{1/2}}$ from theory~$5.6 \times 10^{-31}$~C$^2$~m$^2$~J$^{-1}$~\citep{liprivate17a} and that of the initial state $\alpha_{g}\equiv\alpha_{4{\rm D}_{5/2}}\approx 0$~\citep{jiang09a}.

In the case of properly minimised stray electric fields at the ion position~(Sec.~\ref{subsec:micromotion}), Stark shifts of about few~MHz arising from driven phonon-number-preserving transitions were measured, see Fig.~\ref{fig:SHIFTSPolarisability}(a). In comparison, the quadratic Stark effect leads to more significant frequency shifts for Rydberg ions subjected to excess micromotion, i.e., when the minima of the RF and the static fields do not overlap~(Sec.~\ref{subsec:micromotion}). The reminiscent is the dependency of the ion equilibrium position on state polarisability~(Eqn.~\ref{eq:Xdphononchanging} and Fig.~\ref{fig:phononchanging}(b)).
Therefore, a change in the ion’s motional state is observed when driving electronic states with large polarisabilities (phonon-number-changing transitions). The measured spectrum, see Fig.~\ref{fig:SHIFTSPolarisability}(b), features resonances at integer multiples of $\omega^{\prime}_{x}$, where $\omega^{\prime}_{x}=\omega_{x}+\Delta \omega_{x}$ is the modified secular trapping frequency of the normal mode along the $x$ axis~(Eqn.~\ref{eq:secularfreqX}~and~\ref{eq:omegaXryd}). 

This observation might be used to minimise stray electric fields at the ion position, and hence to mitigate excess micromotion effects with the advantage of being sensitive to ion micromotion in all three spatial directions~\citep{higgins19a}. Stark shifts in neutral Rydberg atoms were used to precisely measure and minimise stray electric fields~\citep{osterwalder99a}.

\subsubsection{Floquet sidebands due to large quadrupole moments of Rydberg ions}\label{subsubsec:trapeffect-Quadru}
\begin{figure}[h!]
\centering
\includegraphics[trim={0cm 2cm 0cm 1cm},clip, width=0.9 \textwidth]{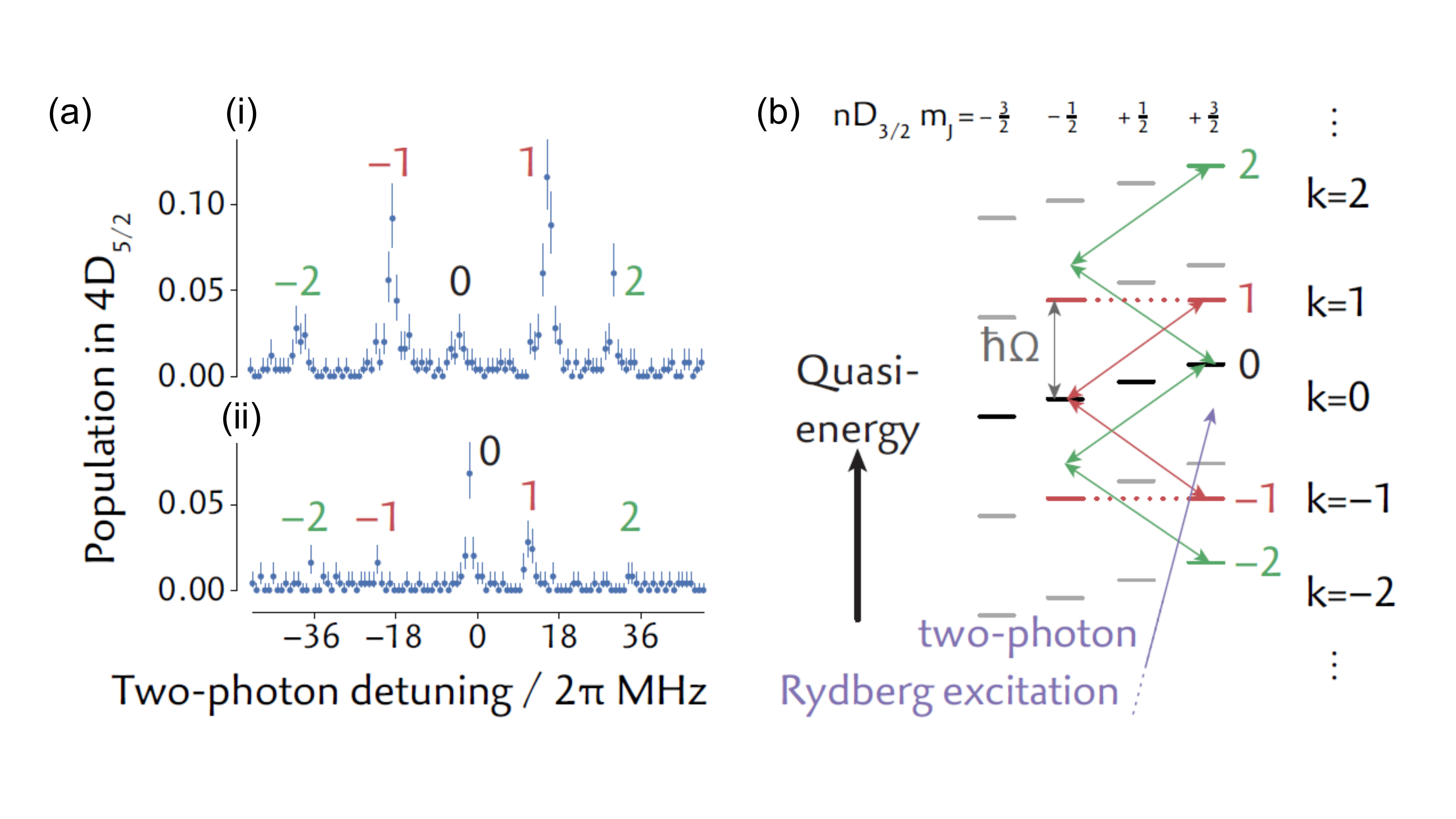}
\caption{(a)~Floquet sidebands observed in the 27D$_{3/2}$ Zeeman manifold of $^{88}$Sr$^{+}$ ions. The frequency of the RF trapping field is $\Omega_{\rm RF}=2 \pi \times 18.2$~MHz, and thus first-order sidebands appear around $2 \pi \times 18$ and second-order ones around~$2 \pi \times 36$ for both strong~(i) and weak trapping fields~(ii), see text for parameters. (b)~Schematic of the relevant coupling for a $n$D$_{3/2}$~($m_{J}=3/2$) state in the Floquet basis. First-order sidebands due to the coupling to the $m_{J}=-1/2$ sublevel and second-order ones resulted from a subsequent coupling to the $m_{J}=-1/2$ and the $m_{J}=3/2$ sublevels are shown. The Floquet eigenstates are labelled with the index~$k$, and states with $\Delta m_{J}=2$ and $\Delta k=1$ are coupled. Adapted from~\protect\citep{higgins18a}.
\label{fig:FloquetSB}}
\end{figure}

--~The coupling between Zeeman sublevels of a $n$D$_{3/2}$ state induced by the time-dependent electric trapping field leads to Floquet sidebands in Rydberg-excitation spectra.
This effect was investigated in the 24D$_{3/2}$ and the 27D$_{3/2}$ states of $^{88}$Sr$^{+}$ ions~\citep{higgins18a}.
As discussed in Sec.~\ref{subsec:rydbergionintrap} and~\ref{subsec:trapeffects}, these sidebands appear only in the excitation spectra of those states with $J>1/2$~(Fig.~\ref{fig:electron_Etrap_coupling}).
A manifold of a Rydberg $n$D$_{3/2}$ state expanded in the Floquet basis is shown in Fig.~\ref{fig:FloquetSB}(b).
 
The Floquet theorem is used to describe the effect of the time-dependent potential in the $H_{\rm{e-Trap}}$ Hamiltonian in Eqn.~\ref{eq:eTrapspherical}. In this case, the effective Rabi frequency of the RF field that couples these states is comparable with the RF drive frequency~(Sec.~\ref{subsec:rydbergionintrap}). For each eigenstate of the time-dependent Hamiltonian, there is a set of eigenstates with eigenenergies shifted by $k \hslash \Omega_{\rm RF}$ with respect to the uncoupled eigenstates, where $k$ is an integer number~(Fig.~\ref{fig:FloquetSB}(b)). 
The measurements were carried out using the strong trapping fields gradients~$\gamma^{\prime}=8.2 \times 10^8$~V~m$^{-2}$ and ~$\gamma=6.8 \times 10^6$~V~m$^{-2}$ and the weak trapping fields gradients~$\gamma^{\prime}=3.3 \times 10^8$~V~m$^{-2}$ and ~$\gamma=5.7 \times 10^6$~V~m$^{-2}$~(Fig.~\ref{fig:FloquetSB}(a)), where $\gamma^{\prime}$ and $\gamma$ are identified by Eqn.~\ref{eq:paultrap}.

\subsection{Stability of Rydberg ions in the trap}\label{subsec:doubleioni}
{\bf Double ionization} might occur in Rydberg excitation of trapped ions and has been observed in single- and two-step excitation experiments~\citep{feldker16a, higgins18a}. This effect has been studied for Ca$^{+}$ ion crystals excited by laser light at 121.26~nm, and thus {\bf above} the ionization limit to generate Ca$^{2+}$ ions~\citep{feldker14a}. For this experiment, the trapping parameters were set to allow for stable trapping of both singly and doubly charged ions. Ca$^{2+}$ ions do not scatter light at the cooling transition and appear as dark voids in a linear ion chain, and thus are identified from an increase of inter ion distances. 
More importantly, they feature different secular frequencies and motional modes because of the double charge-to-mass ratio. 

Only after several hundred excitations were losses or double ionization events observed, which require loading new ions. Even excitations to the 65S and 66F states were observed for Ca$^{+}$ ions~\citep{andrijauskas20a, feldker16a}. This high stability of Rydberg states in the trapping field is remarkable, since a calculation suggested a limit of about $n<50$~\citep{mueller08a}. Several additional effects are conjectured to affect the stability. Blackbody radiation can cause double ionisation particularly for the states with extremely high principal quantum number. The losses might be expected from the time-dependent electric trapping field, but the ionization threshold for such events is far below that of a static electric field. In neutral Rydberg experiments, subsequent Landau-Zener transitions have been observed that can occur between states with different $n$-manifolds. Microwave fields applied to generate Rydberg dressed states (Sec.~\ref{sec:coh_spectroscopy}) might also affect the stability, as MW field ionisation of Ba$^{+}$ ions in free space~\citep{seng98a} and microwave multi-photon transitions between Rydberg states of neutral atoms have been observed~\citep{stoneman88a}. 

\section{Coherent spectroscopy and control of Rydberg ions}\label{sec:coh_spectroscopy}
The previous section describes how the trapping electric fields modify spectral Rydberg lines, and how these effects may be mitigated to achieve narrow Rydberg resonance lines, a starting point for implementing coherent dynamics~\citep{higgins17b, higgins19a, zhang20a}.
The coherent phenomena observed so far include {\bf two-photon Rabi oscillations} (Sec.~\ref{sec:coh_spectroscopy:Rabi}), the {\bf EIT effect} (discussed in Sec.~\ref{subsec:detection} as employed to detect Rydberg resonances), the {\bf Autler-Townes effect} (Sec.~\ref{sec:coh_spectroscopy:AT}) and {\bf stimulated Raman adiabatic passage} (STIRAP) (Sec.~\ref{sec:coh_spectroscopy:STIRAP}).
Coherent control of Rydberg ions has enabled measurement of Rydberg state lifetimes (Sec.~\ref{sec:coh_spectroscopy:lifetime}), a single-qubit gate (Sec.~\ref{sec:coh_spectroscopy:singlequbitgate}), and recently a sub-microsecond two-qubit entangling gate \citep{zhang20a} (Sec.~\ref{sec:coh_spectroscopy:entangle}).
Some of these phenomena rely on the two-photon excitation scheme, and thus we begin this section with a theoretical description of three atomic levels coupled by two laser fields.

\subsection{Three-level system coupled by two laser fields}
\begin{figure}
	\centering
	\includegraphics[width=0.45\textwidth]{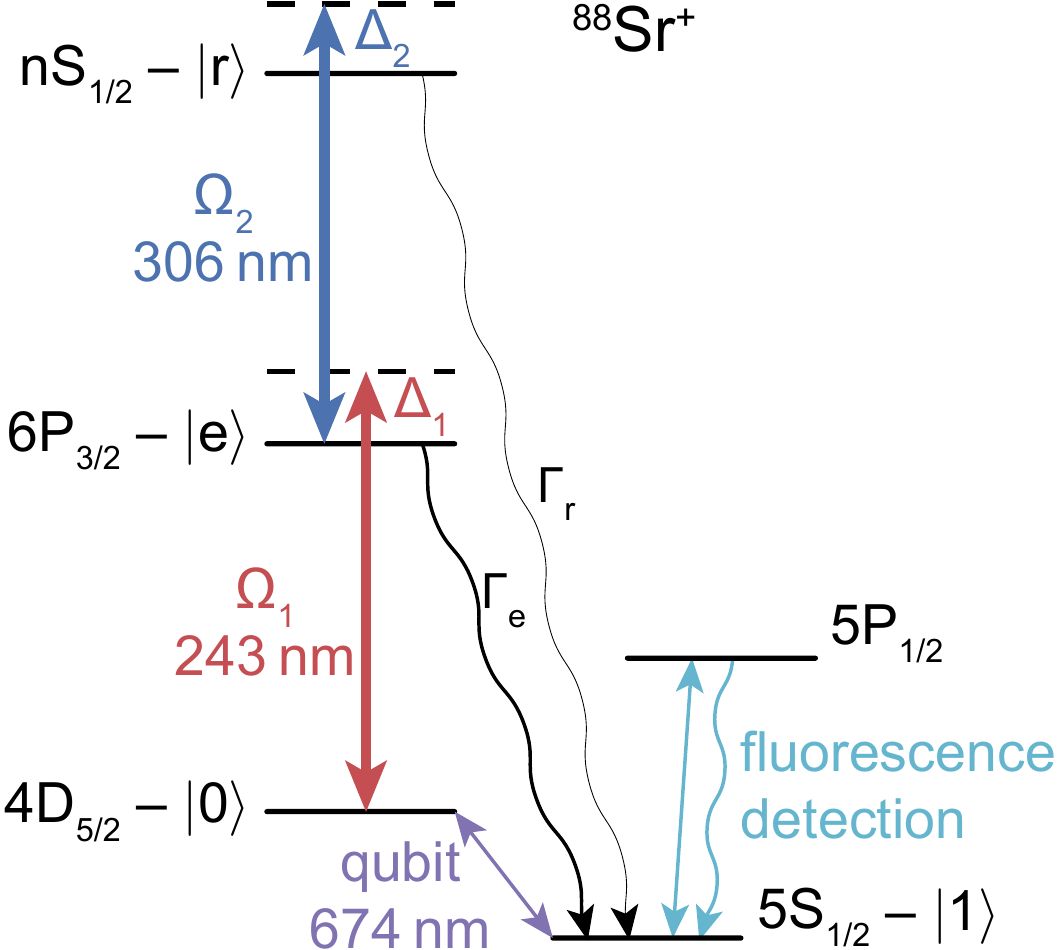}
	\caption{Energy level diagram relevant for coherent manipulation of Rydberg $nS_{1/2}$ states of $^{88}$Sr$^{+}$ ions. Zeeman sublevels of the $4D_{5/2}$, $6P_{3/2}$ and $nS_{1/2}$ states are coupled using laser fields near 243~nm and 306~nm, with Rabi frequencies $\Omega_{1}$ and $\Omega_{2}$ and the corresponding laser frequency detuning $\Delta_{1}$ and $\Delta_{2}$ respectively.
From the $6P_{3/2}$ and the $nS_{1/2}$ states, the population decays to the ground state $5S_{1/2}$. Adapted from~\protect\citep{higgins17b}.}
	\label{fig:coh_level_scheme}
\end{figure}

Sr$^{+}$ ions are excited from the qubit state $|0\rangle = 4D_{5/2},\:m_J=-\frac{5}{2}$ to the Rydberg state $|r\rangle = nS_{1/2},\:m_J=-\frac{1}{2}$ via the intermediate state $|e\rangle = 6P_{3/2},\:m_J=-\frac{3}{2}$ using laser fields at 243\,nm and 306\,nm, as shown in Fig.~\ref{fig:coh_level_scheme}.
The coupling strength between $|0\rangle$ and $|e\rangle$ is $\Omega_1$ and the coupling strength between $|e\rangle$ and $|r\rangle$ is $\Omega_2$, with corresponding laser frequency detuning which is denoted by $\Delta_1$ and $\Delta_2$ respectively. Within the rotating wave approximation, the coupling Hamiltonian $H_c$ for the three levels $\{|0\rangle, |e\rangle, |r\rangle\}$ is given by
\begin{equation} \label{eq:coh_Hc}
H_c=\frac{\hslash}{2}
\begin{pmatrix}
0          & \Omega_{\mathrm{1}}  & 0             \\
\Omega_{\mathrm{1}} & 2\Delta_{\mathrm{1}} & \Omega_{\mathrm{2}} e^{i\phi} \\
0          & \Omega_{\mathrm{2}} e^{-i\phi} & 2\Delta_{\mathrm{1}}+2\Delta_{\mathrm{2}}
\end{pmatrix}.
\end{equation}
Here, $\phi$ is the relative phase between the laser fields within the rotating frame.
The dressed eigenstates are found by diagonalizing $H_c$.

The population in the $|e\rangle$ state decays to the $5S_{1/2}$ state with decay rate $\Gamma_e = 2\pi \times 4.9\,\mathrm{MHz}$ by multi-channel decay processes. Population in $|r\rangle$ also decays by multi-channel processes to $5S_{1/2}$; the rate $\Gamma_r$ at which population leaves $|r\rangle$  depends on the Rydberg state, e.g. $\Gamma_{46S}=2 \pi \times 34\,\mathrm{kHz}$.
Typically $\Gamma_r \ll \Gamma_e$. Finite laser linewidths $\delta_1 \approx \delta_2 \approx 2\pi \times 100\,\mathrm{kHz}$ cause dephasing.

Rabi oscillations may be observed in a two-level system coupled by a single laser field.
In a three-level system coupled by two laser fields the coherent phenomena that emerge are richer, as described in the following sections.

\subsection{Two-photon Rabi oscillations} \label{sec:coh_spectroscopy:Rabi}

\begin{figure}
	\centering
	\includegraphics[width=0.5\textwidth]{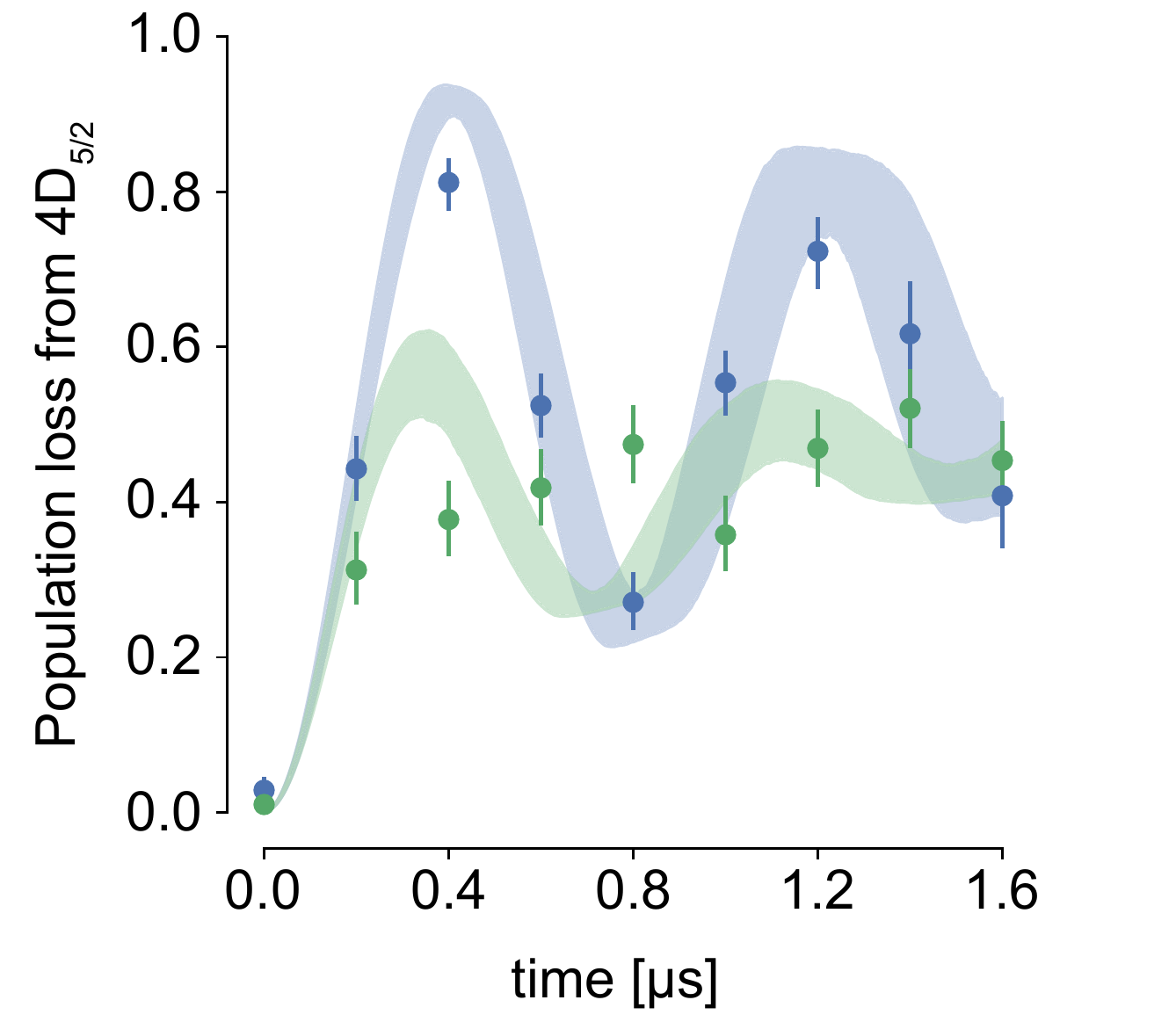}
	\caption{Two-photon Rabi oscillations between $|0\rangle \leftrightarrow |r\rangle$ are observed when the ion is sideband cooled to near the ground state (blue data points).
The oscillations are washed out when only Doppler cooling is employed (green data points).  Here the theory assumes after Doppler cooling $n_{mean} \approx 13$, and after sideband cooling $n_{mean} \approx 0.2$, and $\Omega_{\mathrm{eff}} \approx 2\pi \times 1.2$~MHz.
The shaded areas represent simulated results using experimentally-determined parameters, see text for details. Adapted from~\protect\citep{higgins19a}.}
	\label{fig:coh_Rabi}
\end{figure}

Rabi oscillations between the $|0\rangle$ and $|r\rangle$ states may be driven, while the coupling to the lossy intermediate state $|e\rangle$ needs to be weak $\Omega_1, \Gamma_e, \delta_1 \ll \Delta_1$ and $\Omega_2, \Gamma_e, \delta_2 \ll \Delta_2$, such that little population is transferred to $|e\rangle$. In such way, the state $|e\rangle$ may then be adiabatically eliminated, and $H_c$ simplifies to
\begin{equation}
H_c'=\frac{\hslash}{2}
\begin{pmatrix}
-\frac{\Omega_{\mathrm{1}}^2}{2 \Delta_{\mathrm{1}}} & -\frac{\Omega_{\mathrm{1}}\Omega_{\mathrm{2}}}{2 \Delta_{\mathrm{1}}} \\
-\frac{\Omega_{\mathrm{1}}\Omega_{\mathrm{2}}}{2 \Delta_{\mathrm{1}}} & -\frac{\Omega_{\mathrm{2}}^2}{2 \Delta_{\mathrm{1}}} + 2\Delta_{\mathrm{1}} + 2\Delta_{\mathrm{2}}
\end{pmatrix}
\end{equation}
in the $\{|0\rangle, |r\rangle\}$ basis.
The effective coupling strength between $|0\rangle$ and $|r\rangle$, i.e.\ the two-photon Rabi frequency, is $\Omega_{\mathrm{eff}} = \frac{\Omega_1 \Omega_2}{2\Delta_{1}}$.

To observe high-contrast Rabi oscillations the off-diagonal coupling elements in Eqn.~\ref{equ:40} should exceed the diagonal decay elements, thus a two-photon detuning from the $|0\rangle \leftrightarrow |r\rangle$ resonance should satisfy
\begin{equation}\label{equ:40}
\frac{\Omega_1^2 - \Omega_2^2}{4\Delta_1} + \Delta_1 + \Delta_2 \ll \Omega_{\mathrm{eff}}
\end{equation}
The first term on the left accounts for AC Stark shifts, which cancel for $\Omega_1=\Omega_2$. High-contrast Rabi oscillations additionally require $\Omega_{\mathrm{eff}} \gg \Gamma_r, \delta_1, \delta_2$.

The importance of {\bf mitigating trap effects} is illustrated in Fig.~\ref{fig:coh_Rabi} in which the results for a sideband-cooled ion and a Doppler cooled ion are compared.
Rabi oscillations were observed with the sideband-cooled ion, while with the Doppler-cooled ion the linewidth  $\Gamma_r$ of the $|r\rangle$ state was broadened due to the Stark effect (Sec.~\ref{subsec:micromotion}), and Rabi oscillations were smeared out. The experiment used $|r\rangle = 46S_{1/2},\:m_J=-\frac{1}{2}$.
The oscillation contrast was limited by laser linewidths ($\delta_1 \approx \delta_2 \approx 2\pi \times 100\,\mathrm{kHz}$), laser light intensities and the Rydberg state lifetime.

\subsection{Autler-Townes effect} \label{sec:coh_spectroscopy:AT}
The Autler-Townes effect is observed in spectra when a strong laser field couples $|e\rangle$ and $|r\rangle$ with strength $\Omega_2 > \Gamma_e, \Gamma_r$ such that the new dressed eigenstates of $H_c$ become
\begin{align}
|\phi_0\rangle & = |0\rangle
\\
|\phi_{\pm}\rangle & = \frac{-\Delta_{\mathrm{2}} \pm \sqrt{\Delta_{\mathrm{2}}^2 + \Omega_{\mathrm{2}}^2}}{\Omega_{\mathrm{2}}}|e\rangle + |r\rangle
\end{align}
with eigenvalues
\begin{align} \label{eq:coh_eigenvalues}
E_0 &= 0
\\
E_{\pm} &= \frac{\hslash}{2} \left(\Delta_{\mathrm{2}} \pm \sqrt{\Delta_{\mathrm{2}}^2 + \Omega_{\mathrm{2}}^2}\right).
\end{align}

\begin{figure}
\centering
\includegraphics[width=\textwidth]{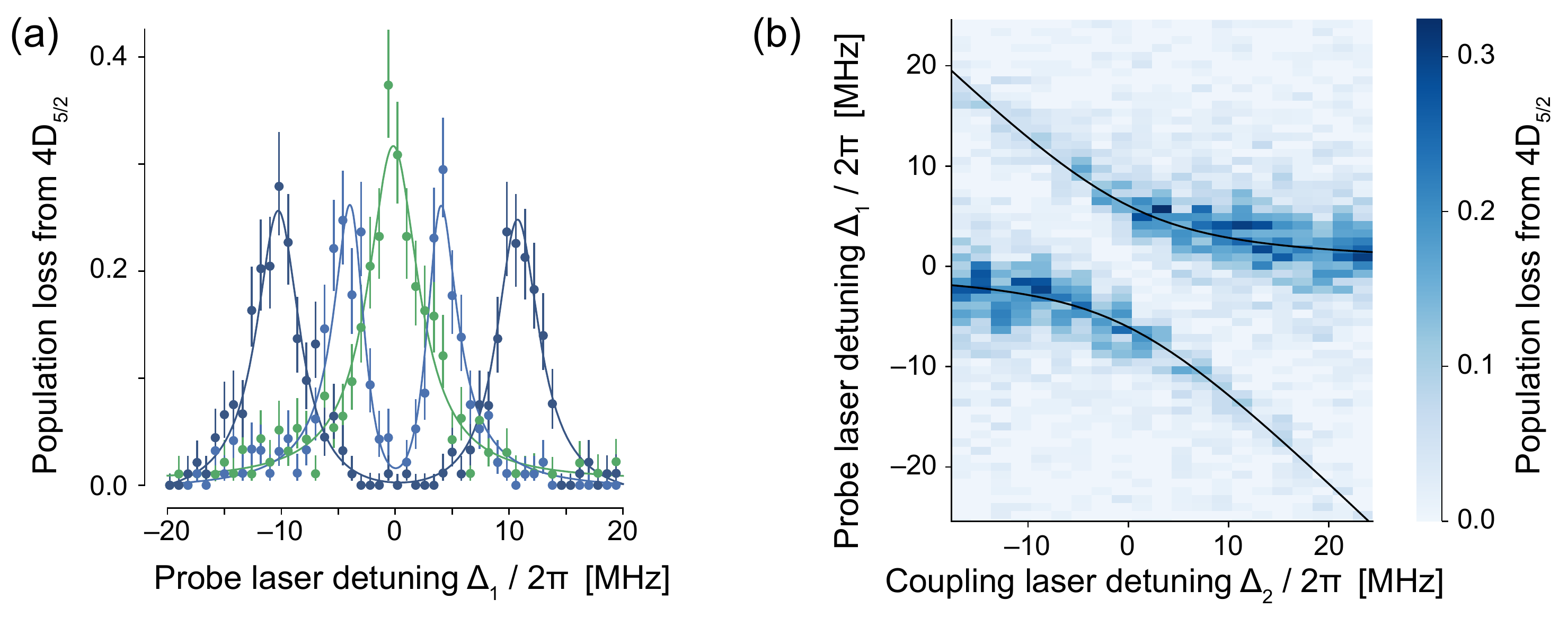}
\caption{Autler-Townes effect.
(a)~A single resonance line corresponding to the $|0\rangle \leftrightarrow |e\rangle$ transition is observed in the absence of the coupling laser field (green data).
With a resonant coupling laser field two resonance lines result, corresponding to excitation of the dressed states $|e\rangle \pm |r\rangle$, which are separated by the coupling strength $\Omega_2$. The separation between the resonance lines increases with the laser field intensity. (b)~Avoided crossing of two resonance lines. The $|0\rangle \leftrightarrow |e\rangle$ transition gives rise to a resonance along $\Delta_1=0$ while the $|0\rangle \leftrightarrow |r\rangle$ transition gives rise to a resonance when the two-photon detuning is zero $\Delta_2 = - \Delta_1$. The coupling laser field couples $|e\rangle$ and $|r\rangle$ such that the resonance lines do not cross. Adapted from~\protect\citep{higgins18a}.
\label{fig:coh_ATS}}
\end{figure}

This eigenstates may be investigated by spectroscopy using a weak probe laser field which couples $|0\rangle \leftrightarrow |e\rangle$ with strength $\Omega_1 \ll \Gamma_e$.
In Fig.~\ref{fig:coh_ATS}(a) the probe laser detuning $\Delta_1$ is scanned while the coupling laser field is resonant $\Delta_2=0$.
The $|0\rangle \leftrightarrow |e\rangle$ resonance is split into two resonances which correspond to excitation of $|\phi_+\rangle= \frac{1}{\sqrt{2}}(|e\rangle + |r\rangle)$ and $|\phi_-\rangle = \frac{1}{\sqrt{2}}(|e\rangle - |r\rangle)$.
The separation between the resonances is a measure of $\Omega_2$, i.e.\ the energy difference between the eigenvalues (Eqn.~\ref{eq:coh_eigenvalues}); thus we use the Autler-Townes effect to calibrate $\Omega_2$.

By conducting spectroscopy as both $\Delta_1$ and $\Delta_2$ are scanned the avoided crossing of $|e\rangle$ and $|r\rangle$ is mapped out, as shown in Fig.~\ref{fig:coh_ATS}(b).
When the dressing is weak $\Omega_2 \ll \Delta_2$ the dressed states resemble the bare atomic states $|e\rangle$ and $|r\rangle$, which are excited when $\Delta_1 \approx 0$ and $\Delta_1 + \Delta_2 \approx 0$ (the non-equality is due to the AC Stark shift $\frac{\Omega_2^2}{4\Delta_2}$ which falls out of Eqn.~\ref{eq:coh_eigenvalues} when $\Delta_2 \gg \Omega_2$). The avoided crossing arises due to the coupling of $|e\rangle$ and $|r\rangle$ by the coupling laser field.
Note that the Autler-Twons effect is distinguished from the EIT effect~(Sec.~\ref{subsec:detection})  by the coupling scheme applied~\citep{hao18a}.

\subsection{Stimulated Raman adiabatic passage} \label{sec:coh_spectroscopy:STIRAP}
Stimulated Raman adiabatic passage (STIRAP)~\citep{bergmann15a} allows for coherent transfer of population  between $|0\rangle$ and $|r\rangle$ when $\Omega_1 \gg \Delta_1$ and $\Omega_2 \gg \Delta_2$ -- this is a completely different parameter regime as compared to that for driving two-photon Rabi oscillations as in Sec.~\ref{sec:coh_spectroscopy:Rabi}. The STIRAP process relies on smoothly changing $\Omega_1$ and $\Omega_2$ such that a dressed state changes its character from $|0\rangle \rightarrow -|r\rangle$ and population adiabatically follows this evolution. Among the advantages of STIRAP transfer is its efficiency which may be insensitive to an imperfect setting or fluctuations of experimental parameters. 

\begin{figure}
\centering
\includegraphics[width=0.5\textwidth]{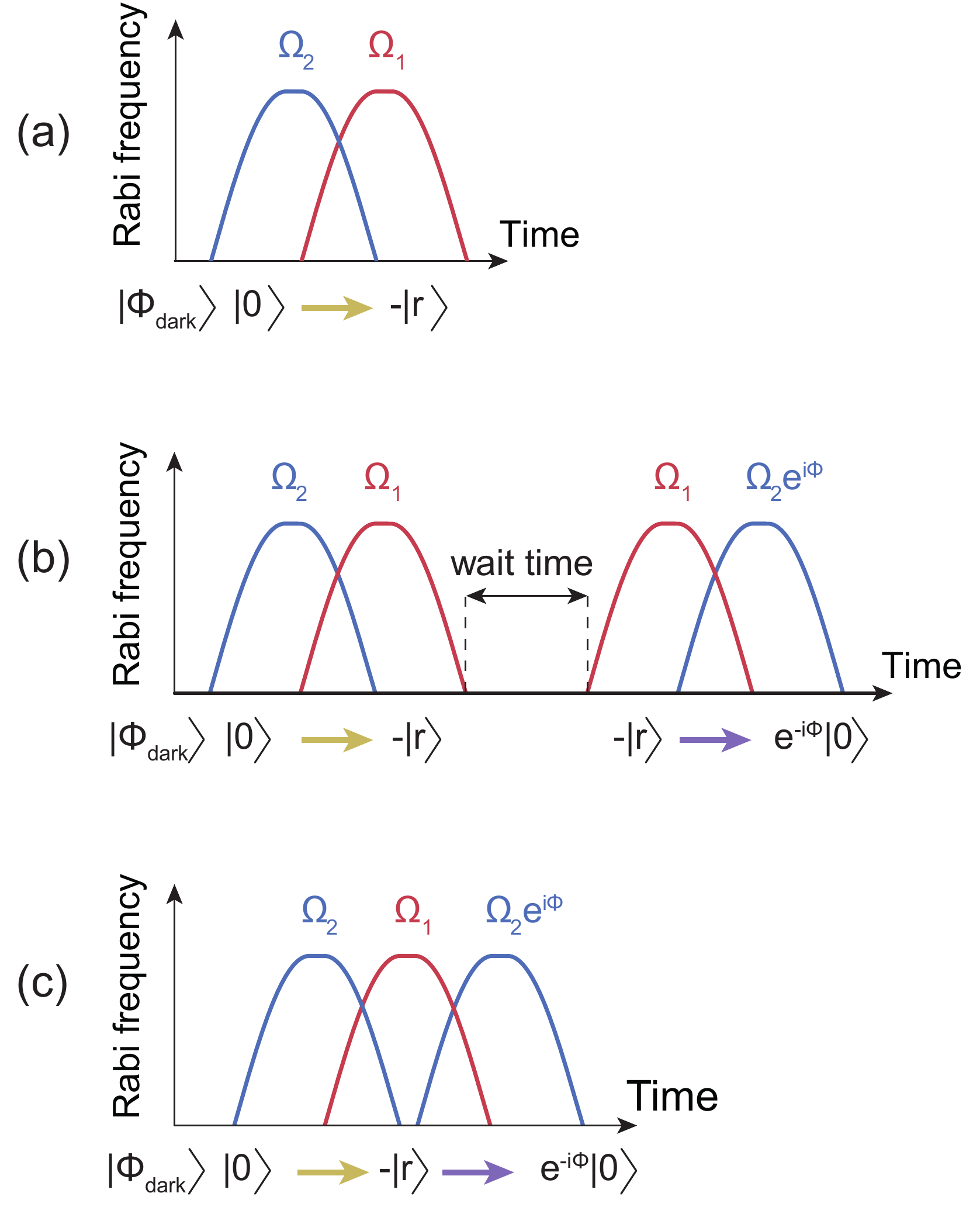}
\caption{STIRAP pulse sequences used for coherent manipulation. The population transfer from $|0\rangle \rightarrow |r\rangle$, $\Omega_2$ is applied before $\Omega_1$. The evolution of the dark state is written below the time axes. The coupling strengths $\Omega_1$ and $\Omega_2$ are varied sinusoidally in the experiment. In (b), population which decays from $|r\rangle$ during the wait time cannot be returned to $|0\rangle$, thus the Rydberg state lifetime is measured by varying the wait time and measuring the population returned to $|0\rangle$. Adapted from~\protect\citep{higgins18a}.}

\label{fig:coh_STIRAP_pulse_seq}
\end{figure}

For resonant two-photon condition $\Delta_1 + \Delta_2 = 0$ the dressed state of $H_c$ is
\begin{equation}
|\phi_{dark}\rangle = \Omega_2 e^{i \phi} |0\rangle - \Omega_1 |r\rangle
\end{equation}
Note that this eigenstate has no component of $|e\rangle$ and it is dubbed the ``dark'' eigenstate because it does not scatter photons (at least on time scales $\ll \Gamma_r^{-1}$).
$\Omega_1$ and $\Omega_2$ are varied according to the counter-intuitive sequences shown in Fig.~\ref{fig:coh_STIRAP_pulse_seq}.
The sequence in (a) transfers population from $|0\rangle \rightarrow -|r\rangle$ while the sequences in (b) and (c) transfer population from $|0\rangle \rightarrow -|r\rangle \rightarrow e^{-i\phi}|0\rangle$.

If the sequence is implemented too quickly, non-adiabatic dynamics cause population to leave the dark eigenstate, decreasing the transfer efficiency.
The character of the dark state is quantified by the mixing angle $\theta$: $\tan{\theta} = \Omega_1 / \Omega_2$.
To reduce losses due to non-adiabaticity we aim to satisfy the adiabaticity criterion $|\dot{\theta}|\ll\sqrt{\Omega_1^2+\Omega_2^2}$.

The description so far has ignored losses due to Rydberg state decay and finite laser linewidths.
A shorter sequence with a higher $|\dot{\theta}|$ is less sensitive to these losses.
Thus a balance must be struck when choosing the appropriate sequence length.


When the $|0\rangle \rightarrow |r\rangle$ transfer is attempted, imperfections and errors cause scattering off $|e\rangle$, which results in population in $5S_{1/2}$.
After ideal transfer from $|0\rangle \rightarrow |r\rangle$ population also ends up in $5S_{1/2}$, due to Rydberg state decay.
Since the experiment does not have the timing resolution to distinguish the decay processes the STIRAP transfer efficiency is instead measured by investigating the process $|0\rangle \rightarrow -|r\rangle \rightarrow |0\rangle$.
After this ``double STIRAP'' process $0.83\substack{+0.05\\-0.06}$ population resided in $|0\rangle$~\citep{higgins17b}, indicating a single STIRAP transfer efficiency of 0.91$\pm$0.03.
This efficiency is limited by the laser light intensities, laser linewidths and Rydberg state decay rates.
Recently, with reduced the laser linewidths, a STIRAP transfer efficiency $\approx$ 0.95 is achieved for the excitation of 42S$_{1/2}$ in Sr$^+$ ions~\citep{zhang20a}.

\subsubsection{Rydberg state lifetime} \label{sec:coh_spectroscopy:lifetime}
\begin{figure}
\centering
\includegraphics[width=0.5\textwidth]{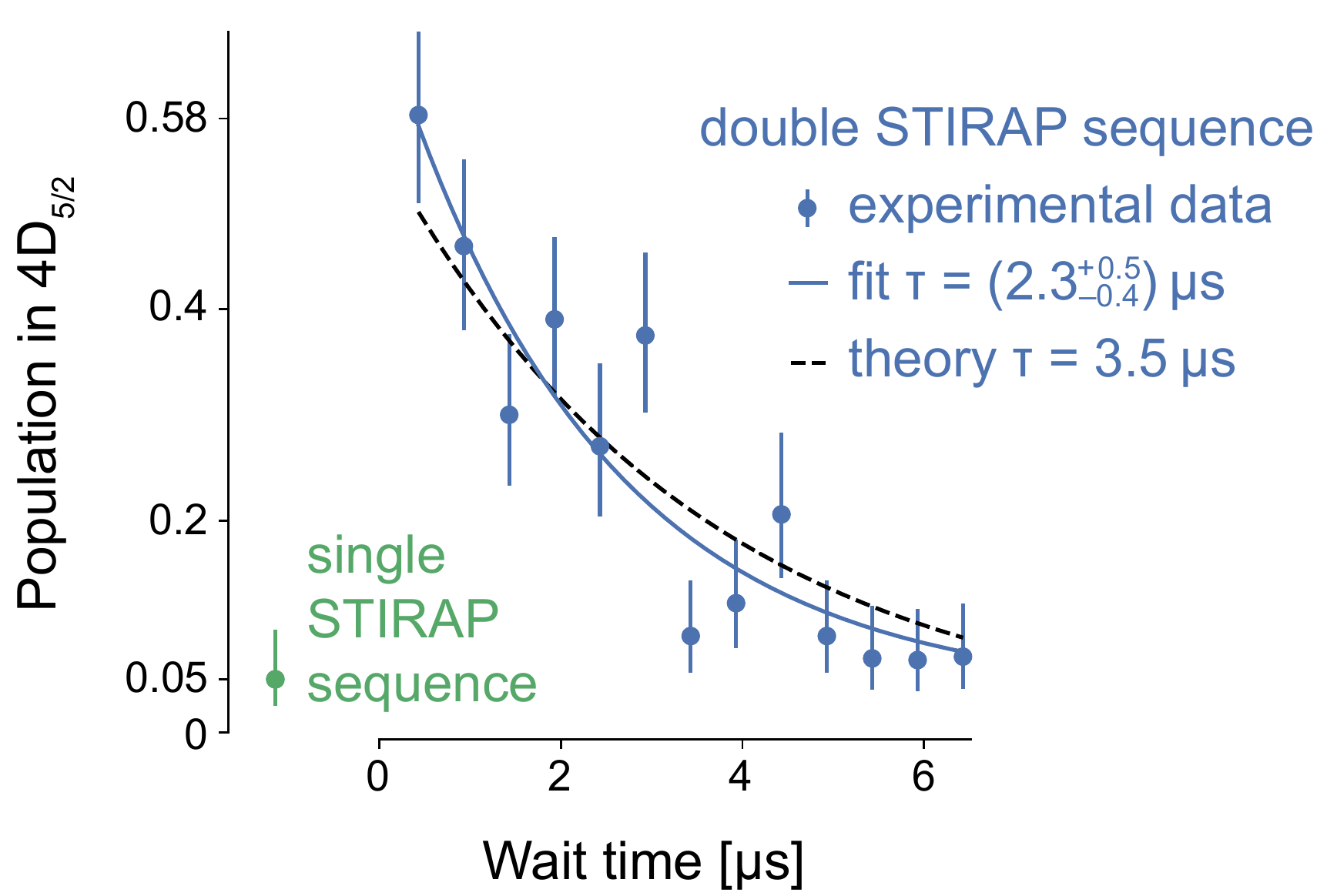}
\caption{Measurement of $42S_{1/2}$ lifetime. The wait time between the STIRAP transfers $|0\rangle \rightarrow -|r\rangle$ and $-|r\rangle \rightarrow |0\rangle$ is varied and the population returned to $|0\rangle$ is measured. As the wait time is increased the returned population falls exponentially, due to Rydberg state decay. Adapted from~\protect\citep{higgins17b}.}
\label{fig:coh_STIRAP_lifetime}
\end{figure}

--~We incorporate a waiting time between both STIRAP pulses such that the population of the Rydberg level $|r\rangle$ may decay, and hence reducing the proportion of population which may be returned to $|0\rangle$, see Fig.~\ref{fig:coh_STIRAP_pulse_seq}(b). By measuring the double STIRAP efficiency when varying the wait time a lifetime of $(2.3\substack{+0.5\\-0.4})\,\mathrm{\mu s}$ was measured for the $42S_{1/2}$ state, see Fig.~\ref{fig:coh_STIRAP_lifetime}.
Theory predicts $3.5\,\mu s$ at room temperature. 

\subsubsection{Imprinting a geometric quantum phase} \label{sec:coh_spectroscopy:singlequbitgate}
\begin{figure}
\centering
\includegraphics[width=\textwidth]{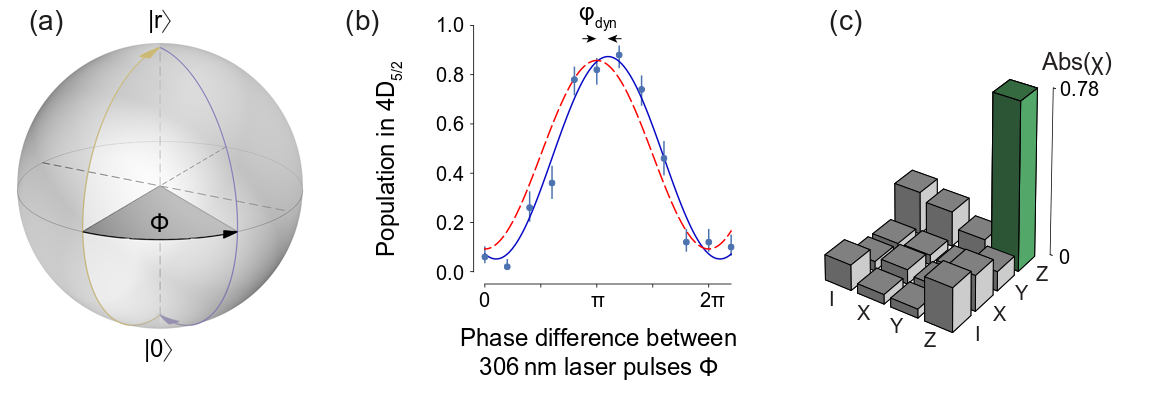}
\caption{Introduction of a geometric phase during STIRAP sequence. (a) During the double STIRAP pulse sequence the dark state traces out a closed path in the parameter space of the mixing angle $\theta$ and the laser phase shift $\phi$, the same space coincides with the Bloch sphere spanned by $|0\rangle$ and $|r\rangle$. The curvature enclosed by the path gives rise to a geometric phase. (b) The geometric phase is detected by conducting a Ramsey experiment between qubit states $|1\rangle$ and $|0\rangle$, with introduction of the geometric phase between the pair of $\frac{\pi}{2}$ pulses. Simulation (red, dashed), measured data and fit (blue) are shown.
As the laser phase shift is varied the resultant geometric phase varies and the final population in $|0\rangle$ oscillates. (c) Using $\phi = \pi$ the double STIRAP pulse sequence is a single-qubit phase gate, which was characterised using process tomography. The process matrix is shown. The gate fidelity was $0.78\pm0.04$. Adapted from~\protect\citep{higgins17b}.
}
\label{fig:coh_STIRAP_Berry}
\end{figure}

--~If a phase difference $\phi$ is introduced between the 306\,nm laser pulses the double STIRAP pulse sequence introduces a geometric phase $|0\rangle \rightarrow -|r\rangle \rightarrow e^{-i \phi} |0\rangle$.
During the sequence the parameters $\theta$ and $\phi$ are varied smoothly.
The sequence can be described as following a loop in this parameter space, as shown in Fig.~\ref{fig:coh_STIRAP_Berry}(a).
The curvature enclosed by this loop gives rise to the geometric phase $\phi$, which is detected by implementing the $|0\rangle \rightarrow e^{-i \phi} |0\rangle$ process inside of a Ramsey experiment between the qubit states $|1\rangle$ and $ |0\rangle $ (Fig.~\ref{fig:coh_STIRAP_Berry}(b)).

With $\phi=\pi$ the double STIRAP process behaves as the single-qubit phase gate $\alpha|1\rangle + \beta|0\rangle \rightarrow \alpha|1\rangle - \beta|0\rangle$.
The fidelity of the gate operation was characterised using quantum process tomography, the reconstructed process matrix is shown in Fig.~\ref{fig:coh_STIRAP_Berry}(c).
We find a fidelity of $0.78\pm0.04$. Dominant limitations are fluctuations in the laser intensities and frequency and the finite Rydberg state lifetime.

\subsection{Two-ion entangling Rydberg interaction} \label{sec:coh_spectroscopy:entangle}
\begin{figure}
\centering
\includegraphics[width=0.9\textwidth]{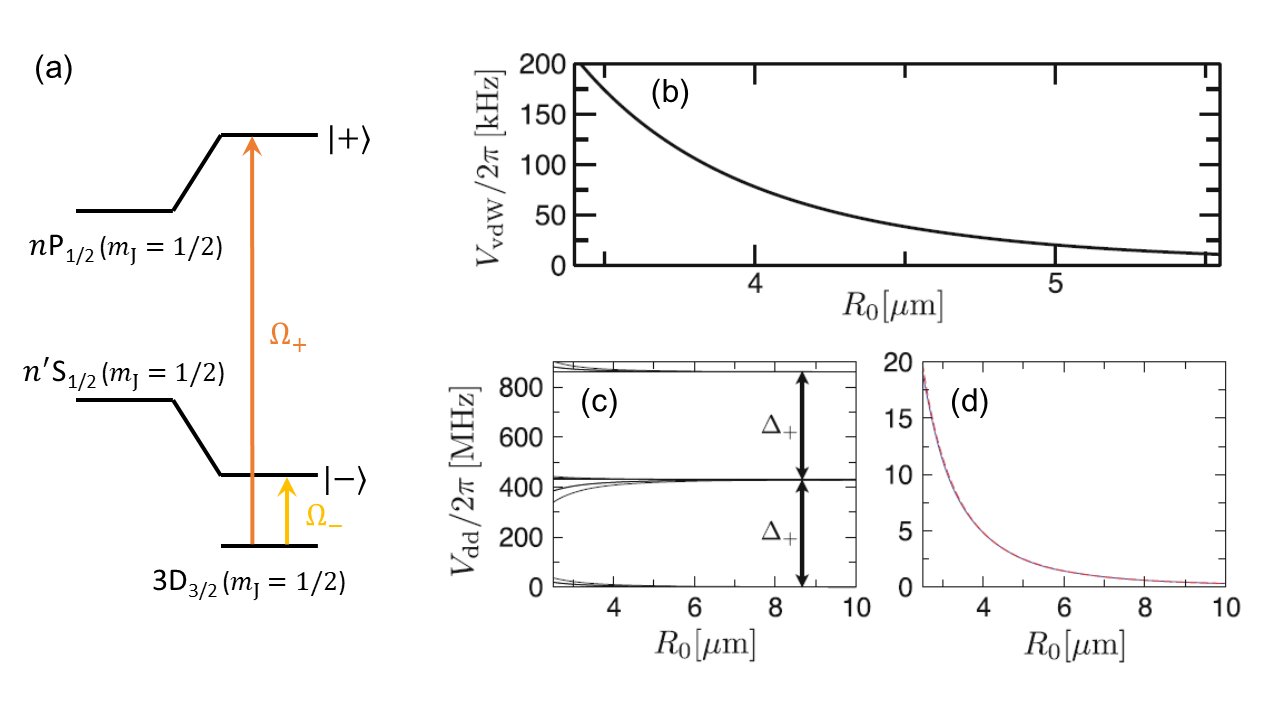}
\caption{MW dressing of ionic Rydberg states. (a) Energy level scheme for MW dressing of Rydberg states of Ca$^{+}$ ions in the strong MW regime, see text for details. (b) Van der Waals interaction between two $^{40}$Ca$^{+}$ ions in the 65P$_{1/2}(m_{J}=1/2)$ state as a function of their distance $R_{0}$. (c) Dipole-dipole interaction between ions in MW-dressed Rydberg states for the 65P$_{1/2}(m_{J}=1/2)\leftrightarrow$~65S$_{1/2}(m_{J}=1/2)$ transition. (d) Dipole-dipole interaction for the electronic pair state $|- -\rangle$. There is a complete overlap between numerical calculations and values obtained from Eqn.~\ref{eq:VddMWdressed}. Adapted from~\protect\citep{li14a}.}
\label{fig:scematicMWdressing}
\end{figure}  

In this section, we discuss gate operations based on the dipole blockade mechanism~\citep{comparat10a, saffman10a} between two trapped ions in MW-dressed Rydberg states.
Rydberg ions do not exhibit permanent dipoles and the van der Waals interaction between a pair of singly-charged ions is 64~times weaker as compared to that between a pair of neutral Rydberg atoms at the same distance~(see Table~\ref{table:properties} and Fig.~\ref{fig:scematicMWdressing}(b-d)). Thus MW fields are used to generate oscillating dipoles and to enable resonant dipole-dipole interactions. Two MW coupling schemes using a single-frequency MW field~\citep{li13a,li14a} and a bichromatic MW field~\citep{mueller08a}~(Sec.~\ref{subsec:excitationtransfer}) have been proposed.
Here, we consider the former in which the interaction potential for the $j$-th ion ($j \in \lbrace 1,2 \rbrace$) is written as $V_{\rm MW}({\boldsymbol r}_{j})=-e E_{1} \hat{\varepsilon}_{1} \cdot {\boldsymbol r}_{j} \cos (\omega_{1} t)$, with $E_{1}$, $\hat{\varepsilon}_{1}$ and $\omega_{1}$ the amplitude, the polarization vector and the frequency of the MW electric field.  
In the strong MW coupling regime considered here, the dynamics of the MW interaction occurs in time scales significantly shorter than the ions' oscillation period and the ion-laser interaction time.
Two Rydberg states with opposite sign of polarisability are considered, e.g., a $n^\prime$S and a $n$P level with $\alpha_{n^{\prime} {\rm S}}>0$ and $\alpha_{n {\rm P}}<0$ respectively (Fig.~\ref{fig:scematicMWdressing}(a)). 
The Hamiltonian of the ion at the position ${\boldsymbol r}_j$ is given by 
\begin{equation}
\label{eq:HMW}
H_{\rm MW}({\boldsymbol r}_{j})= \hslash \Delta_{\rm S} |{\rm S} \rangle _{j} \langle {\rm S}| _{j}+ \hslash \Delta_{\rm P} | {\rm P}\rangle _{j} \langle {\rm P}| _{j}+\frac {\hslash \Omega_{\rm MW}}{2}(|{\rm S} \rangle _{j} \langle {\rm P}| _{j}+|{\rm P} \rangle _{j} \langle {\rm S}| _{j}),
\end{equation}
where $\Omega_{\rm MW}=E_{1}d_{1}$ is the Rabi frequency of the MW driven transition with the $n {\rm P} \leftrightarrow n^{\prime} {\rm S}$ transition dipole moment $d_{1}=-e \langle {\rm P} | y_{i} | {\rm S} \rangle $, where $y_{i}$ is the $y$-coordinate of the Rydberg electron position for the $i$-th ion when the MW electric field polarisation aligned along the $y$ axis. $\Delta_{\rm P}$ and $\Delta_{\rm S}$ denote the detuning of the MW frequency with respect to the $n^\prime$S and the $n$P energy levels respectively. By diagonalizing this Hamiltonian, one obtains the MW-dressed states
\begin{equation}
\label{eq:MWdressed}
|\pm \rangle _{j}= \frac{1}{\sqrt{1+C^2_{\pm}}} (C_{\pm} | {\rm P}\rangle _{j} +| {\rm S}\rangle _{j}),
\end{equation}
where, $C_{\pm}=\frac{\Delta_{-} \pm \sqrt{\Omega^2_{\rm MW}+ \Delta^2_{-}}}{\Omega_{\rm MW}}$ is controlled by the MW field parameters and $\Delta_{\pm}=\Delta_{\rm P}\pm \Delta_{\rm S}$. These dressed states exhibit polarisablility $\alpha_{\pm}=(C^2_{\pm} \alpha_{n {\rm P}}+ \alpha_{n^{\prime} {\rm S}})/(1+C^2_{\pm})$. 
A proper choice of the MW field amplitude allows for tailoring the polarisability of the dressed state such that the energy shifts due to the large polarisability of Rydberg states~(Eqn.~\ref{eq:omegaXryd}-\ref{eq:omegaZryd}) are cancelled out.
For instance, for $n^{\prime}=n$ and $|C_{\pm}| \approx 0.68$, one obtains $\alpha_{\pm} \approx 0$. Under this condition, the secular trapping frequencies of the excited ion are identical to those of the ion in low-lying states, i.e., the shifts due to the large polarisability of Rydberg states given in Eqns.~\ref{eq:omegaXryd}-~\ref{eq:omegaZryd} are suppressed. 

The resonant dipole-dipole interaction between the two ions can be written as~\citep{li14a}
\begin{equation}
\label{eq:VddMWdressed}
V_{\rm dd} (\pm) \approx \frac{e^2}{4 \pi \epsilon_0 R^3_{0}} \left( \frac{d^2_{+} \Pi^{+}+d^2_{-} \Pi^{-}}{d^2_{\pm}} \right),
\end{equation}
where $R_0$ is the inter-ion distance and $\Pi^{+} =\ket{+}_1\bra{+}_1 \otimes \ket{+}_2\bra{+}_2$ and $\Pi^{-} =\ket{-}_1\bra{-}_1 \otimes \ket{-}_2\bra{-}_2$ denote the projection operators. The interaction strength depends on $d_{\pm}= \frac{\vert d_{1} \vert C_{\pm}}{e(1+C^2_{\pm})}$, and thus is tunable by the MW field parameters. 
Figure~\ref{fig:scematicMWdressing}(b-d) shows calculated van der Waals and dipole-dipole interactions for two trapped Rydberg Ca$^{+}$ ions enhanced using the above technique.
In these calculations, the MW Rabi frequency $\Omega_{\rm MW}=2 \pi \times 400$~MHz, $\Delta_{\rm S}=2 \pi \times 136$~MHz and $\Delta_{\rm P}=2 \pi \times 293$~MHz are used, which give rise to zero polarisability of the dressed Rydberg $|-\rangle$ state.
Note that the dipole-dipole interaction does not cause mixing between Rydberg states of different $m_{J}$ in the strong MW driving regime described here. The total angular momentum projection of the two ions $m^{(1)}_{J}+m^{(2)}_{J}= 1$ and the magnetic quantum number is preserved in the 65P$_{1/2}(m_{J}=1/2)\rightarrow$~65S$_{1/2}(m_{J}=1/2)$ transition. 

\begin{figure}
\centering
\includegraphics[width=0.6\textwidth]{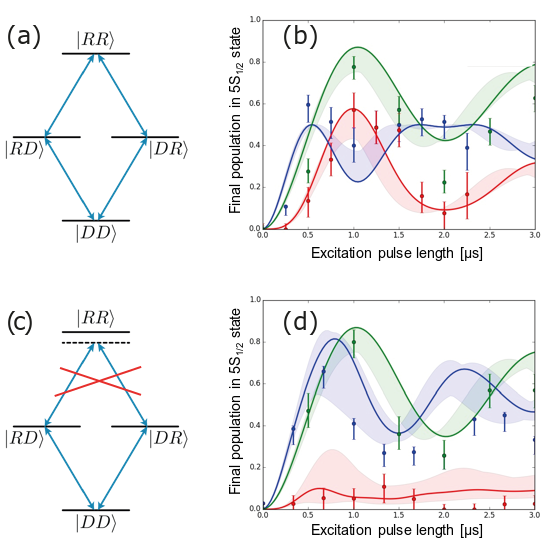}
\caption{Interactions between two Rydberg ions. (a) and (c) The energy of the $| RR \rangle$ is shifted by the dipole-dipole interaction between Rydberg ions. (b) and (d) Rabi oscillations between the ground state and the Rydberg state for one ion (green) and two ions (red and blue) as a function of excitation pulse length. The excitation to the $| RR \rangle$ state is suppressed at certain detuning of the MW field.}
\label{fig:Rydbergblockade}
\end{figure}  

Such a dipole-dipole interaction has been recently measured between two Sr$^{+}$ Rydberg ions using microwave radiation, resonant between $|S\rangle = 46S_{1/2} \leftrightarrow |P\rangle=46P_{1/2}$~\citep{zhang20a}.
In that experiment, the dipole-dipole interaction between the MW-dressed states $|+\rangle = \frac{1}{\sqrt{2}}(|S\rangle + |P\rangle)$ and $|-\rangle = \frac{1}{\sqrt{2}}(|S\rangle - |P\rangle)$ caused a Rydberg blockade effect, which prevented both ions being excited to Rydberg state $|+\rangle$ simultaneously (Fig.~\ref{fig:Rydbergblockade}).
With Rydberg excitation by STIRAP, both ions could be excited to $|+\rangle$ and the interaction allowed an entangling two-ion conditional phase gate to be implemented.
This gate was implemented in 700~ns with $78\%$ fidelity characterised using quantum process tomography.

\section{Future prospects for Rydberg ion crystals}\label{sec:future_prospects}
The prominence of Rydberg ions as a novel platform for quantum optics experiments derives from the possibility for precisely controlling strongly correlated many-body systems.
In this section, we give five specific examples of {\bf theoretical studies that explore such collective effects in trapped Rydberg ions}. We choose these examples such that they cover a few  application fields within the scope of this review, illustrating how techniques and properties described in the previous sections are applied. We note that a wide range of theoretical studies for Rydberg physics in cold atoms and ions can be considered. These include fast gate operations (Sec.~\ref{subsec:fastgatewithEfield}) and mode shaping techniques (Sec.~\ref{subsec:modeshaping}) in the realm of quantum computing, quantum simulators for coherent spin dynamics and excitation transfer~(Sec.~\ref{subsec:excitationtransfer}) and spin-spin interactions~(Sec.~\ref{subsec:spinspinInt}), and finally investigation of non-equilibrium dynamics using Rydberg ions~(Sec.~\ref{subsec:phasetransition}).

\subsection{Fast entangling operations using electric field pulses}\label{subsec:fastgatewithEfield} 	
Beyond the experiment mentioned in Sec.~\ref{sec:coh_spectroscopy:entangle}, an alternative protocol has been proposed to exploit Rydberg ions for fast entangling operations~\citep{vogel19a}.
In this scheme, the state-dependent force is driven by electric pulses that are a few hundred times faster than the period of ion motional modes.
Laser-less gate operations which have been thus far demonstrated use either static~\citep{khromova12a} or dynamic magnetic gradients~\citep{harty16a, warring13a, weidt16a} on the spin states of ions. 
But the scheme proposed in~\citep{vogel19a} takes advantage of large electric field gradients and fast electric pulses as an established technology in Paul traps, as follows.
 
\begin{figure}
\centering
\includegraphics[width=0.7\textwidth]{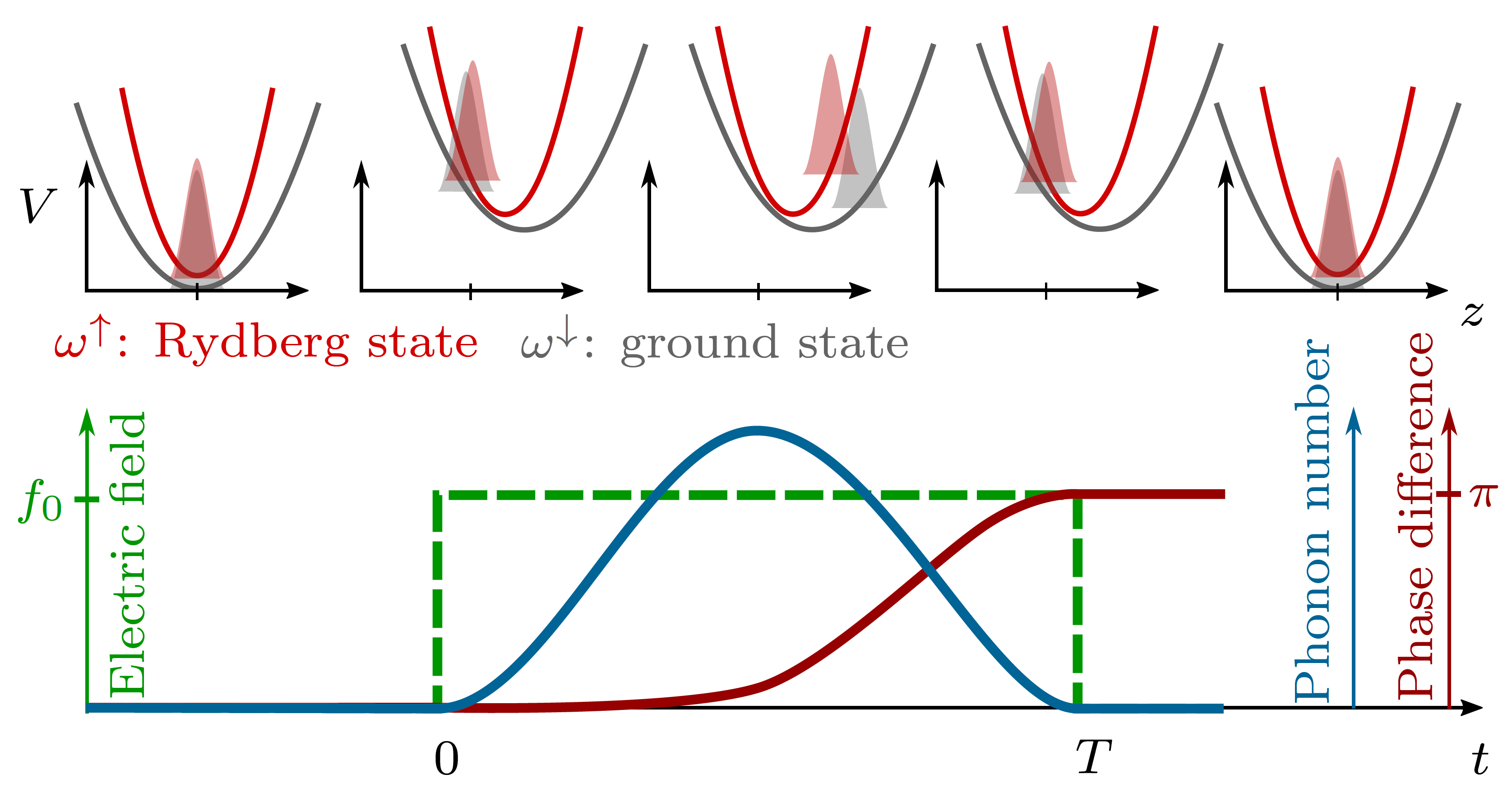}
\caption{Schematic of the state-dependent phase accumulation for Rydberg ions shuttled in a harmonic potential. Time evolution (from left to right) of the ion wavepacket under the act of a fast electric kick with field-sensitive internal states. The trapping frequency for the ion in the Rydberg state $\omega^\uparrow$ (red) is different from that for the ground state $\omega^\downarrow$ (grey). Colour code: green~--~fast electric kick that displaces the ion out of its equilibrium position, dark red~--~accumulated state-dependent phase difference between the Rydberg and the ground state, which is $\pi$ in this case, blue~--~coherent~motional excitation, which can be suppressed using certain pulse shapes. Adapted from~\protect\citep{vogel19a}.}
\label{fig:fastshuttling1}
\end{figure}

Two ions in a Paul trap are acted on by an electric pulse with amplitude $f(t)=f_0$ and pulse duration $T$. This pulse displaces the ions out of their equilibrium positions along the trap axis, introduces an induced electric dipole force and excites vibrational modes of the collective motion~(Fig.~\ref{fig:fastshuttling1}).
The electronic basis states for which the phase is controlled are $\ket{\alpha \beta}$ = $\{\ket{\downdownarrows}$, $\ket{\downuparrows}$, $\ket{\updownarrows}$,  $\ket{\upuparrows}\}$, with {\bf state-dependent collective frequencies} $\omega_j^{\alpha\beta}$ with the mode index $j\in \{ 1,2 \}$, where $\alpha\beta$ denotes the internal states of individual ions, i.e., the ground state or the Rydberg state.
Rydberg excitation in the ion crystal manifests itself in an additional electric potential seen by the adjacent ion, which causes asymmetric vibration around the centre-of-mass of the crystal due to a difference of effective masses~\citep{home16a, morigi01a}.  
The large, state-dependent polarisability of Rydberg states plays a key role by modifying the trapping frequencies~(Eqns.~\ref{eq:omegaXryd}-\ref{eq:omegaZryd}), and thus the phase acquired by each ion under the act of this electric kick depends on the ion internal state. 

\begin{figure}
\centering
\includegraphics[trim={4cm 10cm 4cm 10cm},clip, width=1.0\textwidth]{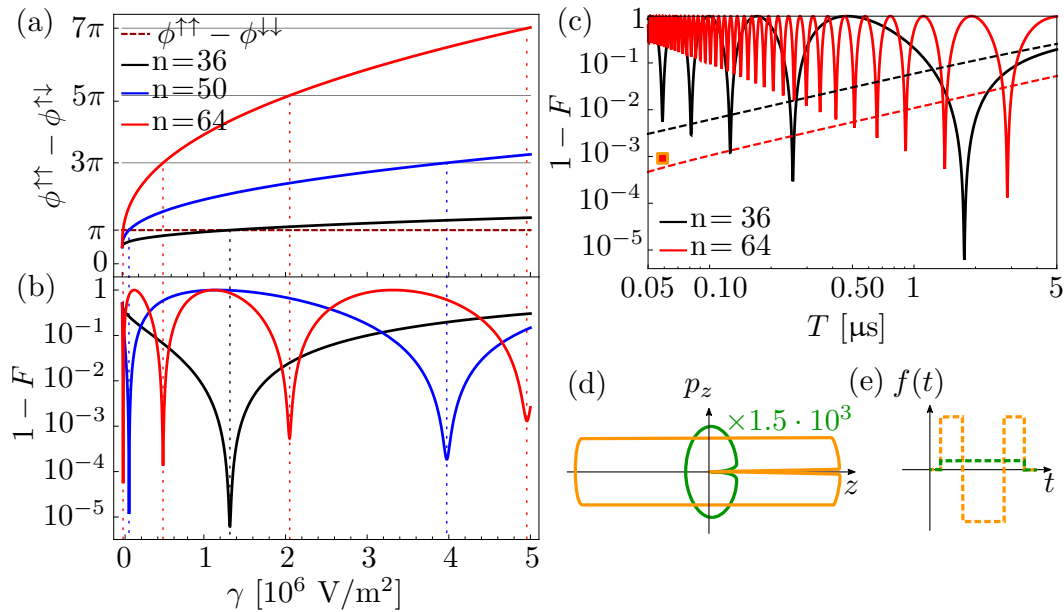}
\caption{Phase and fidelity calculated for two-ion gate operation using fast electric pulses. (a) Relative phase between states $\ket{\upuparrows}$ and $\ket{\updownarrows}$. (b) Infidelity as a function of $\gamma$, the gradient of static electric field of the Paul trap, shown for Rydberg states of $^{40}$Ca$^+$ with the principal quantum numbers $n=36$, 50 and 64. At a given $\gamma$, the electric kick shape can be chosen such that $\phi^{\uparrow\uparrow}-\phi^{\downarrow\downarrow}=\pi$ is satisfied (dashed dark red) and residual phonon numbers is minimised. (c) Entanglement fidelity (solid curves) and Rydberg state lifetime-limited fidelity (dashed lines) as a function of gate duration for $36$P (black) and $64$P (red) states with $65$~$\mu$s and $370$~$\mu$s lifetimes respectively~\citep{glukhov13a}. The red square (with yellow frame) in (c) indicates a bang-bang control: three consecutive kicks improve the fidelity to $99.9\%$ for $n=64$ at $60$~ns operation speed. (d) and (e) Phase space trajectories and field amplitudes for a constant pulse (green, scaled by $1.5\times 10^3$) and the waveform composed of three kicks (yellow). Adapted from~\protect\citep{vogel19a}.}
\label{fig:fastshuttling12}
\end{figure}

The Hamiltonian of the system is written in terms of the state-dependent creation $\tilde{a}_j^\dagger \equiv (a_j^{\alpha\beta})^\dagger$ and annihilation  $\tilde{a}_j \equiv a_j^{\alpha\beta}$ operators~\citep{vogel19a}
\begin{align}
  H_\text{p}= \sum_{\alpha\beta=\uparrow,\downarrow}\left( \sum^2_{j=1} \hslash \omega_j^{\alpha\beta} \tilde{a}_j^\dagger \tilde{a}_j +V_0^{\alpha\beta}\right)\Pi^{\alpha\beta}.
\end{align}
Here, $V_0^{\alpha\beta}$ is a potential term that depends on the equilibrium positions of the ions and $\Pi^{\alpha\beta} =\ket{\alpha}_1\bra{\alpha}_1 \otimes \ket{\beta}_2\bra{\beta}_2$ is the projection operator.
The Hamiltonian of the driven motion due to the electric kick is given by
\begin{equation}
\mspace{-10mu} H_\text{d}(t)=\mspace{-5mu} \sum_{\alpha\beta}\mspace{-5mu}\left[ \sum^2_{j=1} (F_j^{\alpha\beta}(t) \hspace{1mm} \tilde{a}_j + \text{h.c.}) + f(t)\;Z_\text{c}^{\alpha\beta} \right]\Pi^{\alpha\beta}. \label{eq:drivingHamiltonian}
\end{equation}
In the first term, the state-dependent kick $F_j^{\alpha\beta}(t)$ acting on the vibrational mode describes the interaction of the electric pulse with the ion crystal, see also~\citep{cirac00a, garcia-ripoll03a, garcia-ripoll05b}. Coherent excitations of vibrational modes is controlled by applying proper pulse amplitude and duration, as experimentally demonstrated in~\citep{bowler12a, kaushal20a, walther12a}. Moreover, impulsive electric pulses of sub-ns resolution have been used in a ``bang-bang'' control of single ions with up to 10\,000 phonons~\citep{alonso16a}.
The second term in Eqn.~\ref{eq:drivingHamiltonian} is proportional to the crystal centre $Z_\text{c}^{\alpha\beta}$ and leads to non-zero phase evolution only for ion crystals with a single Rydberg excitation.

The total phase accumulated on each of the four basis states is given by $\phi^{\alpha\beta}=\varphi_1^{\alpha\beta}+\varphi_2^{\alpha\beta}+\Phi_\text{e}^{\alpha\beta}$, where $\varphi_1^{\alpha\beta}$ and $\varphi_2^{\alpha\beta}$ denote the contributions of the two vibrational modes and $\Phi_\text{e}^{\alpha\beta}$ results from the crystal centre displacement.
The {\bf entangling operation} is controlled only by the shape of the electric pulse~($f_0$ and $T$) and the common mode frequencies ($\omega_j^{\alpha\beta}$). To realise a {\bf two-ion controlled phase gate}, the phase difference $\phi^{\uparrow\uparrow}-\phi^{\downarrow\downarrow}=\pi$ is required, where $\phi^{\downarrow\downarrow}=\phi^{\uparrow\downarrow}=\phi^{\downarrow\uparrow}$ is satisfied and no residual excitation in phonon modes generated. The result of these calculations for a two-ion crystal of Ca$^{+}$ ions is shown in Fig.~\ref{fig:fastshuttling12}(a-c). By tuning the experimental parameters $\gamma$ and the two collective motional modes, arbitrary phase rotations can be realised.

A remarkable feature of this gate is the possibility for simultaneous improvement of the gate fidelity and its speed using complex electric pulses. This has been calculated for a case of ``bang-bang'' control with three kicks as shown in Fig.~\ref{fig:fastshuttling12}(c-e).
In addition, optimal control algorithms~\citep{caneva11a, fuerst14a, rach15a} can be used for computing required electric pulses as well as laser pulses for exciting Rydberg states. 
It is worth nothing that this scheme uses axial modes and electric kicks along the trap axis, and the extension of the technique for radial modes or combination of radial and axial modes requires synchronization of the electric kick with the phase of the RF drive, as implemented in~\citep{jacob16a}.
 

\subsection{Mode shaping in linear ion crystals by Rydberg excitations} \label{subsec:modeshaping}
\begin{figure}[h!]
\centering
\includegraphics[width=1 \textwidth]{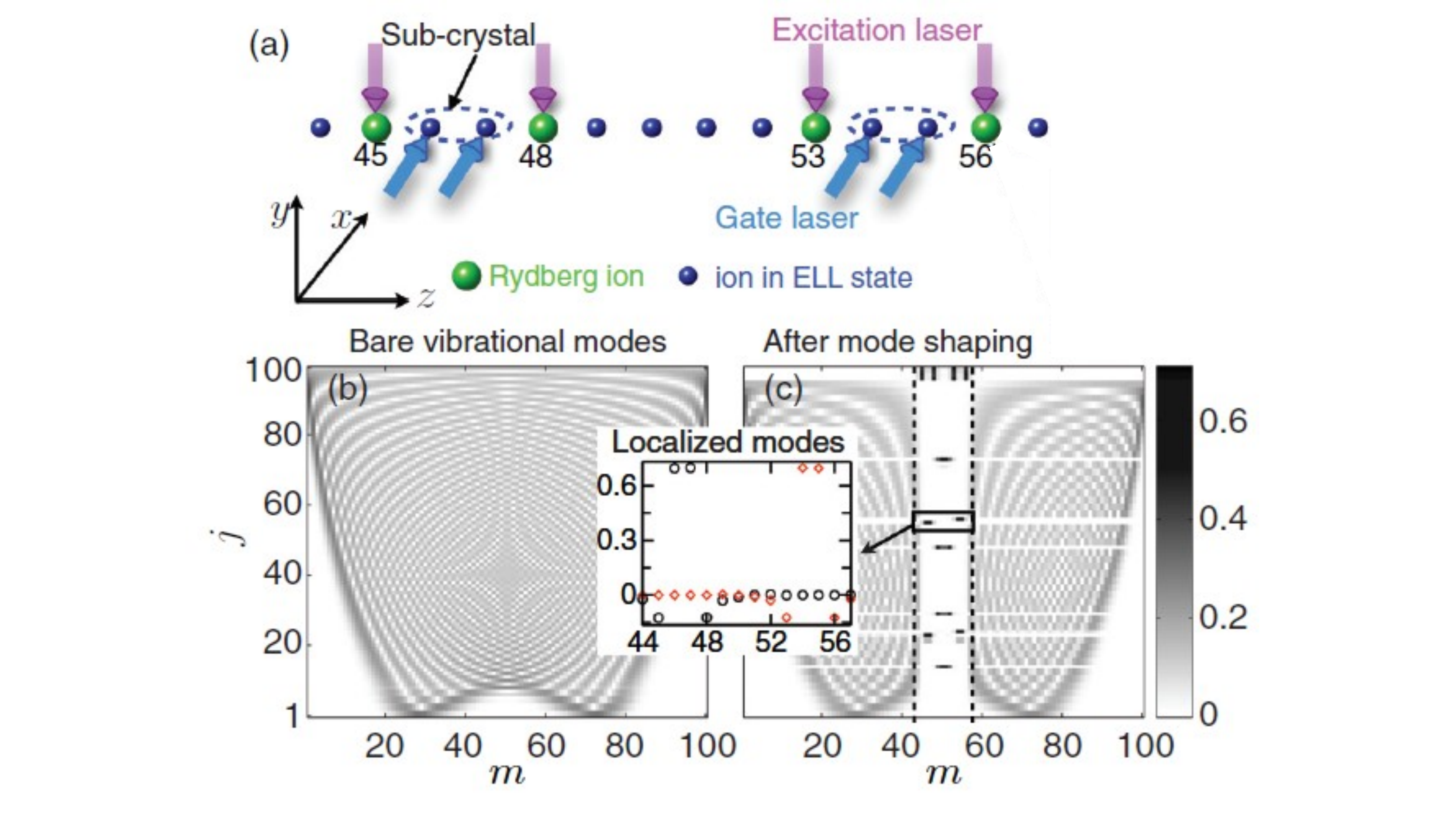}
\caption{(a) Schematic for mode shaping in a linear ion crystal in which sub-crystals of ion pairs in electronically low-lying (blue) states are isolated between ions in Rydberg (green) states. Using laser-induced spin-dependent forces, quantum gate operations can be executed on the two sub-crystals in parallel. (b) and (c) Vibrational modes shown by the modulus of the normal mode eigenvectors~$j$ for the $m$-th ion in a linear crystal with and without mode shaping respectively. Here, in a crystal of 100 Ca$^{+}$ ions the 45-th, 48-th and 53-th, 56-th ions are excited to the Rydberg $n$P$_{1/2}$ state. The black dashed lines in (c) indicates the part of the spectrum related to the ion configuration shown in (a). The Rydberg ions significantly reshape the vibrational mode structure and form localized modes on the two subcrystals as depicted in the inset of (c). Adapted from~\protect\citep{li13a}.}
\label{fig:modeshaping_parallel}
\end{figure} 
Rydberg excitation of an ion in a chain {\bf manifests itself as a drastic change in the vibrational mode structure} of the crystal. Consequently, spatially localised modes are generated in the sub-crystals isolated by excited ions, see Fig.~\ref{fig:modeshaping_parallel}~\citep{li13a}. 
The collective modes are described by the phonon Hamiltonian $H_{\rm ph}=\sum^{N}_{\alpha=1} \hslash \omega_{\alpha}(\hat{a}^{\dagger}_{\alpha}  \hat{a}_{\alpha}+1/2)$, see Sec.~\ref{subsec:normalmodes}, where the eigenfrequency~$\omega_{\alpha}$ is calculated by diagonalising the corresponding Hessian matrix~\citep{li13a}.
Thus, sub-crystals {\bf isolated between the Rydberg ions} can be employed to perform certain gate operations in parallel.
As a particular example, the fidelity of two two-qubit conditional phase flip gates that are executed in parallel on different subcrystals in the same ion chain was explored~\citep{li13a}.

\begin{figure}[h!]
\centering
\includegraphics[width=1.0 \textwidth]{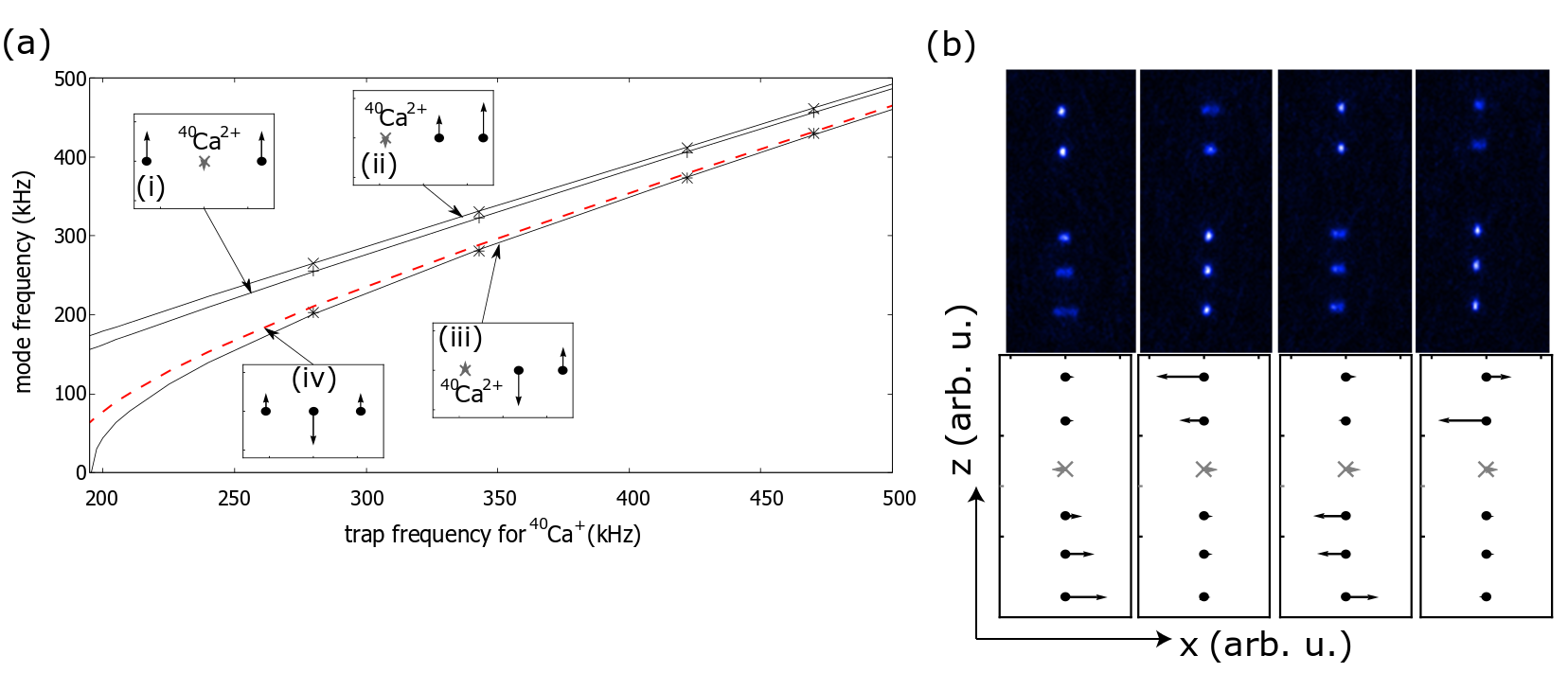}
\caption{(a) Measured (dots) and calculated (lines) motional frequencies for a 3-ion crystal with two $^{40}$Ca$^{+}$ ions and one $^{40}$Ca$^{+2}$ ion with corresponding modes depicted in (i-iii). For comparison, the dashed line shows the simulated frequency of the zigzag mode in a crystal without $^{40}$Ca$^{+2}$ (iv). (b) Fluorescence images of excited local modes in a 6-ion crystal of $^{40}$Ca$^{+}$ ions including one $^{40}$Ca$^{+2}$ ion, which is dark. The data and the simulation demonstrate tailoring of radial modes by the $^{40}$Ca$^{+2}$ ion. Adapted from~\protect\citep{feldker14a}.}
\label{fig:modeshaping_ca2}
\end{figure}   
The basic principles of the above technique were studied in a linear, mixed ion crystal of $^{40}$Ca$^{+}$ and $^{40}$Ca$^{2+}$ ions~\citep{feldker14a, kwapien07a}. The modification of the crystal mode structure owing to different charge-to-mass ratios of these ions leads to motional modes that are not orthogonal to the centre-of-mass modes of a pure crystal of $^{40}$Ca$^{+}$~(Fig.~\ref{fig:modeshaping_ca2}). In a chain consisting of two $^{40}$Ca$^{+}$ ions and one $^{40}$Ca$^{2+}$ ion, this has been observed for the radial modes, except for the radial mode with the $^{40}$Ca$^{2+}$ ion at the centre position~\citep{feldker14a}~(Fig.~\ref{fig:modeshaping_ca2}(a)).
More importantly, it has been shown that in a mixed crystal consisting of five $^{40}$Ca$^{+}$ and one $^{40}$Ca$^{2+}$ ion motional modes of sub-crystals of $^{40}$Ca$^{+}$ ions can be excited separately~(Fig.~\ref{fig:modeshaping_ca2}(b)). 

\subsection{Energy transfer quantum simulation}\label{subsec:excitationtransfer}
{\bf Transport of energy excitations} using dipole-dipole interactions has an important role in light-harvesting processes~\citep{vanamerongen00a}, in neighbouring interactions in solid-state quantum dots~\citep{kagan96a} and in resonant excitation and de-excitation processes in molecular aggregates~\citep{saikin13a}.
In ultracold Rydberg atoms, energetic disorders and decoherences can be introduced by laser interaction with a neutral background gas with potential applications in {\bf quantum simulation of energy excitation dynamics}~\citep{schoenleber15a}. Simulation of energy transfer along a protein chain in biomolecules in DNA systems has been proposed~\citep{plodzien18a} as a specific example of polaron phenomena~\citep{casteels11a}.
The observation of dipole-mediated energy transfer in laser-excited Rydberg atoms has been used to develop a non-destructive imaging technique~\citep{guenter13a}. 

\begin{figure}[h!]
\centering
\includegraphics[width=0.6\textwidth]{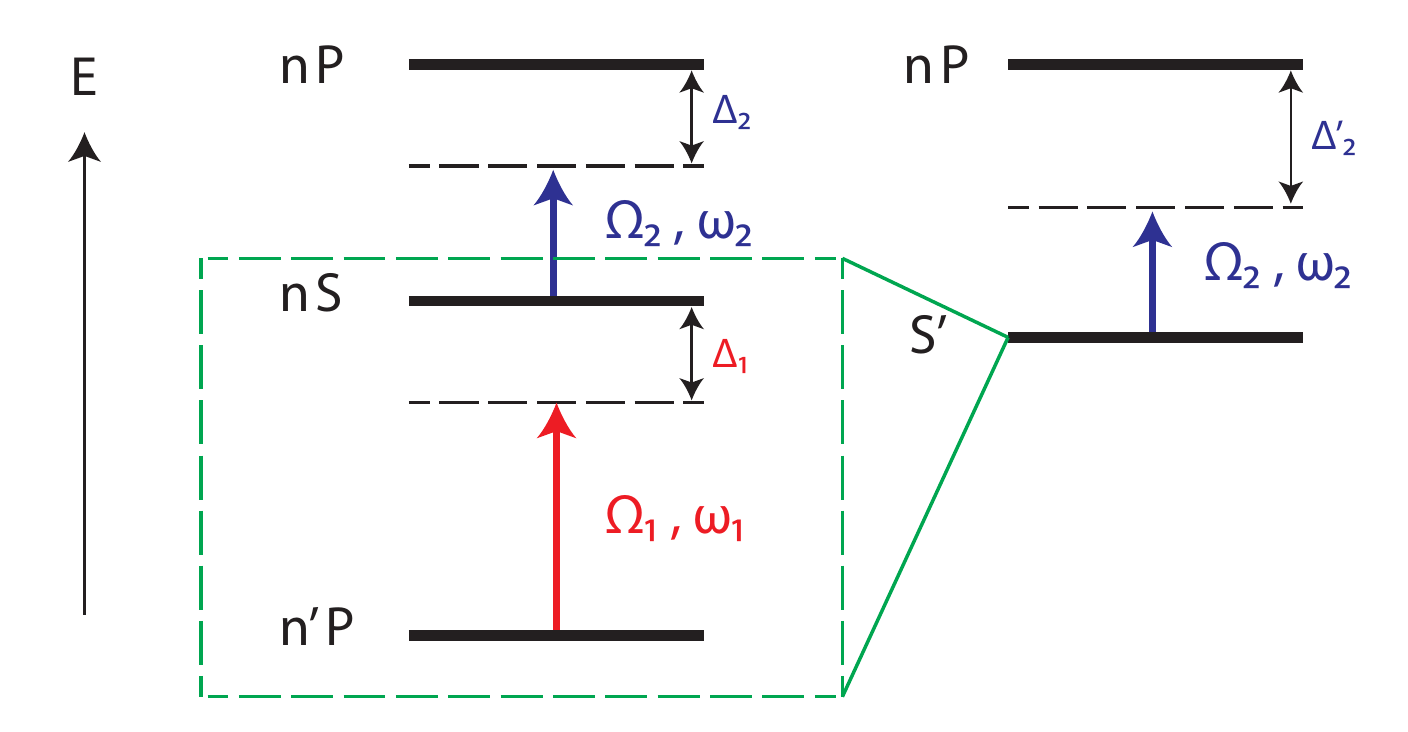}
\caption{Schematic of MW dressing of ionic Rydberg states using a linearly polarised bichromatic MW field that couples one S with two P levels one of which is far detuned. After an adiabatic elimination of the latter (depicted by the green box), one obtains an effective two-level system indicated with S$^{\prime}$ and $n$P states. This scheme can be used for tailoring the polarisability of Rydberg ions, see also the strong coupling scheme in Sec.~\ref{sec:coh_spectroscopy:entangle}. Adapted from~\protect\citep{mueller08a}.}
\label{fig:2MWdressing}
\end{figure}

In a chain of trapped ions excited to Rydberg states, the internal dynamics can be mapped into an effective spin-$1/2$ system, which can be engineered using dipolar interactions between them~\citep{mueller08a}. 
As elaborated in Sec.~\ref{sec:coh_spectroscopy:entangle}, the resonant dipole-dipole interaction between Rydberg ions is achieved by MW-dressing of Rydberg states. Here, we consider the MW scheme using a dichromatic MW potential $V_{\rm MW}({\boldsymbol r}_{j})=-e E_{1} \hat{\varepsilon}_{1} \cdot {\boldsymbol r}_{j} \cos (\omega_{1} t)-e E_{2} \hat{\varepsilon}_{2} \cdot {\boldsymbol r}_{j} \cos (\omega_{2} t)$ to generate an admixture of three Rydberg states as shown in Fig.~\ref{fig:2MWdressing}.
In a reference frame which rotates at $\omega_{1}$ and $\omega_{2}$ and using the rotating-wave approximation, the internal and external dynamics are decoupled, and thus the full Hamiltonian for $N$ interacting ions can be written as
\begin{equation}
H_{\rm ions}= H_{\rm{ph}}+\sum^{N}_{i} H_{{\rm ele}, i}+H_{\rm{int-ext}}+H_{\rm dd}.
\label{eq:HionsExcitation}
\end{equation}
Here, $ H_{\rm{ph}}= \sum_{\alpha=x,y,z} \sum_{m=1}^{N} \hslash \omega_{\alpha, m} \tilde{a}_{\alpha, m}^\dagger \tilde{a}_{\alpha, m}$ denotes the external oscillation dynamics~(Sec.~\ref{subsec:normalmodes}), in which generalised forms of state-dependent creation~$\tilde{a}_{\alpha, m}^\dagger$ and annihilation~$\tilde{a}_{\alpha, m}$ operators are used. The charge-dipole interaction manifests itself in the position-dependent electric field seen by each ion along the trap axis $z$. This gives rise to additional terms in the Hamiltonian of a single ion located at the minimum of the static electric field (Eqn.~\ref{eq:Helectron}). 
\begin{equation}
\begin{aligned}
H_{{\rm ele}, i}=&\frac{\boldsymbol{p}^2}{2m_{\rm{e}}}+V(|{\boldsymbol r}|)+V_{ls}({\boldsymbol r})-e \gamma^{\prime} \cos(\Omega_{\rm RF} t) (x^2_i -y^2_i)\\ & + e \left( \gamma+\frac{e}{8 \pi \epsilon_{0}} \sum_{j \neq i}^{N} \frac{1}{\vert Z_i^{(0)}-Z_j^{(0)} \vert ^3} \right)(x^2_i+y^2_i-2 z^2_i)
\label{eq:Heledeltagamma}
\end{aligned}
\end{equation}

The term $H_{\rm{int-ext}}$ in Eqn.~\ref{eq:HionsExcitation} describes the {\bf coupling of external and internal dynamics}, which arises from dipole-charge interactions and the quadrupolar static field of the Paul trap. The shifts due to state-dependent trapping frequencies that are used for implementing fast gate operations in Sec.~\ref{subsec:fastgatewithEfield} are negligible for the total interaction $H_{\rm ions}$~(Eqn.~\ref{eq:HionsExcitation}). 
The dipole-dipole interaction in Eqn.~\ref{eq:HionsExcitation} is given by~\citep{mueller08a}
\begin{equation}
H_{\rm dd}=\frac{-1}{4 \pi \epsilon_{0}} \sum_{j \neq i}^{N} \frac{d^{(i)}_z d^{(j)}_z}{\vert Z_i^{(0)}-Z_j^{(0)} \vert ^3},
\label{eq:HddExcitation} 
\end{equation}
with the dipole operator $d^{(j)}_z$ for the $j$-th ion in the set of S$^{\prime}$ and $n$P states given by 
\begin{equation}
d_z=\frac{1}{3}
\begin{pmatrix}
-\frac{\Omega_{\mathrm{1}}}{\Delta_{\mathrm{1}}} d_1 \cos (\omega_1 t) & d_2 {\rm e}^{-i \omega_2 t} \\
d_2 {\rm e}^{i \omega_2 t} & 0
\end{pmatrix}.
\end{equation}
Note that the electric field at the ions' equilibrium positions denoted by $Z^{(0)}_j$ for the $j$-th ion is cancelled out by the act of the Coulomb force and the trapping force, but the position-dependent electronic interactions stem from the couplings described by Eqns.~\ref{eq:Heledeltagamma}~and~\ref{eq:HddExcitation}.

\begin{figure}[h!]
\centering
\includegraphics[width=0.85\textwidth]{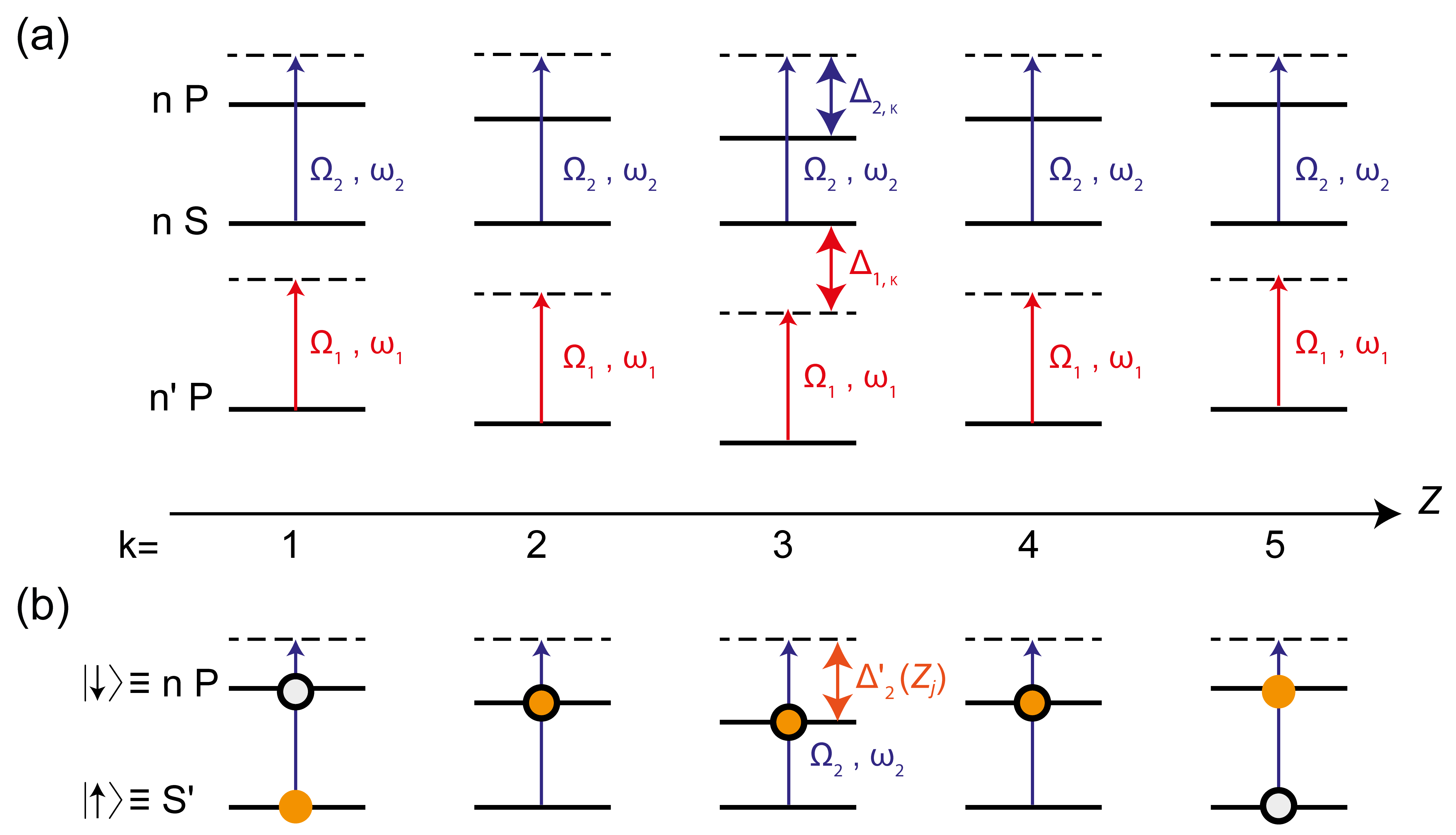}
\caption{Illustration of the excitation transport along a linear chain of ions. (a) Schematic level diagrams for a five-ion crystal with position-dependent energy shifts for each MW-coupled Rydberg ion. The states are labelled as $\ket{\downarrow} \equiv n$P and $\ket{\uparrow} \equiv n$S. The energy shifts for the state $\ket{\downarrow}$ manifest themselves in the inhomogeneous distribution of the MW-field detuning $\Delta^{\prime}_{2} (Z^{(0)}_{j})$, where $Z^{(0)}_{j}$ is the equilibrium position of the $j$-th ion along the trap axis, see Fig.~\ref{fig:2MWdressing} for the MW coupling scheme. (b) The initial configuration (orange circles) with the first ion in the state $\ket{\uparrow}$ while all the other ions are in the state $\ket{\downarrow}$. After a certain time, the Rydberg excitation has transported to the right end of the chain (black open circles). $k$ indicates the ions' position in the chain. Adapted from~\protect\citep{mueller08a}.}
\label{fig:excitationtrasfer1}
\end{figure}

\begin{figure}[h!]
\centering
\includegraphics[width=0.85\textwidth]{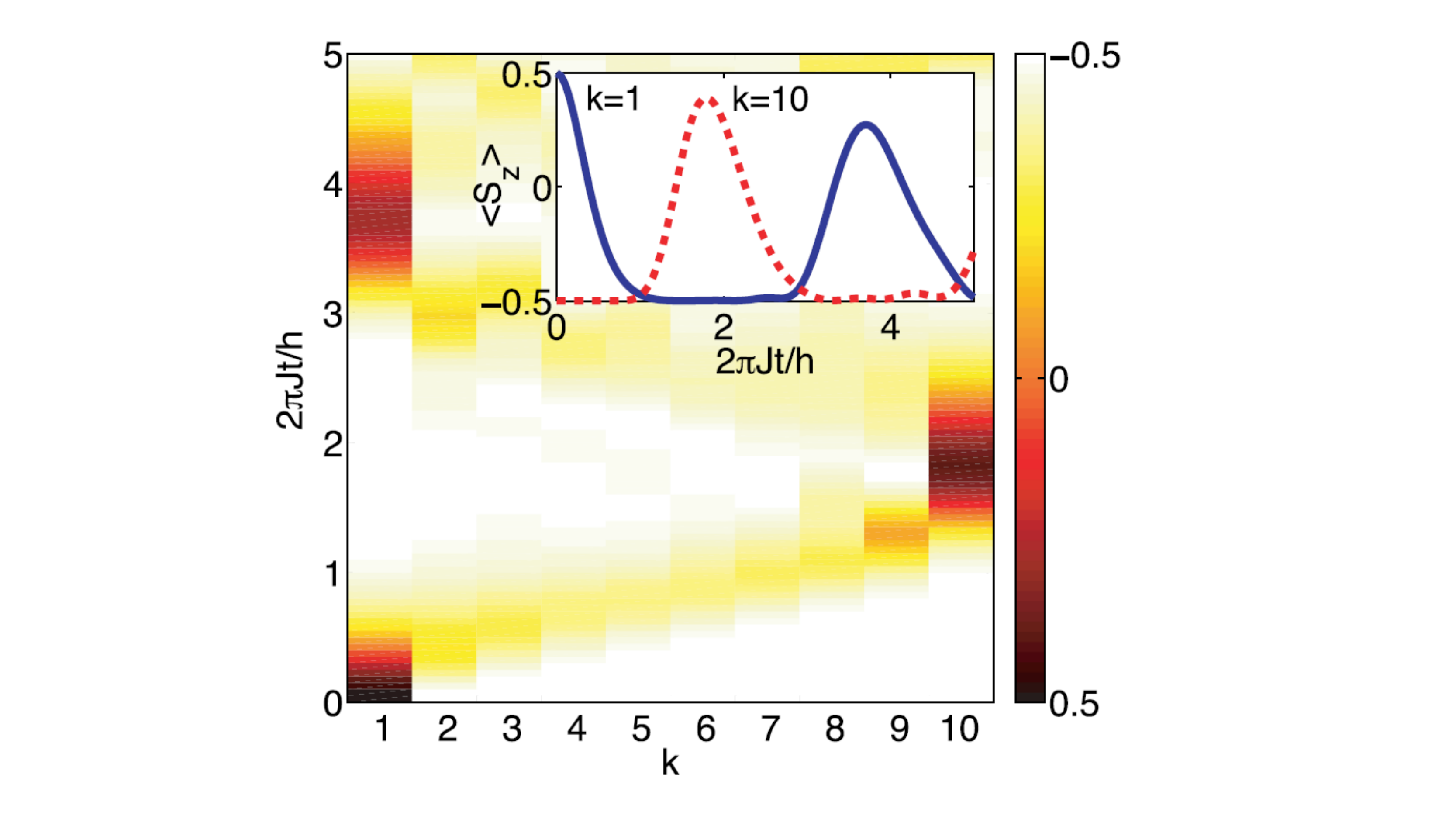}
\caption{Calculations for a     spin excitation transport in a chain of ten trapped ions. The ion located at the site ${\rm k}=1$ is initially excited to the state $\ket{\uparrow}$, while all others are in the state $\ket{\downarrow}$. The excitation is coherently transferred from the first ion to the last one. The scale of the interaction energy is given by $J=-2M \omega^2_{Z}d^2_{2}/9e^2$, with the Rydberg transition dipole matrix element $d_{2}$. The time evolution of the expectation values $\langle S^{\rm }_{z} \rangle$ for the first and the last ions after time $t=1.8 \hslash /J$ are shown in the inset. Adapted from~\protect\citep{mueller08a}.}
\label{fig:excitationtrasfer2}
\end{figure}

The position-dependent energies of the total Hamiltonian (Eqns.~\ref{eq:HionsExcitation}-\ref{eq:HddExcitation}) represent inhomogeneity both in the exchange couplings and effective magnetic field in the spin dynamics language. This feature manifests itself in the position-dependent MW detuning $\Delta^{\prime}_{2} (Z^{(0)}_{j})$ for each ion in the chain~(Fig.~\ref{fig:excitationtrasfer1}(a)). 
A linear chain of ions is initially prepared by a series of $\pi$-laser pulses such that the ion located at the most left position is in the $\ket{\uparrow} \equiv |n, S \rangle$ state while all others are in the $\ket{\downarrow} \equiv |n, P \rangle $ state~(Fig.~\ref{fig:excitationtrasfer1}(b)).
For simplicity, cases in which only resonant dipole-dipole interactions contribute were studied.
The temporal evolution of the internal states of ions under the Hamiltonian $H_{\rm int}=H_{{\rm ele},i}+H_{\rm dd}$ given in Eqn.~\ref{eq:Heledeltagamma} and \ref{eq:HddExcitation} leads to the excitation transfer after a given time~(Fig.~\ref{fig:excitationtrasfer1}(b)). Numerical calculations for energy transfer dynamics in a chain of ten ions are shown in Fig.~\ref{fig:excitationtrasfer2}.   

\subsection{Planar ion crystals with Rydberg ions for quantum simulation of frustrated quantum magnets}
\label{subsec:spinspinInt}
\begin{figure}[h!]
\centering
\includegraphics[width=0.9\textwidth]{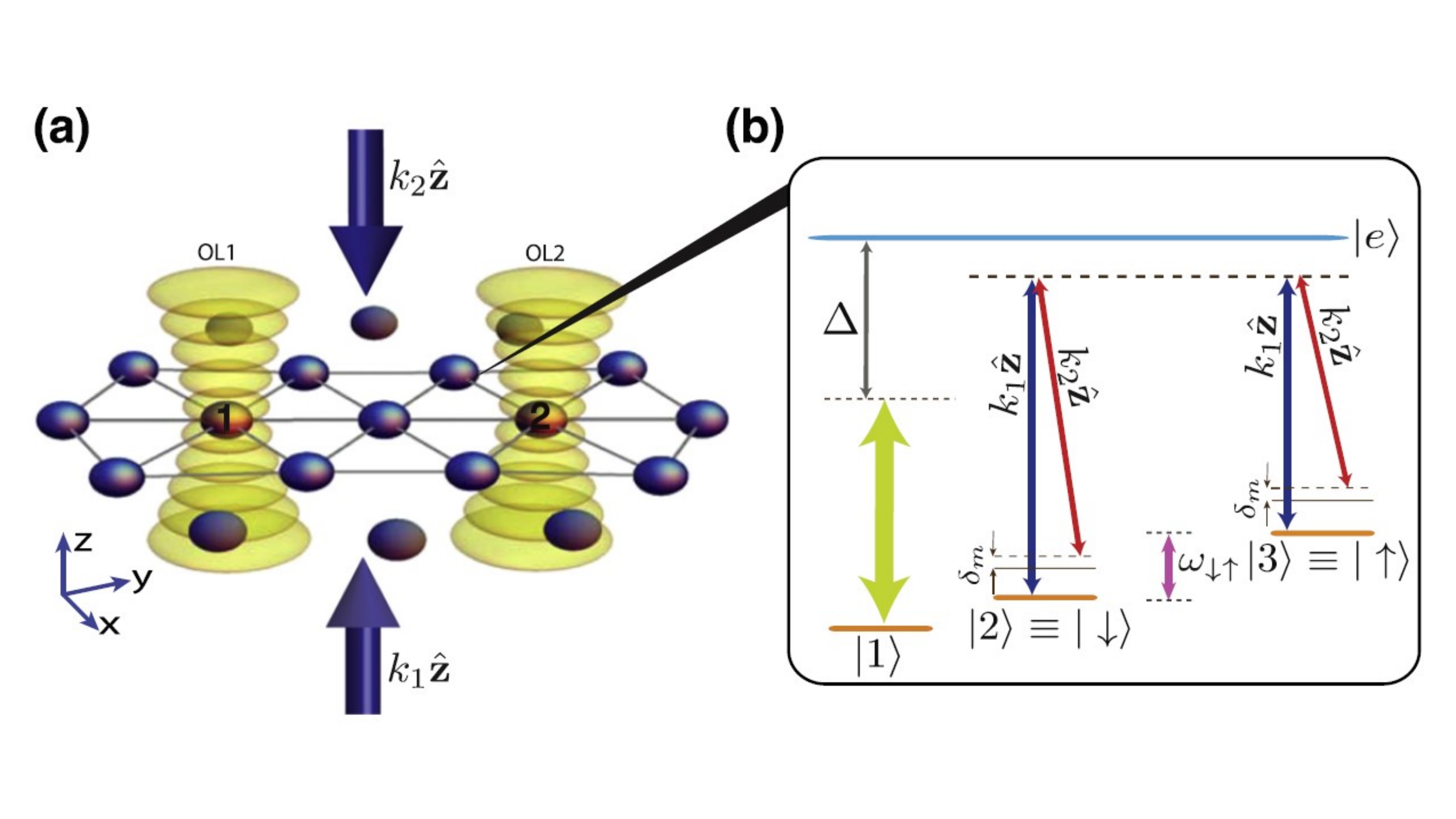}
\caption{(a) Schematic of the setup for hexagonal-plaquette spin-spin interactions. Laser beams with wavevectors $k_1 \hat{\bf z}$ and $k_2 \hat{\bf z}$ and frequencies $\omega_1$ and $\omega_2$ are used for driving Raman transitions shown in (b). Ions at sites 1 and 2 are pinned using far-detuned laser fields (yellow arrow) either by optical lattices OL1 and OL2 or by Rydberg excitations. (b) Energy level diagram and laser driven transitions. Raman lasers (blue and red arrows) are detuned by $\delta_m$ with respect to the frequency of the $m$-th motional mode, and generate state-dependent optical shifts in the spin states $|2\rangle\equiv \ket{\downarrow}$ and $|3\rangle\equiv \ket{\uparrow}$ with an energy separation of $\hslash\omega_{\downarrow\uparrow}$. Adapted from~\protect\citep{nath15a}.}
\label{fig:PLQsetup}
\end{figure}

Having discussed mode shaping techniques using Rydberg excitations in a linear ion crystal (Sec.~\ref{subsec:modeshaping}), we now turn to their extension to two-dimensional ion crystals for quantum simulation of spin-spin interactions.
A particular example of such an experiment is proposed in~\citep{nath15a} to emulate topological quantum spin liquids using the spin-spin iterations between ions in hexagonal plaquettes in a 2D ion crystal~(Fig.~\ref{fig:PLQsetup}(a)).
The role of a Rydberg ion is to modify the phonon mode spectrum such that constrained dynamics required for realising the specific Hamiltonian of the Balents-Fisher-Girvin (BFG) model~\citep{balents02a} using a Kagome lattice are reached. This model is described by~\citep{nath15a} 
\begin{equation}
H_{\rm s}=J_z\left(\sum_{i\in \hexagon}S_z^i \right)^2+\frac{J_{\perp}}{2}\sum_{\langle ij \rangle}\left(S_+^{i}S_-^{j}+ S_-^{i}S_+^{j} \right),
\label{ham1}
\end{equation}
where $S_\alpha^j$ ($\alpha \in \lbrace +,-,z \rbrace$) is the spin operator acting on the site $j$ with one spin-$1/2$ degree of freedom. The first term accounts for Ising interactions for each hexagonal-plaquette with $J_z(>0)$ and the second term denotes the nearest-neighbour spin exchange interaction. In this model, the source of the frustration is a macroscopic classical ground-state degeneracy caused by local hexagonal plaquette constraints, i.e.,~$\sum_{i\in\hexagon}S_z^i=0$ \citep{nath15a}. 

To generate such hexagonal-plaquette spin-spin interactions, spin-dependent optical dipole forces interact with engineered collective vibrational modes of a 2D ion crystal. 
The internal level structure of ions is identified by three long-lived states: the state $|1\rangle$ is coupled to an excited state $|e\rangle$, which is used to pin the ion as required for tailoring phononic modes, the states $|2\rangle\equiv \ket{\downarrow}$ and $|3\rangle\equiv \ket{\uparrow}$ encode the spin-$1/2$ states of the BFG model. Laser-mediated spin-spin interactions in trapped ions have been extensively studied, for instance see~\citep{lanyon11a}. 
A pair of counter-propagating Raman laser beams with wavevectors $k_1 \hat{\bf z}$ and $k_2 \hat{\bf z}$ and frequencies $\omega_1$ and $\omega_2$ are used to generate spin-spin couplings~(Fig.~\ref{fig:PLQsetup}(a)) and state-dependent optical shifts in the spin states $\ket{\downarrow}$ and $\ket{\uparrow}$ with an energy separation of $\hslash\omega_{\downarrow\uparrow}$.

The schematic of such an experiment is illustrated in Fig.~\ref{fig:PLQsetup}. Note that here the secular trapping frequencies conform $\omega_{X, Y} \ll \omega_{Z}$. Ions at sites 1 and 2 are prepared in the state $|1\rangle$ and are superimposed with additional laser fields which either generate optical lattices or excite the ions to Rydberg states. In either way, the result is the local modification of ions' secular trapping frequencies in the direction perpendicular to the crystal plane. 
In the former case, the transversal trapping frequency $\tilde{\omega}_{Z}$ is modified by the harmonic frequency of the local potential of the optical lattices applied, whereas in the latter, the modification is given in terms of the polarisability of the Rydberg state excited as given in Eqn.~\ref{eq:omegaZryd}.

The effective spin-spin interaction for this hexagonal plaquette is written as~\citep{nath15a} 
\begin{equation}
H_{SS}= \sum_{i<j}J_z^{ij}S_z^i\otimes S_z^j+\sum_{i<j}J_{\perp}^{ij}\left(S_x^i\otimes S_x^j+S_y^i\otimes S_y^j\right).
\end{equation}
Here, the coupling matrix is $J_z^{ij}=\sum_{m=1}^{N}(4\Omega_I^i\Omega_I^j  \eta_m^i\eta_m^j)/{\delta_m}$, where $\Omega_I^i=\Omega_I^i(\ket{\downarrow})-\Omega_I^i(\ket{\uparrow})$ and the two photon Rabi frequency for the Raman beams is $\Omega_I=\Omega_1\Omega_2/\Delta$ with the detuning of $\Delta_1 \approx \Delta_2 \gg \Omega_{1,2}$ for each laser, and $\eta^i_{m}$ is the Lamb-Dicke parameter for the $i$-th ion for the motional mode $m$~(Eqn.~\ref{eq:Lamb-Dicke effective}). 
For the phonon mode $m$, the frequency difference of the Raman beams $\omega_I$ is tuned close to the phonon frequencies $\omega_m$ and is shown with $\delta_m=\omega_I-\omega_m$ in Fig.~\ref{fig:PLQsetup}(b).
Note that plaquette pattern implies that each of the six spins occupying the corners of a hexagon interact with every other spin in the hexagon with the same strength regardless of their separations.

\begin{figure}[h!]
\centering
\includegraphics[trim={6cm 1cm 6cm 0cm},clip, width=1.0 \textwidth]{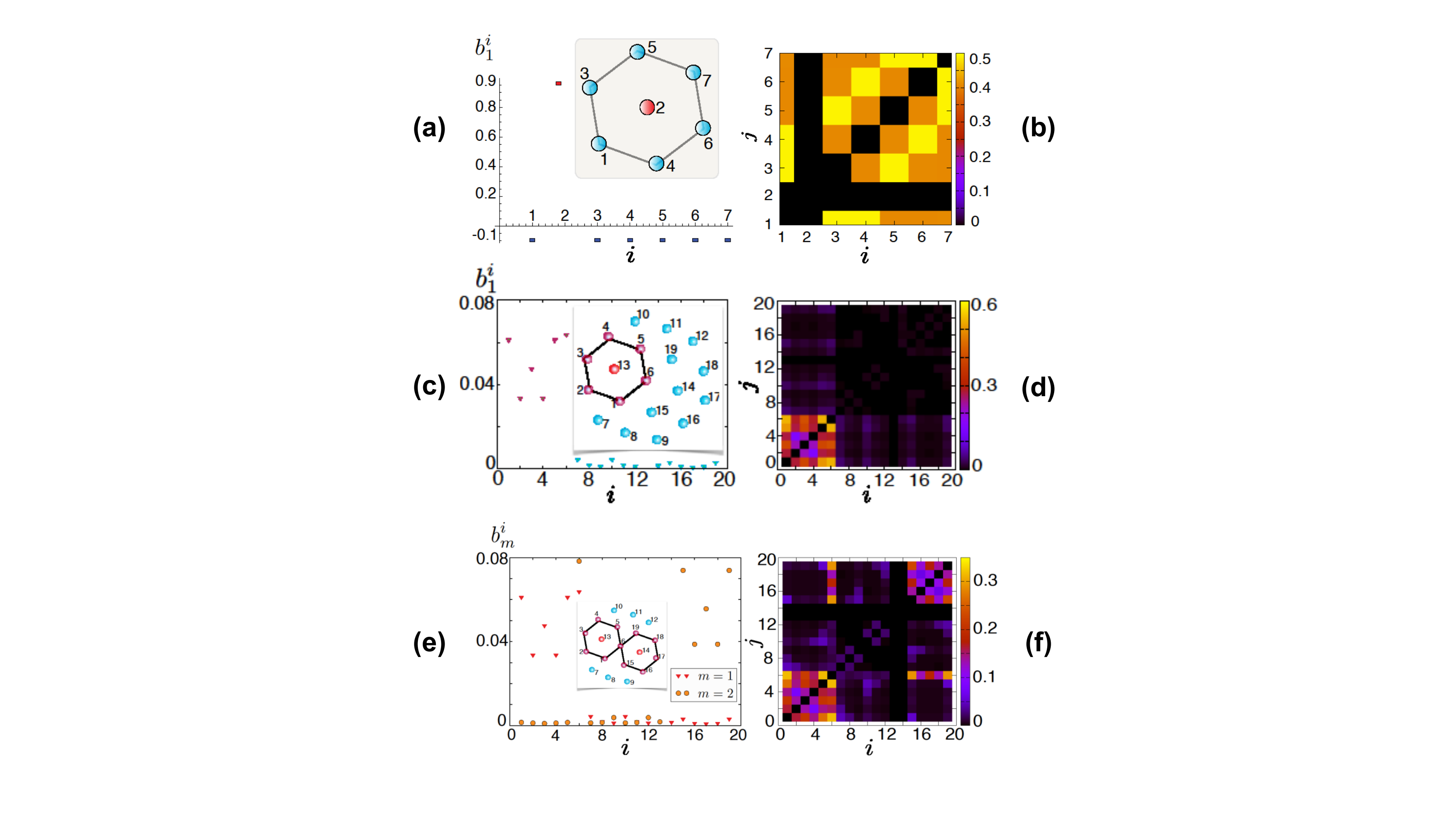}
\caption{(a) The eigenvector $b_1^i$ for the lowest transversal mode (TM) of the ion crystal. The inset shows the equilibrium position of 7 ions in a Paul trap with trapping frequencies $\omega_{X}=\omega_{Y}=2\pi\times 1$ MHz and $\omega_Z=2\pi\times 3$ MHz. The central ion, labelled with 2, experiences an additional force due to the optical lattices or Rydberg excitation and hence modified trapping frequency $\omega^{\prime}_Z=2\pi\times 2.7$ MHz. (b) The dimensionless spin-spin couplings $J_{ij}/\omega_0$, where $\omega_0=\left(\hslash k_I^2/8M\right)\left(\omega_x/\Omega_I\right)^2$, with $\delta_{1}=2\pi\times 10$ kHz, the detuning from the lowest TM. (c) The eigenvector for the lowest transversal mode in a 19-ion crystal in the presence of a pinning lattice on ion 13. (d) $J_z^{ij}/\omega_0$ calculated for trapping frequencies $\omega_X=\omega_Y=2\pi\times1$ MHz and $\omega_Z=2\pi\times3.5$ MHz and the pinned ion experiences an effective shallow trap along the $z$ axis with a frequency of $\omega_Z^{\prime}(1)=2\pi\times2.1$ MHz. (e) The eigenvectors of the two lowest transversal modes in a trapped 19-ion crystal with trapping frequencies $\omega_X=\omega_Y=2\pi\times1$ MHz and $\omega_Z=2\pi\times3.5$ MHz in the presence of pinning lattices or Rydberg excitations on ions 13 and 14. The modified trapping frequencies of ions 13  and 14 along the $z$ axis are $\omega_Z^{\prime}(1)=2\pi\times 2.1$ MHz and $\omega_Z^{\prime}(2)=2\pi\times 2.45$ MHz respectively. (f) $J_z^{ij}/\omega_0$ calculated for $\Omega_I^2/\Omega_I^1=1.1$ and $\delta=2\pi\times 20$kHz. Adapted from~\protect\citep{nath15a}.}
\label{fig:PLQresults}
\end{figure} 
Normalised phonon eigenvectors ${\bf b}_m=\{b_m^i\}$ and interaction strengths calculated for single and two plaquette structures are shown in Fig.~\ref{fig:PLQresults}.
The eigenvector $b_1^i$ for the lowest transversal mode (TM) of the ion crystal accounts for the oscillation of the central ion, where the ions in the outer hexagonal ring oscillate with same amplitude and are in-phase.
Note that in these calculations it has been assumed that the Raman lasers excite phonons only transiently and the ion-light field interaction remains inside the Lamb-Dicke regime~(Sec.~\ref{subsec:lamb_dicke}). 
The slight imperfections from the BFG plaquette interactions arise from the off-resonant coupling to higher TMs. 

\subsection{Structural phase transitions controlled by Rydberg excitations} \label{subsec:phasetransition}
Transitions from linear to zigzag configurations in an ion crystal are second-order phase transitions in the thermodynamic limit, and they can be controlled by the trapping frequencies and the inter-ion distance~\citep{fishman08a}. 
Experimental observations were enabled by adiabatically ramping trap parameters~\citep{pyka13a, ulm13a}, and defect formations due to the quenching of the system are well described by the Kibble-Zurek mechanism~\citep{delcampo10a, kibble76a, zurek85a}. 
  
An exciting alternative for such classical control is using Rydberg ions for triggering structural phase transitions with two advantages: the phase transition dynamics can be controlled fast and coherently as compared to the ion motion inside the trap. 
In addition, the strong coupling between the internal and external dynamics of trapped Rydberg manifest itself in collective effects.
Here, we consider the theory proposal in which Rydberg excitation of a single ion in a chain of three ions lead to a structural phase transition from a linear to zigzag configuration~\citep{li12a}.
A large kick due to the Rydberg-induced structural change affects the Franck-Condon factors that quantify couplings between different vibrational states of the two potential surfaces. 
\begin{figure}[h!]
\centering
\includegraphics[trim={0cm 3.5cm 0cm 1.8cm},clip, width=1.0 \textwidth]{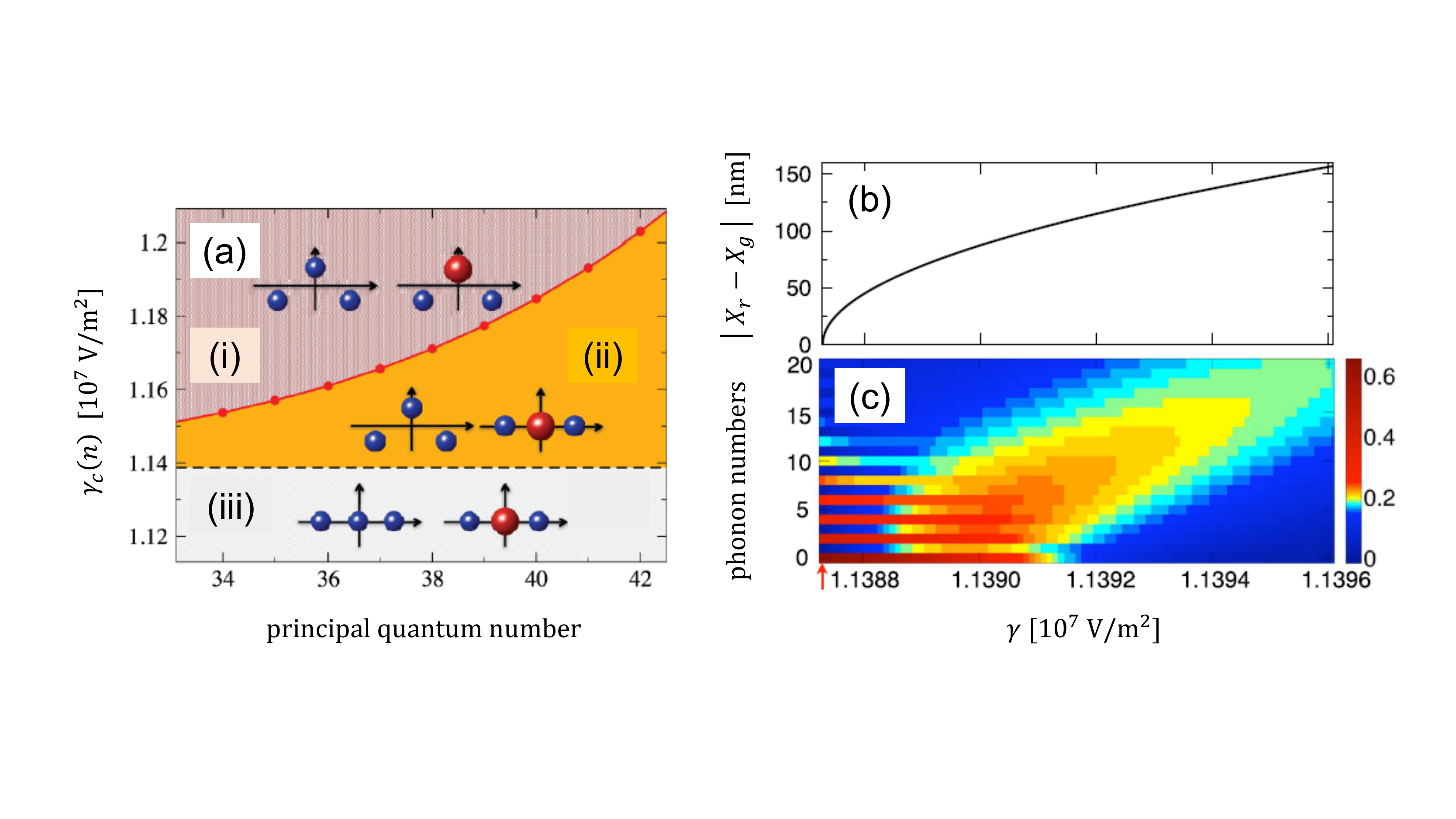}
\caption{(a) Structural configurations for a three-ion crystal with and without central ion excited to Rydberg state as indicated in red. The crystal configuration depends on the principal quantum number of the excited state and the gradient of the trapping static electric field~$\gamma$. Note that $\gamma_{\rm c}$ denotes a critical value at which linear to zig-zag phase transition occurs. The configurations in (i) and (iii) are identical, whereas they are state-dependent in (ii). (b) Displacement $\vert X_{r}-X_{g} \vert$ of the centre-of-mass potential for a ground state and excited
three-ion crystal as shown in (a) versus the trapping field gradient~$\gamma$. (c) Franck-Condon factors for transitions from the vibrational ground state of the ground state potential surface. Adapted from~\protect\citep{li12a}.}
\label{fig:structural_phase_li}
\end{figure} 


\section{Outlook}\label{sec:outlook}
Experiments with trapped Rydberg ions have shown potential of this new platform for quantum computing.
At the time of this publication, a two-ion gate using energy shifts between microwave-dressed Rydberg states was demonstrated with a gate time of about 0.7~$\mu$s and fidelity of 78\% (Sec.~\ref{sec:coh_spectroscopy:entangle}), which can in principle be improved by overcoming some technical issues~\citep{zhang20a}. 
Such a promising performance is to be compared with those achievable in trapped ions in their low-lying states and in neutral cold atomic gases in Rydberg states.
In state-of-the-art trapped ion experiments, two-ion gate fidelities better than 99.9\% with gate times of about 10~$\mu$s were demonstrated~\citep{ballance16a}, followed by progresses in multipartite entangling operations~\citep{kaufmann17a} and in reducing the gate time~\citep{schaefer18a, shapira18a, wongcampos17a}. In cold atomic gases, the Rydberg blockade effect has enabled two-atom gate fidelities of about 80\%  for a few~$\mu$s gate time~\citep{jau16a, maller15a} with the advantage of qubit number scalability in optical arrays~\citep{saffman16a}.  
These two systems are thus far more advanced for such an application as compared to Rydberg ions; however, novel methods in which the interplay between Coulomb and Rydberg interactions are used will possibly close this gap in the future, e.g., see the proposal in Sec.~\ref{subsec:fastgatewithEfield}.
 
Preparing trapped {\bf Rydberg ions in circular states}~\citep{hulet83a}, which feature significantly longer lifetimes and excessively large magnetic dipole moments, can be regarded as the next experimental milestone.
The long coherence time of neutral circular Rydberg states in the range of a few ms has enabled fundamental cavity quantum electrodynamics experiments~\citep{haroche06a}, and has been used to implement quantum sensors for magnetic and electric fields with unprecedented accuracy~\citep{dietsche19a, facon16a}. 
In such experiments, suppression of thermal phonon transitions can be achieved by using cryogenic traps, leading to even longer lifetimes of Rydberg circular states.

One obvious direction to explore is using Rydberg ions as {\bf sensitive detectors} for static and RF electric fields as well as MW fields.
In addition, using circular states allows for measurements in which the phase shift due to the interaction of ions with these external fields can be accumulated for an extremely long time. To improve signal-to-noise ratios of such sensors, dynamical decoupling pulse sequences or dressing techniques can be designed for a given measurement using single or an entangled pair of ions.
For instance, a single Rydberg ion inside a segmented micro trap can be used for mapping out a MW field emitted from a trap-integrated waveguide antenna. 
Micro-fabricated ion traps with segmented electrodes will allow for high resolution sensing by accurately characterising the position of ions.
In addition, sensing fields outside of a trapping apparatus is of technological relevance. For this pupose, a single-ion-nanoscope can be used~\citep{jacob16a} to extract ions from a Paul trap, steer them in the vicinity of a probe surface and focus the ion beam into to a few nm area. Exciting ions to Rydberg states before this extraction procedure results in enhancement in force detection, measured by ion-energy-loss spectroscopy.

{\bf Hybrid systems of trapped ions immersed in ultracold atomic gases}~\citep{tomza19a} have led to remarkable advances in collision studies~\citep{hall11a, ratschbacher12a}, controlled chemistry~\citep{sikorsky18a, willitsch17a, wolf17a} and quantum simulation~\citep{bissbort13a, hempel18a}.
In this way, a quantum interface may be built between ultracold atoms and trapped ions~\citep{secker17a}. It may even be possible to engineer repulsive interactions between atoms and ions, that prevent Langevin collisions between the two and thus prevent so-called micromotion induced heating in hybrid atom-ion systems~\citep{meir16a, rouse17a, secker16a}. Recently, interactions between ultracold Rydberg atoms and trapped ions have been observed in experiments~\citep{ewald19a, haze19a}.
A different hybrid ion-atom system directly produced from an ultracold ensemble of Rb atoms was employed for the observation of a single-ion-induced Rydberg blockade effect for Rydberg atoms~\citep{engel18a}. 
Quantum control at the single-particle level enabled by the Rydberg excitation of an atom or an ion in a hybrid system~\citep{wang19a} can be extended to trapped Rydberg ions superimposed by neutral atomic gases.

{\bf Rydberg excitation in cold trapped ions has been experimentally and theoretically explored over the last decade and opens doors to novel applications in quantum computing, quantum simulation and sensing}.

\hfill \break
\section*{Acknowledgements}
A. M. acknowledges the funding from the European Union's Horizon 2020 research and innovation programme under the Marie Sk{\l}odowska-Curie grant agreement No.~796866 (Rydion). Additional funding from DFG SPP 1929 ``Giant interactions in Rydberg Systems'' (GiRyd) and the ERA-Net QuantERA for ERyQSenS project is acknowledged. We thank G. Higgins for his contributions, R. Gerritsma for helpful comments, and B. Lekitsch and T. A. Sutherland for careful reading of the manuscript.

\newcommand{\newblock}{}

\bibliographystyle{abbrvnat.bst}
\bibliography{myreferences}
\end{document}